\documentclass[a4paper,noindent]{JHEP3}
\usepackage[dvips]{graphicx}
\usepackage{amsmath}
\usepackage{multirow}
\usepackage{feynarts}
\usepackage{array}
\usepackage{cite}
\usepackage{color}
\let\normalcolor\relax
\usepackage{afterpage}

\newcommand{\be}{\begin{eqnarray*}}
\newcommand{\ee}{\end{eqnarray*}}
\newcommand{\bee}{\begin{eqnarray}}
\newcommand{\eee}{\end{eqnarray}}
\renewcommand{\eqref}[1]{Eq.~(\ref{#1})}

\newcommand{\figref}[1]{Figure~\ref{#1}}

\newcommand{\SLASH}[2]{\makebox[#2ex][l]{$#1$}/}
\newcommand{\pslash}{\SLASH{p}{.2}}
\newcommand{\I}{{\rm i}}

\newcommand{\cf}{cf.\/}
\newcommand{\eg}{e.g.\/}
\newcommand{\ie}{i.e.\/}
\newcommand{\cm}{c.m.\/}

\newcommand{\alphas}{\alpha_s}

\def \mm{\mathcal{M}}
\def \Order{\mathcal{O}}

\def \msbar{\overline{\text{MS}}}
\def \drbar{\overline{\text{DR}}}
\def \os{\text{OS}}

\renewcommand{\Re}{\mbox{Re}}

\newcommand {\Born} {\sigma^{\rm Born}} 
\newcommand {\TreeEW} {\Delta\sigma^{\rm tree~EW}} 
\newcommand {\Treegy} {\Delta\sigma^{\rm g\gamma}}
\newcommand {\NLOEW} {\Delta\sigma^{\rm NLO~EW}} 
\newcommand {\EW} {\Delta\sigma^{\rm EW}}
\newcommand {\NLO} {\sigma^{\rm NLO}}
\newcommand {\dTreeEW}{\delta^{\rm tree~EW}}

\newcommand {\dNLOEW}{\delta^{\rm NLO~EW}}
\newcommand {\dNLO}{\delta^{\rm EW}}

\newcommand {\tilq}{\tilde{q}}

\newcommand {\bR}{\tilde{b}_{R}}
\newcommand {\bx}{\tilde{b}_{1}}
\newcommand {\by}{\tilde{b}_{2}}

\newcommand {\tR}{\tilde{t}_{R}}

\newcommand {\ty}{\tilde{t}_{2}}

\newcommand{\tb}{\tilde{t}_{\beta}}
\newcommand{\ba}{\tilde{b}_{\alpha}}
\newcommand{\bb}{\tilde{b}_{\beta}}

\newcommand{\qa}{\tilde{q}_{\alpha}}

\newcommand{\St}{\tilde{t}}
\newcommand{\Sb}{\tilde{b}}
\newcommand{\cha}{\tilde{\chi}^{\pm}}
\newcommand{\neu}{\tilde{\chi}^{0}}
\newcommand{\gluino}{\tilde{g}}
\newcommand{\MeV}{{\rm Me\kern -1pt V}}
\newcommand{\GeV}{{\rm Ge\kern -1pt V}}
\newcommand{\TeV}{{\rm Te\kern -1pt V}}

\newcommand {\SPA} {SPS1a${}^\prime$}

\newcommand{\trule}{\rule[-1.5mm]{0mm}{6mm}}

\title{Hadronic production of bottom-squark pairs with electroweak contributions}
\author{Jan Germer, Wolfgang Hollik  \\
Max-Planck-Institut f\"ur Physik, F\"ohringer Ring 6, D-80805 M\"unchen, Germany\\
Email: \email{germer@mppmu.mpg.de, hollik@mppmu.mpg.de}}
\author{Edoardo Mirabella\\
Institut de Physique Th\'eorique, CEA-Saclay, F-91191, Gif-sur-Yvette
cedex, France \\
Email: \email{edoardo.mirabella@cea.fr}}

\abstract{ We present the complete computation of the tree-level and
  the next-to-leading order electroweak contributions to
  bottom-squark pair production at the LHC.  The computation
  is performed within the minimal supersymmetric extension of the
  Standard Model.  We discuss the numerical impact of these
  contributions in several supersymmetric scenarios.}

\preprint{MPP-2011-157\\IPhT-t11/021}

\begin{document}

\section{Introduction}
\label{sect_Intro}
Supersymmetry (SUSY), is a well motivated Beyond the Standard Model
(BSM) scenario.  Electroweak precision data indicates that SUSY should
be realized at the \TeV{} scale or below. If this is the case, it will
be accessible to direct experimental measurements at the Large Hadron
Collider (LHC) through the production of SUSY particles.  Monte Carlo
simulations have shown the possibility of the discovery of \TeV-scale
SUSY with $1$~fb$^{-1}$ of integrated
luminosity~\cite{0954-3899-34-6-S01,:2008zzm}.  First measurements
involving supersymmetry-sensitive variables have been performed by
both ATLAS~\cite{Atlas1, Atlas2, Atlas3, Atlas4} and CMS~\cite{CMS1}
collaborations.
\medskip

At the LHC, colored particles like the SUSY partners of quarks and
gluons, i.e.\ squarks and gluinos, will be copiously produced.
Theoretically, these processes are extensively studied within the
Minimal Supersymmetric extension of the Standard Model (MSSM).  The
leading-order (LO) contributions, of $\Order(\alpha_s)$, 
have been known for a  long
time~\cite{Kane:1982hw,Harrison:1982yi,Reya:1984yz,Dawson:1983fw,Baer:1985xz}.
The next-to leading order (NLO) QCD corrections have been
computed~\cite{Beenakker:1996ch, Beenakker:1997ut} and implemented in
the public code \verb+PROSPINO+~\cite{Beenakker:1996ed}. They affect
the LO predictions substantially and they reduce the scale dependence
considerably.  More recent is the estimation of the logarithmically
enhanced next-to-next-to-leading order (NNLO) QCD contributions to
squark hadroproduction, the resummation of the QCD Sudakov logarithms
at the next-to-leading-logarithmic (NLL) accuracy, and the resummation
of the leading Coulomb corrections~\cite{Langenfeld:2009eg,
  Kulesza:2008jb,Kulesza:2009kq,Beneke:2009nr,Beenakker:2009ha,Beenakker:2010nq,Beneke:2010da}.
Their contribution amounts up to 10\% for squark and gluino masses of
the order of $1$~\TeV{}. They further stabilize the prediction against
scale variation.
 
The electroweak (EW) contributions have been computed for several
processes producing colored SUSY particles. They exhibit an extremely
rich and complicate pattern, in particular when the EW contributions
appear already at tree level.  The tree-level EW contributions are of
$\Order(\alpha^2+ \alpha_s\alpha)$. At the parton level, they arise
from $q\bar q$-annihilation, $q q$-scattering or photon-induced
processes, depending on the final state considered.  They are
known~\cite{Hollik:2007wf,Bornhauser:2007bf,Hollik:2008vm,Bornhauser:2009ru,Arhrib:2009sb,Germer:2010vn}
and they can increase the LO cross section by up to 20\%.  The impact
of these contributions in the context of non-minimal flavor violation
and explicit CP violation has been investigated as
well~\cite{Bozzi:2005sy,Alan:2007rp}. The NLO EW corrections
contribute at $\Order(\alpha^2_s \alpha)$ and have been computed for
stop--anti-stop~\cite{Hollik:2007wf,Beccaria:2008mi},
squark--(anti-)squark~\cite{Hollik:2008yi,Germer:2010vn},
gluino--squark~\cite{Hollik:2008vm}, and
gluino--gluino~\cite{Mirabella:2009ap} production.  Their size is
comparable with that of the tree-level EW and NNLO QCD contributions,
and their impact strongly depend on the SUSY scenario considered.
\medskip

The production of third generation squarks is special. The
non-negligible mixing in the stop and sbottom sector could lead to
relatively low masses for the lightest bottom and top squarks,
favoring their direct production at the LHC.  Moreover $b$-tagging
makes bottom- and top-squark production experimentally distinguishable
from the production of the squarks of the first two
generations~\cite{Paige:2003sh,Chiorboli:2004jk,Kawagoe:2004rz}.  A
dedicated analysis looking for third generation squark production at
the LHC is already available~\cite{Atlas4}.  This kind of searches are
particularly important in SUSY scenarios such as the ATLAS benchmark
scenario $SU6$, where inclusive searches with jets, missing transverse
energy, and leptons are problematic~\cite{:2008zzm}.

In this paper we focus on the hadronic production of bottom-squark
pairs
\begin{equation}
  \label{eq:tree}
  P\, P \rightarrow  \tilde{b}_\alpha  \tilde{b}^\ast_\beta,\; 
 \tilde{b}_\alpha  \tilde{b}_\beta,\;
  \tilde{b}^\ast_\alpha  \tilde{b}^\ast_\beta, \qquad  \alpha,\beta \in \{1,2\}.
\end{equation}
In particular we present the first complete computation of the NLO~EW
corrections to diagonal sbottom--anti-sbottom pair production,
\begin{equation}
  \label{eq:tree1}
  P\, P \rightarrow \ba \, \ba^\ast,  \qquad  \alpha\in \{1,2\}.
\end{equation}
%
%
The contribution of the remaining processes is small (\cf\ Section 5), hence we
will not include them in our discussion on the EW corrections.
The process~(\ref{eq:tree1}) exhibits specific features like the
mixing between left- and right-handed b-squarks, the renormalization
of the sbottom sector~\cite{Heinemeyer:2004xw,Heinemeyer:2010mm}, the
non-negligible Higgs-boson contributions, and the enhanced Yukawa
couplings for large values of $\tan \beta$ with the related need of
resummation~\cite{Carena:1999py}. These features make the computations
of the electroweak contributions to the processes~(\ref{eq:tree1})
substantially different from those for
squark--anti-squark~\cite{Hollik:2008yi} and
stop--anti-stop~\cite{Hollik:2007wf} production, and justify 
a specific investigation, which is reported in this paper.
\medskip

The outline of the paper is as follows. In Section~\ref{sect_tree} we
summarize the tree-level contributions to the
processes~(\ref{eq:tree1}). Section~\ref{sect_NLO} describes the
various partonic processes contributing at $\Order(\alpha_s^2\alpha)$
and the strategy of the calculation.  The numerical impact of the NLO
EW contributions at the LHC with $\sqrt{S}=14$~\TeV{} and
$\sqrt{S}=7$~\TeV{} is presented in Section~\ref{sect_results}.  In
Section~\ref{sect_other} we discuss the numerical impact of the
subleading bottom-squark pair production processes.  The Feynman
diagrams and the technical details of the renormalization of the
sbottom sector are collected in the Appendix.


\section{Tree-level cross section}
\label{sect_tree}
In this section we describe the tree-level contributions to the
process~(\ref{eq:tree1}), which are of order $\Order(\alpha_s^2)$,
$\Order(\alpha_s \alpha)$, and $\Order(\alpha^2)$.  We will
conventionally denote the cross section (amplitude) of a partonic
process $X$ at a given order $\Order(\alphas^a\alpha^b)$ as ${\rm d} \hat
\sigma^{a\,,b}_X$ ($\mm^{a,\,b}_{X}$).  The parton luminosities are
defined as
\begin{equation}
  \frac{{\rm d}L_{ij}}{{\rm d}\tau}(\tau)=  \frac{1}{1+\delta_{ij}}
  \int_\tau^1   \frac{{\rm d}x}{x} 
  \left[ f^A_i\left(\frac{\tau}{x},\mu_F\right)\, f^B_{j}(x, \mu_F) + 
    f^A_j(x, \mu_F)\,  f^B_{i}\left(\frac{\tau}{x}, \mu_F\right)\right],
\label{eq:Lumi}
\end{equation}
where $f^{A}_i(x, \mu_F)$ is the parton distribution function (PDF)
of the parton~$i$ inside the hadron~$A$.

\subsection{Tree-level QCD contributions}
The leading-order cross section, of the order $\Order(\alpha_s^2)$, is
given by
\begin{align}
\begin{split}
  {\rm d}\sigma^{\mbox{\tiny LO QCD}}_{PP \to  \ba\ba^\ast}(S) &=
   \int_{\tau_0}^1 {\rm d}\tau \; \frac{{\rm d}L_{gg}}{{\rm d}\tau}
  {\rm d} \hat  \sigma^{2,\, 0}_{gg \to \ba\ba^\ast}(\hat s) +
   \sum_{q} \;
  \int_{\tau_0}^1 {\rm d}\tau \; \frac{{\rm d}L_{q \bar q}}{{\rm d}\tau}
  {\rm d} \hat \sigma^{2,\, 0}_{q \bar q \to \ba\ba^\ast}(\hat s) \\
  &+
   \int_{\tau_0}^1 {\rm d}\tau \; \frac{{\rm d}L_{b \bar b}}{{\rm d}\tau}
  {\rm d} \hat \sigma^{2,\, 0}_{b \bar b \to \ba\ba^\ast}(\hat s),
\end{split}
\end{align}
where $\tau_0= 4 m_{\ba}^2/S$ is the production threshold. $S$ and
$\hat s=\tau S$ are the squared center-of-mass (\cm) energies of the
hadronic and partonic processes, respectively. The sum runs over
$q=u,\,d,\,c,\,s$. The three classes of partonic processes
contributing are
\begin{subequations}
\begin{align}
  \label{eq:LO1}
    g(p_1) \, g(p_2) &\rightarrow \ba(p_3) \, \ba^\ast(p_4),\\ 
\label{eq:LO2}
    q(p_1)  \, \bar{q}(p_2) &\rightarrow \ba(p_3) \, \ba^\ast(p_4),\\  
\label{eq:LO3}
    b(p_1)  \,  \bar{b}(p_2) &\rightarrow \ba(p_3) \, \ba^\ast(p_4).  
\end{align}
\end{subequations}
The corresponding partonic cross section can be obtained from the
Feynman diagrams in \figref{fig:feynman_tree}. In terms of the
Mandelstam variables,
\begin{equation}
  \label{eq:Mandelstam}
  \hat s = (p_1 + p_2)^2, \quad  \hat t = (p_1 - p_3)^2, \quad \hat u
  = (p_1 - p_4)^2,
\end{equation}
the differential partonic cross section for a given subprocess $\xi
\xi^\prime\to \ba \ba^\ast$ can be written as
\begin{equation}
    {\rm d}\hat{\sigma}^{2,\,0}_{\xi\xi^\prime \to \ba\ba^\ast}(\hat s) =
    \overline{\sum} \Bigl\lvert{\mm}^{1,\,0}_{\xi\xi^\prime \to
      \ba\ba^\ast} \Bigr\rvert^2 \frac{{\rm d}\hat{t}}{16 \pi \hat s^2},
\end{equation}
with the squared lowest order matrix element averaged (summed)
over initial (final) state spin and color.
\subsection{Tree-level EW contributions}
The tree-level electroweak (EW) contributions, which are of the order
$\Order(\alpha_s \alpha)$ and $\Order(\alpha^2)$, read as follows,
\begin{align}
\begin{split}
\label{eq:LOEW}
  {\rm d}\sigma^{\mbox{\tiny LO EW}}_{PP \to  \ba\ba^\ast}(S) &=
   \int_{\tau_0}^1 {\rm d}\tau \; \frac{{\rm d}L_{g\gamma}}{{\rm d}\tau}
  {\rm d} \hat \sigma^{1,\, 1}_{g\gamma \to \ba\ba^\ast}(\hat s) +
   \sum_{q} \;
  \int_{\tau_0}^1 {\rm d}\tau \; \frac{{\rm d}L_{q \bar q}}{{\rm d}\tau}
  {\rm d} \hat \sigma^{0,\, 2}_{q \bar q \to \ba\ba^\ast}(\hat s) \\
  &+
   \int_{\tau_0}^1 {\rm d}\tau \; \frac{{\rm d}L_{b \bar b}}{{\rm d}\tau} \left [ 
    {\rm d} \hat \sigma^{1,\, 1}_{b \bar b \to \ba\ba^\ast}(\hat s) + 
  {\rm d} \hat \sigma^{0,\, 2}_{b \bar b \to \ba\ba^\ast}(\hat s) 
  \right ].
\end{split}
\end{align}
The contributions of $\Order(\alpha^2)$ arise from the
processes~(\ref{eq:LO2}) and~(\ref{eq:LO3}). The partonic cross
sections,
\begin{equation}
  \label{eq:TreeEW1}
  {\rm d}\hat{\sigma}^{0,\,2}_{q\bar{q} \to \ba\ba^\ast}(\hat s) =
  \overline{\sum} \Bigl\lvert{\mm}^{0,\,1}_{q \bar q \to \ba\ba^\ast}
  \Bigr\rvert^2 \frac{{\rm d}\hat{t}}{16 \pi \hat s^2},  ~~~
  {\rm d}\hat{\sigma}^{0,\,2}_{b\bar{b} \to \tilde b_a\tilde b^*_a}(\hat s)
  = \overline{\sum} \Bigl\lvert{\mm}^{0,\,1}_{b \bar b \to
    \ba\ba^\ast} \Bigr\rvert^2 \frac{{\rm d}\hat{t}}{16 \pi \hat s^2},    
\end{equation}
are obtained from the diagrams in \figref{fig:feynman_tree}.
In the case of process~(\ref{eq:LO3}), the diagrams with $t$-channel
gluino and neutralino exchange further allow for a non-vanishing
QCD--EW interference term of $\Order(\alpha_s \alpha)$,
\begin{align}
  \label{eq:TreeEW2}
    {\rm d}\hat{\sigma}^{1,\,1}_{b \bar b \to \ba\ba^\ast}(\hat s) &= 2\, \overline{\sum} \Re \left \{ 
    {\mm}^{1,\,0}_{b \bar b \to \ba\ba^\ast} \left({\mm}^{0,\,1}_{b \bar b \to \ba\ba^\ast}\right)^\ast \right \}
  \frac{{\rm d}\hat{t}}{16 \pi \hat s^2} \,.
\end{align}
We approximate the CKM matrix $V$ with the unity matrix, thus we
consistently neglect the diagram depicted in
\figref{fig:feynman_tree}(d). The latter diagram, together with the
diagrams in \figref{fig:feynman_tree}(b), gives rise to
$\Order(\alpha_s\alpha+\alpha^2)$ contributions. These contributions
are at least quadratic in $|V_{cb}| \approx 10 \cdot |V_{ub}|\approx
4\cdot 10^{-2}$ and numerically negligible (see
Section~\ref{sec:parameterscan}).  \medskip

The first $\Order(\alpha_s \alpha)$ contribution in \eqref{eq:LOEW}
arises from the photon--gluon induced process
\begin{equation}
  \label{eq:photonind}
    g(p_1) \, \gamma(p_2) \rightarrow \ba(p_3) \, \ba^\ast(p_4).
\end{equation}
The corresponding partonic cross section can be obtained from the
diagrams in \figref{fig:feynman_ggamma} and reads as follows
\begin{align}
    {\rm d}\hat{\sigma}^{1,\,1}_{g \gamma \to \ba\ba^\ast}(\hat s) &= \overline{\sum} \left |
    {\mm}^{1/2, 1/2}_{g \gamma  \to \ba\ba^\ast}  \right |^2 \frac{{\rm d}\hat{t}}{16 \pi \hat s^2} \, .   
\end{align}

\section{Next-to-leading order  EW contributions}
\label{sect_NLO}
In this section we list the NLO EW corrections to the
process~(\ref{eq:tree1}). These contributions are of the order
$\Order(\alpha^2_s \alpha)$ and arise from virtual corrections and
bremsstrahlung processes. Using an obvious notation, the corresponding
contributions to the total cross section read as follows

\begin{align}
  {\rm d}\sigma^{\mbox{\tiny NLO EW}}_{PP \to \ba\ba^\ast}(S) &=
   \int_{\tau_0}^1 {\rm d}\tau \;    \frac{{\rm d}L_{gg}}{{\rm d}\tau} \, \left [
  {\rm d} \hat \sigma^{2,\, 1}_{gg \to \ba\ba^\ast}(\hat s) +
   {\rm d} \hat \sigma^{2,\, 1}_{gg \to \ba\ba^\ast\gamma}(\hat s) 
  \right ]  \nonumber \\ 
  &+  \sum_{q} \;
  \int_{\tau_0}^1 {\rm d}\tau \; \frac{{\rm d}L_{q \bar q}}{{\rm d}\tau} \, \left [
  {\rm d} \hat \sigma^{2,\, 1}_{q \bar q \to \ba\ba^\ast}(\hat s) +
  {\rm d} \hat \sigma^{2,\, 1}_{q \bar q \to \ba\ba^\ast \gamma}(\hat s) + 
  {\rm d} \hat \sigma^{2,\, 1}_{q \bar q \to \ba\ba^\ast g}(\hat s)  \right ] \nonumber \\
  &+ \sum_{q} \;
   \int_{\tau_0}^1 {\rm d}\tau \; \left [  \frac{{\rm d}L_{q g}}{{\rm d}\tau}  
    {\rm d} \hat \sigma^{2,\, 1}_{g q  \to \ba\ba^\ast q}(\hat s) + 
     \frac{{\rm d}L_{\bar q g}}{{\rm d}\tau}  
  {\rm d} \hat \sigma^{2,\, 1}_{g \bar q \to \ba\ba^\ast \bar q}(\hat s) 
  \right ].
  \label{eq:NLOEW}
\end{align}
We do not consider the contributions arising from the bremsstrahlung processes
\begin{equation}
\gamma(p_1)\, q(p_2) \to  \ba(p_3) \, \ba^\ast(p_4) \,  q(p_5),~~~
\gamma(p_1)\, \bar q(p_2) \to  \ba(p_3) \, \ba^\ast(p_4) \,  \bar q(p_5).
\end{equation} 
As already pointed out in Ref.~\cite{Hollik:2008yi}, they are
suppressed because of the $\Order(\alpha)$ suppression of the photon
PDF inside the proton. Moreover, these processes are further
suppressed by an additional factor $\alpha_s$ with respect to the
process~(\ref{eq:photonind}) and thus negligible.  The
$\Order(\alpha_s^2 \alpha)$ contributions of the partonic processes
with a bottom quark in the initial state are neglected as well. The
reason is twofold. First of all these contributions are suppressed by
the bottom PDF with respect to the contributions in \eqref{eq:NLOEW}.
In addition they have an additional factor $\alpha_s$ with respect to
the $\Order(\alpha_s \alpha + \alpha^2)$ contributions of the
process~(\ref{eq:LO3}), which turn out to be small (\cf\
Section~\ref{sec:parameterscan}).

The amplitudes are generated and algebraically simplified with support
of \verb+FeynArts+~\cite{Hahn:2000kx,Hahn:2006qw} and
\verb+FormCalc+~\cite{Hahn:2006qw,Hahn:2001rv}, while the numerical
evaluation of the one-loop integrals has been performed using
\verb+LoopTools+~\cite{Hahn:2001rv}. Infrared (IR) singularities are
regularized giving a small mass $\lambda_\gamma$ and $\lambda_g$ to
the photon and to the gluon, respectively. The mass of the light
quarks is kept in order to regularize the collinear singularities.

\subsection{Virtual corrections}
The $\Order(\alpha_s^2 \alpha)$ contributions to the partonic
process~(\ref{eq:LO1}) are given by
\begin{align}
  \label{eq:LoopGG}
    {\rm d}\hat{\sigma}^{2,\,1}_{g g \to \ba\ba^\ast}(\hat s) &= 2\, \overline{\sum} \Re \left \{ 
    {\mm}^{1,\,0}_{g  g \to \ba\ba^\ast} \left({\mm}^{1,\,1}_{g  g \to \ba\ba^\ast}\right)^\ast \right \}
  \frac{{\rm d}\hat{t}}{16 \pi \hat s^2} \,,
\end{align}
where ${\mm}^{1,\,0}$ is the tree-level amplitude while
${\mm}^{1,\,1}$ is the one-loop amplitude obtained from the diagrams
depicted in \figref{fig:feynman_ggvirt}. The virtual corrections to the
process~(\ref{eq:LO2}) read as follows
\begin{align}
  \label{eq:LoopQQ}
    {\rm d}\hat{\sigma}^{2,\,1}_{q \bar q \to \ba\ba^\ast}(\hat s) &= 2\, \overline{\sum} \Re \left \{ 
    {\mm}^{1,\,0}_{q \bar q \to \ba\ba^\ast} \left({\mm}^{1,\,1}_{q \bar q \to \ba\ba^\ast}\right)^\ast  +  {\mm}^{0,\,1}_{q \bar q \to \ba\ba^\ast} \left({\mm}^{2,\,0}_{q\bar q \to \ba\ba^\ast}\right)^\ast
    \right \}
  \frac{{\rm d}\hat{t}}{16 \pi \hat s^2}. 
\end{align}
${\mm}^{0,\,1}$ and ${\mm}^{1,\,0}$ are the tree-level EW and the
tree-level QCD amplitudes, respectively. ${\mm}^{1,\,1}$ is the
one-loop amplitude obtained from the EW insertions to the
leading-order diagrams and from the QCD corrections to the tree-level
EW diagrams (\figref{fig:feynman_qqvirtEW}).  ${\mm}^{2,\,0}$ is the
amplitude corresponding to the QCD box diagrams depicted in
\figref{fig:feynman_qqvirtQCD}.

In order to cancel the UV divergences we need the $\Order(\alpha)$
renormalization of the wavefunction of the light quarks and of the
sbottom sector.  The field renormalization constants are fixed in the
on-shell scheme, in analogy to that described in
Ref.~\cite{Hollik:2008yi}. The renormalization of the sbottom sector
has to be performed together with that of the stop sector.  In order
to avoid numerical instabilities and artificially big contributions
from the counterterms, care has to be taken in choosing the
renormalization scheme~\cite{Heinemeyer:2004xw, Heinemeyer:2010mm}. We
use the ``$\overline{\mbox{DR}}$ bottom-quark mass'' scheme introduced
in Ref.~\cite{Heinemeyer:2004xw}.  Since in particular regions of the
MSSM parameter space this scheme can give rise to numerical
instabilities, we have explicitly checked its reliability in the SUSY
scenarios considered in this paper (\cf\
Section~\ref{sect_results}). The explicit expression of the
renormalization constants in the ``$\overline{\mbox{DR}}$ bottom-quark
mass'' scheme are collected in Appendix~\ref{sec:stopsbot}.

\subsection{Real corrections}
The $\Order(\alpha^2_s \alpha)$ contributions to the partonic
processes with a photon in the final state,
\begin{subequations}
\begin{align}
\label{eq:gamma1}
    g(p_1) \, g(p_2) &\rightarrow \ba(p_3) \, \ba^\ast(p_4) \, \gamma(p_5),\\ 
\label{eq:gamma2}
    q(p_1)  \, \bar{q}(p_2) &\rightarrow \ba(p_3) \, \ba^\ast(p_4) \, \gamma(p_5),   
\end{align}
\end{subequations}
are obtained from the tree-level diagrams in
\figref{fig:feynman_ggRE}. The phase space integration is divergent in
the soft photon region, \ie\ in the region $p^0_5 \to 0$. In the
case of the process~(\ref{eq:gamma2}) further singularities arise in
the collinear region, \ie\  $p_{1,2} \cdot p_5 \to 0$.  IR and
collinear singularities are treated using the phase-space slicing
method. The description of the method and the relevant formulae are
collected in Ref.~\cite{Germer:2010vn}.

The gluon bremsstrahlung process,
\begin{align}
\label{eq:gluon1}
    q(p_1)  \, \bar{q}(p_2) &\rightarrow \ba(p_3) \, \ba^\ast(p_4) \, g(p_5),   
\end{align}
contributes at $\Order(\alpha^2_s \alpha)$ via the interference of
QCD-based and EW-based Feynman diagrams depicted in
\figref{fig:feynman_gluonRE}. IR singularities of gluonic origin are
treated in close analogy to the photonic case. Color correlations are
taken into account using the formulae collected in Appendix~B of
Ref.~\cite{Germer:2010vn}.  Due to the color structure, the
interference term of a QCD-based and an EW-based diagram vanishes if
both gluons are emitted from an initial-state or a final-state
particle.

Real quark radiation contributes at $\Order(\alpha^2_s \alpha)$  as well,
\begin{subequations}
\begin{align}
  \label{eq:quark1}
    g(p_1) \, q(p_2) &\rightarrow \ba(p_3) \, \ba^\ast(p_4) \, q(p_5),\\ 
\label{eq:quark2}
    g(p_1)  \, \bar{q}(p_2) &\rightarrow \ba(p_3) \, \ba^\ast(p_4) \, \bar q(p_5).  
\end{align}
\end{subequations}
This IR- and collinear-finite set is given by the interference of QCD
and EW tree-level diagrams (\cf\ \figref{fig:feynman_quarkRE}). Only
the interference from initial-state and final-state radiation
contributes.

The IR singularities arising in the $gg$ channel cancel in the sum of
virtual corrections, process~(\ref{eq:LoopGG}), and real photon
radiation~(\ref{eq:gamma1}). In the $q\bar{q}$ channel the sum of
the virtual corrections~(\ref{eq:LoopQQ}) and of the contributions
of real photon radiation~(\ref{eq:gamma2}) and real gluon
radiation~(\ref{eq:gluon1}) is IR finite. 
This sum is affected by universal collinear singularities of photonic
origin that can be absorbed in the PDFs. This can be achieved by means
of the following substitution~\cite{Baur:1998kt},
\begin{equation}
  f_q(x,\mu_F)\, \to  \, f_q(x,\mu_F) \left( 1-\frac{\alpha e_q^2}{\pi}  \kappa_{v+s}  \right) 
  -  \frac{\alpha e_q^2}{2\pi}  \, \int_{x}^{1-\delta_s} \frac{{\rm d}z}{z}
  f_q\left(\frac{x}{z},\mu_F\right) \kappa_c(z)  
\label{eq_PDFredef}.
\end{equation}
$e_q$ is the electric charge of quark $q$ expressed in units of the
positron charge, while
\begin{align}
\begin{split}
  \kappa_{v+s} &= 1 -\ln\delta_s -\ln^2\delta_s
  +\left(\ln\delta_s+\frac{3}{4}\right) 
  \ln\left(\frac{\mu_F^2}{m_{q}^2}\right)+
  \frac{1}{4} \left (
  9 +\frac{2\pi^2}{3} +3\ln\delta_s -2\ln^2\delta_s
  \right ),
\\
  \kappa_c(z) &= P_{qq}(z)\, \ln\left( \frac{\mu_F^2}{m_q^2}
  \frac{1}{(1-z)^2} -1 \right)- \left [  P_{qq}(z) \,\ln\left( \frac{1-z}{z}\right) -\frac{3}{2}
  \frac{1}{1-z} + 2z +3 \right  ],
\end{split}
\end{align}
with the splitting function $P_{qq}(z) = (1+z^2)/(1-z)$. The
factorization is performed in the DIS scheme. The replacement of the
PDFs in \eqref{eq:Lumi} gives further contributions of
$\mathcal{O}(\alphas^2\alpha)$ to the total cross section. As already
mentioned, they cancel the collinear singularities affecting the
$\Order(\alpha_s^2 \alpha)$ contributions~(\ref{eq:NLOEW}).

\subsection[Resummation in the $b / \tilde b$ sector]
{Resummation in the $\mathbf{b / \tilde b}$ sector}
The Higgs sector in the MSSM corresponds to a type-II two-Higgs
doublet model, \ie\ the down-type quarks couple to $H_1$ and the
up-type quarks to $H_2$. After spontaneous symmetry breaking, the up-
(down-)type quarks get their mass by their coupling to the vacuum
expectation value $v_2$ ($v_1$) of $H_2$ ($H_1$). At tree-level, the
bottom-quark mass $m_b$ is related to the $H_1 b \bar b$ Yukawa
coupling $\lambda_b$ via
\begin{equation}
m_b = \lambda_b v_1.
\label{eq:treeMBLB}
\end{equation}
Radiative corrections induces and effective $H_2 b \bar b$ coupling
that can significantly alter the tree-level
relation~(\ref{eq:treeMBLB})~\cite{Hall:1993gn,Hempfling:1993kv,Carena:1994bv,Pierce:1996zz,Carena:1999py}.
This higher-order contributions do not decouple at low energies and
are enhanced by a factor $\tan \beta =v_2 /v_1$. As shown in
Ref.~\cite{Carena:1999py}, the leading $\tan \beta$ enhanced terms can
be resummed by using an appropriate effective bottom-quark Yukawa
coupling.  We follow Ref.~\cite{Heinemeyer:2004xw} and use an
effective Yukawa coupling defined as follows,
\begin{equation}
  \label{eq:eff1}
\bar \lambda_b  = \frac{1}{v_1}\frac{m_b^{\drbar}(\mu_R) + m_b \Delta m_b}{1+\Delta m_b} \equiv \frac{m_b^{\drbar,\text{eff}}}{v_1}  \, ,
\end{equation}
where $\Delta m_b$ is given by
\begin{align}
  \label{eq:eff2}
  \Delta m_b &=  \frac{2\alphas}{3\pi} M_{\gluino} \mu \tan\beta\;
  I(m_{\Sb_1}, m_{\Sb_2}, m_{\gluino}) + \frac{\lambda_t^2}{16\pi^2} \mu A_t \tan \beta\;
  I(m_{\St_1}, m_{\St_2},\mu) \nonumber \\
&-\frac{g^2}{16\pi^2} \mu M_2 \tan\beta
  \bigg [
    \cos^2\theta_{\St} I(m_{\St_1},M_2,\mu) + \sin^2 \theta_{\St_2}
    I(m_{\St_2},M_2,\mu) \nonumber\\
  &+ \frac{1}{2} \cos^2\theta_{\Sb}
    I(m_{\Sb_1},M_2,\mu)+\frac{1}{2} \sin^2 \theta_{\Sb_2}
    I(m_{\Sb_2},M_2,\mu) \,  \bigg ],  \\  
  I(a,b,c) &= \frac{1}{(a^2-b^2)(b^2-c^2)(a^2-c^2)}
  \left[ a^2b^2 \log\frac{a^2}{b^2} + b^2c^2 \log\frac{b^2}{c^2} +
    c^2a^2 \log\frac{c^2}{a^2} \right]. \nonumber 
\end{align}
Large logarithms from the running of the Yukawa coupling $\lambda_b$
at the renormalization scale $\mu_R$ are resummed using the
$\overline{\mbox{DR}}$ bottom-quark mass,
 \begin{equation}
 m_b^{\drbar}(\mu_R) =m_b^{\text{OS}} +\frac{m_b}{2} \left (  \Sigma^{\text{fin.}}_{bL}(m_b)+ 
 \Sigma^{\text{fin.}}_{bR}(m_b) + 2  \Sigma^{\text{fin.}}_{bS}(m_b) \right ).  
\end{equation}
$m_b^{\text{OS}}$ is the on-shell bottom-quark mass defined according to
\begin{equation}
m_b^{\text{OS}} = m_b^{\msbar}(m_Z) b^{\text{shift}},
\qquad
 b^{\text{shift}} = 1+ \frac{\alpha_s}{\pi} \left(\frac{4}{3} -\log \frac{(m_b^{\msbar})^2}{m_Z^2} \right).
\end{equation}
$\Sigma^{\text{fin.}}$ is the finite part of the scalar self-energies
defined according to the Lorentz decomposition~(\ref{eq:LorDec}) of
the Appendix~\ref{sec:stopsbot}.  The term proportional to $\Delta
m_b$ in the numerator of \eqref{eq:eff1} has to be inserted to avoid
double counting of the one-loop contributions of the resummed terms.

Further $\tan\beta$ enhancement effects arise from three-point
functions involving Higgs--bottom vertices. The $\tan\beta$-enhanced
terms can be taken into account by modifying the $\mathcal{H}b \bar b$
coupling $g_{\mathcal{H}b \bar b}$.\footnote{$\mathcal{H}$ stands for
  any of the neutral Higgs and Goldstone bosons, \ie\
  $\mathcal{H}=h^0,H^0,A^0,G^0$.} The combined effect of the
resummation in the relation between $\lambda_b$ and $m_b$ and of the
resummation in the Higgs--bottom vertices is accounted for by
performing the following substitutions,
\begin{subequations}
 \label{eq:eff5}
\begin{align}
  g_{h^0bb} &\rightarrow g_{h^0bb}\big|_{\lambda_b \rightarrow
    \bar\lambda_b} \left(1-\frac{\Delta
      m_b}{\tan\beta\tan\alpha}\right) , 
  &   g_{A^0bb} &\rightarrow g_{A^0bb}\big|_{\lambda_b \rightarrow \bar\lambda_b} \left(1-\frac{\Delta m_b}{\tan\beta^2}\right ) ,
  \label{eq:eff5a}\\
  g_{H^0bb} &\rightarrow g_{H^0bb}\big|_{\lambda_b \rightarrow
    \bar\lambda_b} \left(1+\Delta
    m_b\frac{\tan\alpha}{\tan\beta}\right) ,
  & g_{G^0 bb} &\rightarrow g_{G^0 bb}\,.
  \label{eq:eff5b}
\end{align}
\end{subequations}
The coupling involving  the Goldstone boson $G^0$ is  
not modified since the contribution from the vertex corrections exactly
compensates the contribution of the bottom-Yukawa coupling
resummation.

\section{Numerical results}
\label{sect_results}
In this section we perform a detailed numerical analysis for diagonal
sbottom--pair production at NLO EW. We stick to the notation introduced in 
Ref.~\cite{Germer:2010vn}. The leading order cross section, the tree-level EW and 
the NLO EW contributions to the cross section are labeled by
\begin{equation}
\Born = \sigma^{2,\,0}, \qquad
\TreeEW = (\sigma^{1,\,1}+\sigma^{0,\,2}), \qquad
\NLOEW = \sigma^{2,\,1},
\end{equation}
 respectively. $\EW = \TreeEW + \NLOEW$ will be referred to as the EW contribution.
The total sum of the LO cross section with the EW contributions is denoted by
$\NLO=\Born+\EW$. Relative EW contributions are defined by
\begin{align}
\label{eq:delta}
\dTreeEW = \TreeEW / \Born, \qquad
\dNLOEW = \NLOEW / \Born, \qquad
\dNLO = \EW /\Born.
\end{align}
In distributions $\delta$ denotes the relative EW contribution defined
as $\delta = (\mathcal{O}_{\rm NLO}-\mathcal{O}_{\rm Born}) /
\mathcal{O}_{\rm Born}$, where $\mathcal{O}$ is a generic observable
and $\mathcal{O}_{\rm NLO}$ is the sum of the Born and the EW
contributions.

\subsection{Input parameters}
The Standard Model input parameters are chosen in correspondence with
\cite{Nakamura:2010zzi,CDFtopmass},
\begin{align}
M_Z & = 91.1876~{\GeV}, & M_W &= 80.399~\GeV, \nonumber\\
\alpha^{-1} & = 137.036, & \alphas(M_Z) &= 0.119,\\
m_t & = 170.9~{\GeV}, & m_b^{\msbar}(m_Z) &= 2.94~\GeV. \nonumber
\end{align}
The strong coupling constant $\alphas$ has been defined in the
$\msbar$ scheme using the two-loop renormalization group equation with
five active flavors.

For the numerical analysis we consider the mSUGRA scenarios \SPA{} and
SPS4. The first one is a ``typical'' SUSY scenario proposed by the SPA
convention for comparison with other
calculations~\cite{AguilarSaavedra:2005pw}. The scenario SPS4 is
characterized by a large value of $\tan \beta$.  Within this scenario
we study the dependence of the total cross section on the squark
masses and on $\tan\beta$. The third scenario considered is the GMSB
scenario SPS8.

The particle spectrum is determined following the procedure described
in Ref.~\cite{Germer:2010vn}.  Starting from GUT-scale parameters,
\cf~Table~\ref{tab_SPSinput}, we use the program
\verb+Softsusy+~\cite{Allanach:2001kg} to evolve the soft-breaking
parameters down to the SUSY scale $M_{\rm SUSY}$. In accordance to the
SPA convention, a common SUSY scale $M_{\rm SUSY} = 1$\,\TeV{} has
been chosen.  To get the right mixing in the sbottom sector, we first
compute the low energy masses and use these to calculate the effective
bottom-quark mass $m_b^{\drbar,\text{eff}}$, \eqref{eq:eff1}. This
mass is then used in the bottom-squark mass matrix to calculate the
sbottom-mass eigenstates.
\TABULAR[t]{cccccc}{
  \hline\trule
  & $\boldsymbol{m_0}$
  & $\boldsymbol{m_{1/2}}$
  & $\boldsymbol{A_0}$
  & $\boldsymbol{\tan\beta}$
  & $\boldsymbol{{\rm sign}(\mu)}$\\
  \SPA{} & 70~\GeV & 250~\GeV & $-$300~\GeV & 10 & $+$\\
  SPS4 & 400~\GeV & 300~\GeV & 0 & 49.4 & $+$\\
  \hline\hline\trule
  & $\boldsymbol{\Lambda}$
  & $\boldsymbol{M_{\text{mess}}}$
  & $\boldsymbol{N_{\text{mess}}}$
  & $\boldsymbol{\tan\beta}$
  & $\boldsymbol{{\rm sign}(\mu)}$\\
  SPS8 & 100~\TeV & 200~\TeV & 1 & 15 & $+$\\
  \hline
}{High-energy input parameters for the different SUSY scenarios
  considered. The mass parameters $m_0$, $m_{1/2}$ and $A_0$ are
  given at the GUT scale, $\tan\beta$ is evaluated at
  $M_{\rm SUSY}=1$\,\TeV. \label{tab_SPSinput}}
\TABULAR[t]{cccccccc}{
  \hline\trule
  & $\boldsymbol{\Delta m_b}$
  & $\boldsymbol{m_b^{\drbar,\text{eff}}}$
  & $\boldsymbol{\Sb_1}$
  & $\boldsymbol{\Sb_2}$
  & $\boldsymbol{\gluino}$
  & $\boldsymbol{\neu_1}$
  & $\boldsymbol{\cha_1}$\\
  \hline
  \SPA{} & 0.037 & 2.38 & 500 & 533 & 609 & 101 & 180\\
  SPS4 & 0.23 & 2.05 & 428 & 633 & 736 & 123 & 217\\
  SPS8 & 0.03 & 2.42 & 1070 & 1085 & 141 & 253\\
  \hline}
{The shift $\Delta m_b$ and the resulting effective bottom-quark mass
  as well as the on-shell masses of the bottom squarks, of the gluino,
  and of the lightest neutralino and chargino within the different
  SUSY scenarios considered. $\Delta m_b$ and
  $m_b^{\drbar,\text{eff}}$ are evaluated at the scale used for
  $\bx\bx^\ast$ production. All masses are given in \GeV.\label{tab_SPSval}
}
The lighter of the two bottom-squarks is taken as the dependent
squark. Its mass is therefore fixed by SU(2) invariance.
Table~\ref{tab_SPSval} collects the shift $\Delta m_b$,
\cf\ \eqref{eq:eff2} and the effective bottom-quark mass together
with the on-shell mass of the bottom squarks, of the gluino and of the
lightest neutralino and chargino.

Unless otherwise stated, the results presented in this section are
computed setting the hadronic center of mass energy to
$\sqrt{S}=14~\TeV$ and using the MRST2004QED parton distribution
functions \cite{Martin:2004dh}. The factorization and renormalization
scales are set to a common value, $\mu = \mu_{R}=\mu_{F}=m_{\ba}$,
\ie\ to the mass of the produced bottom squark.

\subsection{Total hadronic cross section}
\label{sec:total_had_CS}
Table~\ref{tab:totalCS_14TeV} shows the hadronic cross section for
diagonal bottom-squark production within the three considered
scenarios for $\sqrt{S}=14$~\TeV{}. 
\TABULAR[t]{cccccc}{
  \hline\trule
  \multirow{2}{*}{\bf 14~Te\kern -1pt V}
  & {$\boldsymbol{\Born}$}
  & {$\boldsymbol{\TreeEW}$}
  & {$\boldsymbol{\Treegy}$}
  & {$\boldsymbol{\NLOEW}$}
  & {$\boldsymbol{\EW}$}\\
  & {$\mathcal{O}(\alphas^2)$} &
  {$\mathcal{O}(\alphas\alpha+\alpha^2)$} &
  {$\mathcal{O}(\alphas\alpha)$} &
  {$\mathcal{O}(\alphas^2\alpha)$} &
  {$\mathcal{O}(\alphas\alpha+\alpha^2+\alphas^2\alpha)$}
  \\
  \hline\trule
  {\bf \SPA{}}&&&&&\\
$\Sb_1\Sb_1^\ast$  & 444.3(3) & 0.8 & 2.0 & $-$6.0 & $-$3.2\\
 &  &     0.2\% &     0.5\% &    $-$1.4\% &    $-$0.7\% \\
 \cline{2-6}
$\Sb_2\Sb_2^\ast$  & 310.3(1) & $\ll$0.1 & 1.5 & $-$2.9 & $-$1.4\\
 &  &     $\approx$0\% &     0.5\% &    $-$0.9\% &    $-$0.5\% \\
 \hline\trule
{\bf SPS4}&&&&&\\
$\Sb_1\Sb_1^\ast$  & 1050.9(3) & $-$0.4 & 4.3 & $-$19.4 & $-$15.5\\
 &  &    $\approx$0\% &     0.4\% &    $-$1.8\% &    $-$1.5\% \\
\cline{2-6}
$\Sb_2\Sb_2^\ast$  & 112.36(6) & 0.27 & 0.61 & $-$2.85 & $-$1.97\\
 &  &     0.2\% &     0.5\% &    $-$2.5\% &    $-$1.8\% \\
\hline\trule
{\bf SPS8}&&&&&\\
$\Sb_1\Sb_1^\ast$  & 3.405(1) & 0.002 & 0.029 & $-$0.003 & 0.028\\
 &  &     0.1\% &     0.9\% &    $-$0.1\% &     0.8\% \\
\cline{2-6}
$\Sb_2\Sb_2^\ast$  & 3.042(1) & 0.007 & 0.026 & 0.008 & 0.042\\
 &  &     0.2\% &     0.9\% &     0.3\% &     1.4\% \\
\hline
}
{Hadronic cross section for diagonal $\ba\ba^\ast$ production at the
14~\TeV{} LHC within three different scenarios. Shown are the LO cross
section, the tree-level EW as well as NLO EW contributions and the
relative corrections as defined in the text. The numbers in brackets
refer to the integration uncertainty in the last digit and are omitted
if the uncertainty is at least one order of magnitude smaller than the
quoted precision. All cross sections are given in femtobarn (fb).
\label{tab:totalCS_14TeV}}

\TABULAR[t]{cccccc}{
  \hline\trule
  \multirow{2}{*}{\bf 7~Te\kern -1pt V}
  & {$\boldsymbol{\Born}$}
  & {$\boldsymbol{\TreeEW}$}
  & {$\boldsymbol{\Treegy}$}
  & {$\boldsymbol{\NLOEW}$}
  & {$\boldsymbol{\EW}$}\\
  & {$\mathcal{O}(\alphas^2)$} &
  {$\mathcal{O}(\alphas\alpha+\alpha^2)$} &
  {$\mathcal{O}(\alphas\alpha)$} &
  {$\mathcal{O}(\alphas^2\alpha)$} &
  {$\mathcal{O}(\alphas\alpha+\alpha^2+\alphas^2\alpha)$}
  \\
  \hline\trule
{\bf \SPA{}}&&&&&\\
$\Sb_1\Sb_1^\ast$  & 30.42(2) & 0.10 & 0.20 & $-$0.35 & $-$0.05\\
 &  &     0.3\% &     0.7\% &    $-$1.1\% &    $-$0.2\% \\
\cline{2-6}
$\Sb_2\Sb_2^\ast$  & 19.286(6) & 0.004 & 0.136 & $-$0.203 & $-$0.064\\
 &  &     $\approx$0\% &     0.7\% &    $-$1.1\% &    $-$0.3\% \\
\hline\trule
{\bf SPS4}&&&&&\\
$\Sb_1\Sb_1^\ast$  & 89.10(2) & $-$0.01 & 0.52 & $-$1.69 & $-$1.18\\
 &  &    $\approx$0\% &     0.6\% &    $-$1.9\% &    $-$1.3\% \\
\cline{2-6}
$\Sb_2\Sb_2^\ast$  & 5.175(2) & 0.023 & 0.043 & $-$0.125 & $-$0.059\\
 &  &     0.5\% &     0.8\% &    $-$2.4\% &    $-$1.1\% \\
\hline\trule
{\bf SPS8}&&&&&\\
$\Sb_1\Sb_1^\ast$  & 0.03706(1) & 0.00004 & 0.00059 & 0.00010 & 0.00073\\
 &  &     0.1\% &     1.6\% &     0.3\% &     2.0\% \\
\cline{2-6}
$\Sb_2\Sb_2^\ast$  & 0.03118(1) & 0.00011 & 0.00051 & 0.00030 & 0.00092\\
 &  &     0.4\% &     1.6\% &     1.0\% &     3.0\% \\
\hline
}
{Same as Table~\ref{tab:totalCS_14TeV} but considering 
diagonal $\ba\ba^\ast$ production at the 7~\TeV{} LHC.
\label{tab:totalCS_7TeV}}
As expected, the total cross section is dominated by the LO QCD
contribution of $\Order(\alphas^2)$. The contribution of the
photon-induced process is independent of the mixing angle. In each
scenario, its yield relative to the leading order cross section is
similar for the two processes considered.  Although formally
suppressed by a factor $\alpha_s$, the NLO EW corrections are
typically bigger than the tree-level EW contributions.  In the \SPA{}
(SPS4) scenario the tree level and NLO EW contributions are more
important in case of $\tilde b_1 \tilde b_1^\ast$ $(\tilde b_2 \tilde
b_2^\ast)$ production.  This can be explained by the chirality
dependence of the SU(2) coupling and by the fact that in the \SPA{}
(SPS4) scenario $\tilde b_1$ $(\tilde b_2)$ is
mostly left-handed. \\
In the SPS8 scenario the bottom squarks are twice as heavy as in the
aforementioned scenarios, thus the Born cross section is about two
orders of magnitude smaller.  Further, the mixing between left- and
right-handed squarks is more important and the sbottom masses are
nearly degenerate.  These features partially soften the differences
among the tree-level EW contributions to $\tilde b_1 \tilde b_1^\ast$
production and the ones to $\tilde b_2 \tilde b_2^\ast$
production.\footnote{In the no-mixing limit the tree-level EW
  contributions to $\tilde b_L$ production is one order of magnitude
  larger than the one contributing to $\tilde b_R$ production.}  Huge
cancellations between the $q \bar q$ and the $gg$ channel amplify the
dependence of the NLO EW contribution on the production process
considered. As a result the NLO EW contributions to $b_1 b_1^\ast$
production and the ones to $b_2 b_2^\ast$ production have opposite
sign, the latter being three times bigger than the former. Summing up
the various contributions, the relative yield in the scenario
considered is below $2 \%$.

Table~\ref{tab:totalCS_7TeV} collects the hadronic cross section for
$\sqrt{S}=7$~\TeV{}. The leading order total cross sections are
reduced proportionally to the mass of the produced squark. They amount
to $1-10 \%$ of their value at $\sqrt{S}=14$~\TeV{}.  In all the
scenarios considered, the contribution of the photon-induced channel
is enhanced with respect to the $14$~\TeV{} case.  In the \SPA{}
(SPS4) scenario the impact of the NLO EW contributions to the
production of the mostly left-handed bottom-squark, $\tilde b_1 \tilde
b_1^\ast$ ($\tilde b_2\tilde b_2^\ast$), is reduced. In contrast, the
NLO EW contributions become more important in case of the production
of the mostly right-handed sbottom.  In the SPS8 scenario the EW
contributions of the various channels are enhanced. In particular the
NLO EW contributions are positive for both production processes. This
is a consequence of the enhancement of the NLO EW contributions to the
$gg$ channel at $\sqrt{S}=7$~\TeV{} (\cf\
Section~\ref{sec:Differential}).

\subsection{Parameter scan}
\label{sec:parameterscan}
The impact of $\tan\beta$ and of the sbottom masses have been studied
performing a parameter scan on these parameters. In this scan, the
soft breaking parameters $M_L$, $M_{\bR}$, and $M_{\tR}$ appearing the
squared mass matrix, \eqref{Eq:MassMatrix2}, are set to a common value
$m_{\text{squark}}$. All other parameters are set to their SPS4
values. The scans presented in this section are obtained for three
different values of
$m_{\text{squark}}=\{300,600,900\}~\GeV$. $\tan\beta$ is varied from
10 to 50.
%
%

In all scenarios considered, we have verified the smallness of the
bottom-initiated tree-level contributions, justifying our procedure of
neglecting the $\Order(\alpha_s^2 \alpha)$ contributions to this
channel. In \figref{fig:tree-level-yield} we show the relevance of the
various production channels at tree-level. For
$\tilde{b}_1\tilde{b}_1^\ast$ and $\tilde{b}_2\tilde{b}_2^\ast$
production the gluon fusion channel contributes $70-90\%$ of the total
cross section. The remaining $10-30\%$ of the total cross section is
given by the $q\bar{q}$ channel, its relative yield increasing with
$\tan\beta$. The contribution of the $b\bar{b}$ channel is at the 
permille level, in accordance with the analysis in
Ref.~\cite{Beenakker:2010nq}.\footnote{The big contributions from the
  $b\bar{b}$ channel quoted in Ref.~\cite{Arhrib:2009sb} are a
  consequence of two enhancement factors. First of all, the Higgs
  exchanged in the $s$-channel is resonant. Moreover, the
  $\mathcal{H}bb$ Yukawa couplings are enhanced by the choice of a
  negative value for the parameter $\mu$, such that $\Delta m_b
  \approx -0.76$. In our analysis the Higgs bosons are not resonant
  and we do not consider negative values of $\mu$, which are
  disfavored by the measured value of the anomalous magnetic moment of
  the muon $(g-2)_{\mu}$~\cite{Stockinger:2006zn}.}  Owing to the
small yield of the $b \bar b$ channel at tree-level, we will safely
neglect the NLO EW contributions to this channel.
\FIGURE[t]{
  \includegraphics[width=.9\textwidth]
  {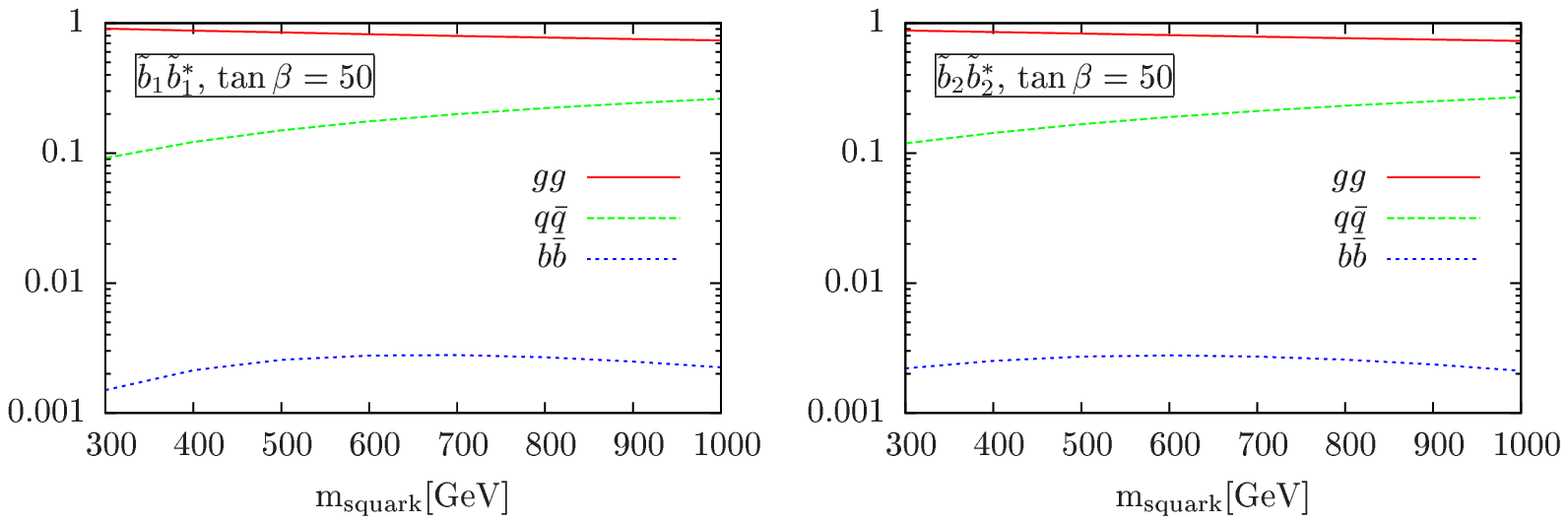}
  \caption{Relative yield of the various tree-level production
    channels as a function of the common squark mass breaking
    parameter $m_{\text{squark}}$. The left (right) plot shows
    $\tilde{b}_1\tilde{b}_1^\ast$ ($\tilde{b}_2\tilde{b}_2^\ast$)
    production. }
  \label{fig:tree-level-yield}
}
%

\FIGURE[t]{
  \includegraphics[width=.95\textwidth]
  {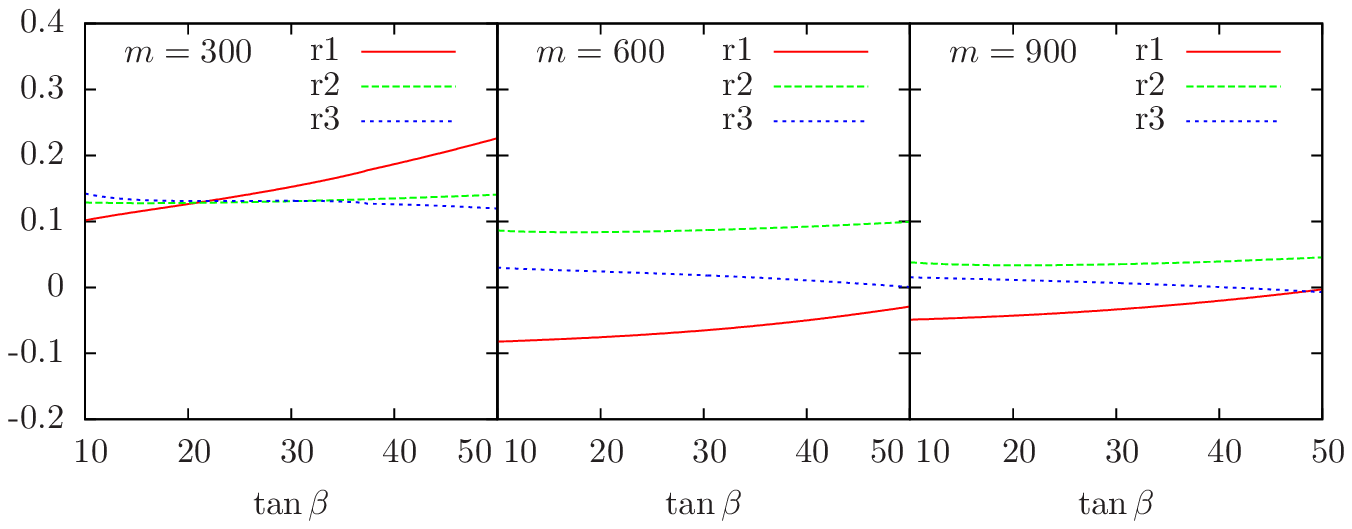}
  \caption{The ratios $r_1$,$r_2$, and $r_3$ as defined in
    \eqref{Eq:r1r2r3} as a function of $\tan \beta$ for different
    values of the common squark mass breaking parameter
    $m=m_{\text{squark}}$ in \GeV.}
  \label{fig:reliability}} 
The reliability of the renormalization scheme in the scenario
considered has been verified by investigating the behavior of the
dependent parameters since these can potentially get large finite
shifts by their renormalization constants. In the case of
$m^2_{\Sb_1}$ and $A_{\tilde t}$ we have checked that the finite part
of their renormalization constant is smaller than the parameter
itself.  The mixing angle renormalization constant enters only the
counterterm of the $\Sb^\ast_1\Sb_2$ and $\Sb^\ast_2\Sb_1$ quadratic
terms via the combination
\begin{equation}
\delta  Y_{b} \equiv  \left (m^2_{\Sb_2} - m^2_{\Sb_1} \right ) \delta \theta_b.  
\end{equation}
We have checked that the finite part of $\delta Y_b$ is smaller than
the mass of the bottom squarks. Figure~\ref{fig:reliability} shows
the ratios
\begin{equation}
r_1= \frac{\delta m^{2\; \text{fin}}_{\Sb_1}}{m^2_{\Sb_1}}, ~~~~~~
r_2=\frac{\delta A^\text{fin}_{\tilde t} }{A_{\tilde t}},~~~~~~
r_3=\frac{\delta Y^{\text{fin}}_b}{m^2_{\Sb_1}},
\label{Eq:r1r2r3}
\end{equation}
as a function of $\tan \beta$ for the various values of
$m_{\text{squark}}$.  The value of $r_1$, $r_2$, and $r_3$ is below 0.2
in all the scenarios.

We have explicitly checked the impact of the $\Order(\alpha_s\alpha+
\alpha^2)$ contributions arising from the diagram
depicted in Figure~\ref{fig:feynman_tree}(d).  In the region
of the parameter space considered in this paper these contributions
are  negligible. Indeed its numerical value is six orders
of magnitudes smaller than the Born cross section, justifying the
approximation of identifying the CKM matrix with the unit matrix.
\medskip

\FIGURE{
  \includegraphics[width=\textwidth]
  {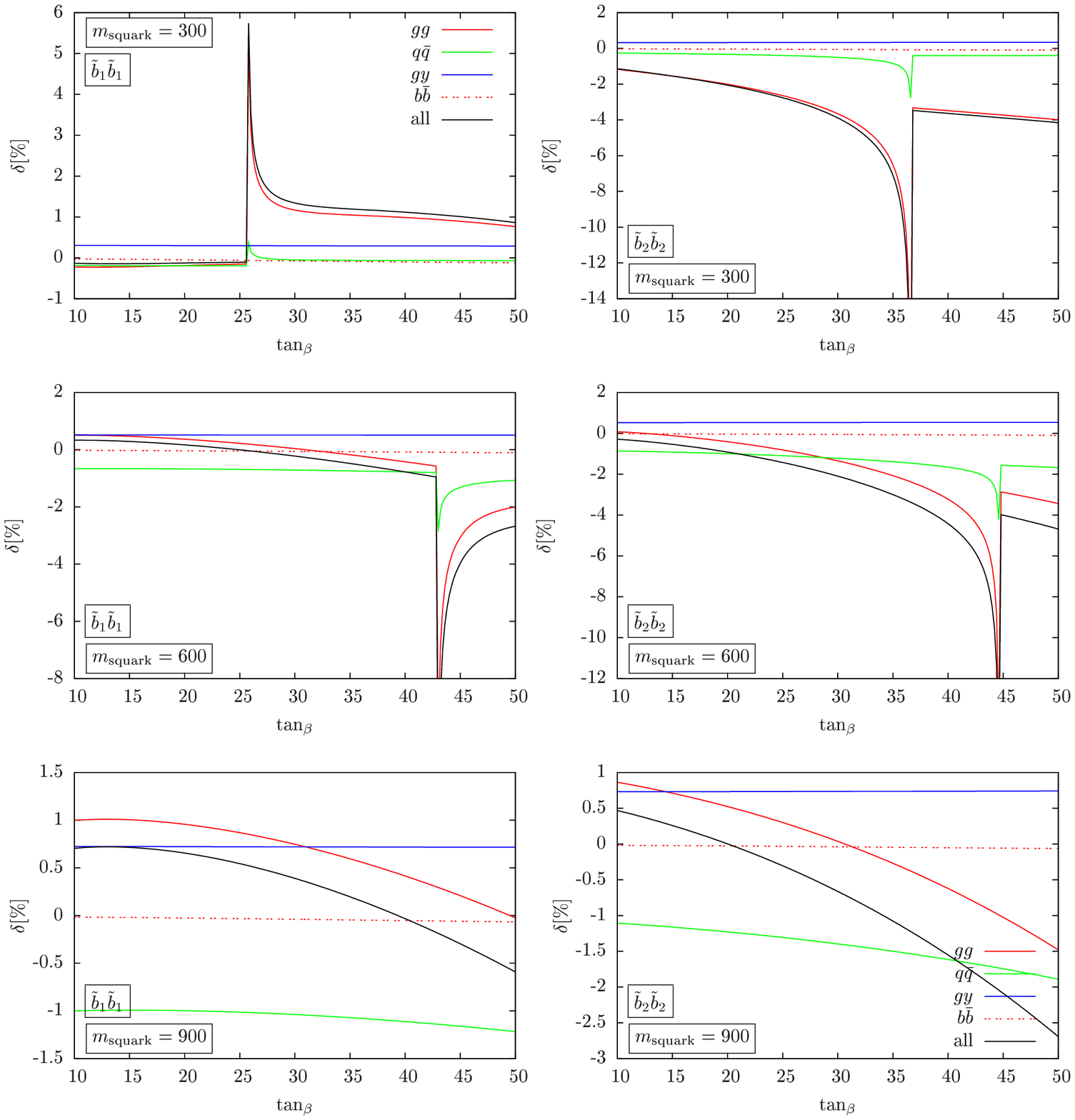}
  \caption{Relative yield of the EW contribution for $\Sb_1\Sb^\ast_1$
    production (left panels) and for $\Sb_2\Sb^\ast_2$ production
    (right panels) as a function of $\tan \beta$. Shown are the
    various production channels as well as the combinded effect. The
    soft breaking parameter $m_{\mbox{\tiny squark}}$ is defined according to
    Section~\ref{sec:parameterscan} and is expressed in \GeV. 
    $\tan\beta$ is varied between 10 and 50.}
  \label{fig:SPS4_b1b1b2b2_scan}
}
\afterpage{\clearpage}
The results of the scan are collected in
Figure~\ref{fig:SPS4_b1b1b2b2_scan}. The $q \bar q$ contribution is
the sum of the tree-level EW and of the NLO EW contributions from the
$q \bar q$ annihilation channel.  The peaks in the corrections
correspond to neutralino, chargino, or sfermion thresholds. These
unphysical singularities affect the self-energy of the produced
sbottom and can be regularized by taking into account the finite
widths of the unstable particle~\cite{Kniehl:2002wn}.  The curves in
Figure~\ref{fig:SPS4_b1b1b2b2_scan} exhibit a step-function like
behavior in the region where the mass of the produced sbottom is above
threshold. Both, for $\tilde b_1 \tilde b_1^\ast$ and for $\tilde b_2
\tilde b_2^\ast$ production the behavior of the various contributions
strongly depend on the size of $\tan \beta$. In the following we will
distinguish between the low and the high $\tan \beta$ region, which
are separated by the threshold.

The case of $\tilde b_1 \tilde b_1^\ast$ production is shown in the
left panels of Figure~\ref{fig:SPS4_b1b1b2b2_scan}. The EW
contributions of the $q\bar{q}$ and the $gg$ channel have
substantially the same $\tan\beta$ dependence close to threshold. In
the low $\tan \beta$ region the contributions from the $g\gamma$
channel cancel against the one coming from the $q \bar q$
annihilation. The overall effect of the EW contribution is below $1\%$
of the tree-level QCD contribution. In the high $\tan \beta$ region
the leading contributions come from the $gg$ channel and are only
partly canceled by the other channels. The EW contributions are of the
order of a few percents. In the $m_{\text{squark}}=900~\GeV$ case
partial cancellations between the $g\gamma$ and the $q\bar{q}$
channels further suppress the EW contributions. In this case they are
below $1\%$.  The EW contributions to $\tilde b_2 \tilde b_2^\ast$
production are depicted in the right panels of
Figure~\ref{fig:SPS4_b1b1b2b2_scan} and exhibit similar features. The
NLO EW contributions are more pronounced since the corrections from
the $gg$ channel are more important. They are of the order of several
percents, \eg\ $5\%$ for $m_{\text{squark}} =600~\GeV{}$ and $\tan
\beta \ge 45$.

\FIGURE[b]{
  \includegraphics[width=.95\textwidth]
  {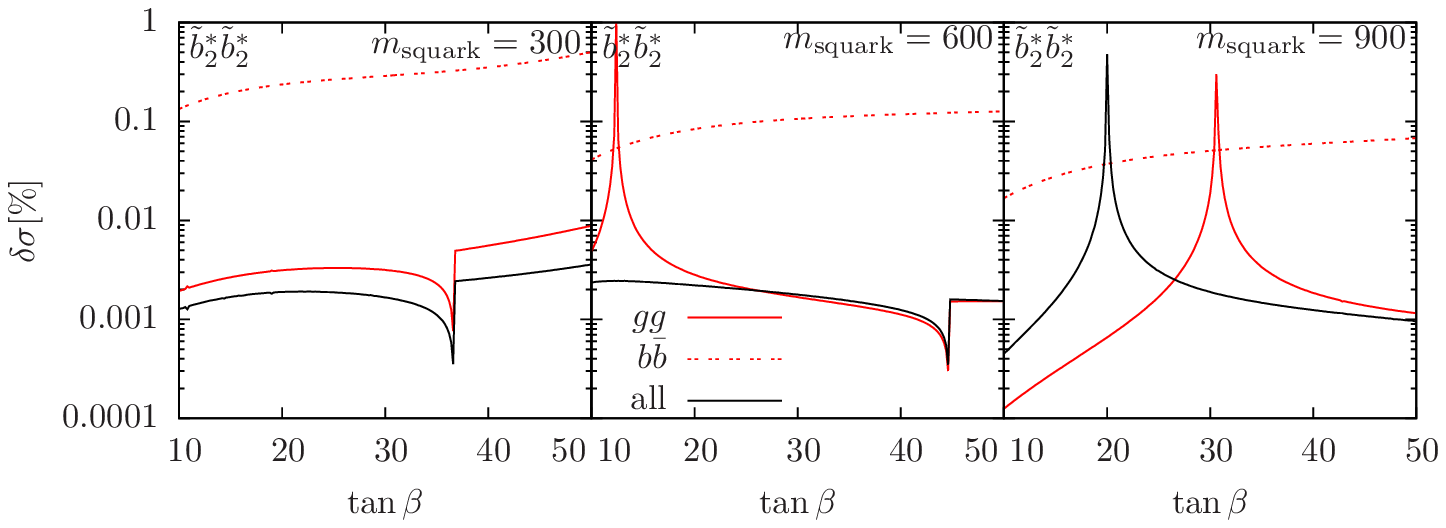}
  \caption{Relative difference between the EW contribution to the
    cross section for $\by\by^\ast$ production with and without
    enhanced Yukawa couplings (black lines). The individual effect on
    the $gg$ ($b\bar{b}$) channel is depicted by the red(-dotted)
    lines. The $q \bar q$ channel is not affected by the resummation
    and hence not shown. $m_{\mbox{\tiny squark}}$ is given in \GeV.}
  \label{fig:enh_diff}
}
It is worth noticing that the only practical effect of the
resummation in the $b / \tilde b$ sector is to change the value of the
mass of the bottom squark.  Indeed the
impact of the effective Yukawa couplings in the computation of the
amplitudes is negligible. 
This can be inferred from
Figure~\ref{fig:enh_diff},  where we plot
\begin{equation}
  \label{eq:dsigma_enh}
  \delta\sigma \equiv \frac{\EW_{\text{eff}}
    -\EW_{\text{no-eff}}} {\EW_{\text{eff}}}
\end{equation}
in the case of $\tilde b_2 \tilde b^\ast_2$
production. $\EW_{\text{eff}}$ and $\EW_{\text{no-eff}}$ denote the EW
contributions with and without the effective Yukawa couplings,
respectively.  The overall effect of the effective Yukawa coupling,
given by the black line, is in most regions below the permille level.
The positive peaks for $m_{\text{squark}}=600~\GeV$ and
$m_{\text{squark}}=900~\GeV$ correspond to the zeros of the
denominator in the definition~(\ref{eq:dsigma_enh}), \cf\
Figure~\ref{fig:SPS4_b1b1b2b2_scan}.

\subsection{Differential distributions}
\label{sec:Differential}
Even though the EW contributions seem to have a rather small impact on
the total cross section, they can become important in specific phase
space regions. In Figure~\ref{fig:b1b1_SPS1aPrime_dist}
and~\ref{fig:b2b2_SPS1aPrime_dist} we consider differential
distributions with respect to three different kinematical variables in
the case of $\bx\bx^\ast$ and $\by\by^\ast$ production,
respectively. We focus on the the \SPA{} scenario and on the 14~\TeV{}
LHC.  The left panels show to total EW contributions to the
differential cross section.  The tree-level EW contribution and the
NLO EW contribution of the various production channels are depicted as
well.  The right panels show the impact of the EW contributions
relative to the tree-level QCD cross section for each production
channel. In contrast to the left panels, in the right panels the
$q\bar{q}$ contribution is given by the sum of the tree-level and NLO
EW contributions.
\medskip

Figure~\ref{fig:b1b1_SPS1aPrime_dist}(a) refers to the transverse
momentum distribution of the sbottom with highest $p_T$.
Close to the threshold, \ie\ in the region $p_T<300~\GeV$, the
contribution of the $gg$ channel is positive. Far from the threshold,
this contribution becomes negative and relatively more important.  In
the low $p_T$ region the two contributions from the $q\bar{q}$ channel
are different in sign, and their partial cancellation reduces the
overall effect of this channel.  In the high $p_T$ region the $q\bar
q$ channel increases its importance.  The photon induced channel peaks
at low $p_T$ and it is almost proportional to the LO QCD cross
section, \ie\ its relative yield is constant in $p_T$. As expected,
the $b\bar{b}$ channel is irrelevant in the whole $p_T$ region.  The
total EW contributions have a small positive yield of the order of
$1-2\%$ in the low $p_T$ region, while for $p_T>500~\GeV$ the cross
section is altered by $5-10\%$.
It is interesting to note that a lower cut $p_{T\,\text{min}}$ on the
transverse momentum can significantly rise the impact of the EW
contributions.  For instance the cut $p_{T\,\text{min}}=320~\GeV$
would discard the positive yield of the $gg$ channel in the low $p_T$
region. As a consequence the relative yield of the EW contribution to
the total cross section would become of the order of $-3.2\%$.

The invariant mass distribution is displayed in
Figure~\ref{fig:b1b1_SPS1aPrime_dist}(b).  The EW contributions
exhibit the same high-energy behavior they have in the case of the
$p_T$ distribution. In this energy region they alter the leading-order
prediction up to $10\%$.  The peaks in the $gg$ channel correspond to
$\by$ and $\ty$ thresholds.  Although in the low invariant mass region
no remarkable cancellations occur among the various channels, the
overall positive contribution is small, of the order of $2 \%$.

In order to study how the EW contributions to the total cross section
are altered by a lower cut $M_{\text{inv,min}}$ on the invariant mass,
we consider $\sigma(M_{\text{inv,min}})$ defined as the total cross
section integrated from the value $M_{\text{inv,min}}$ of the
invariant mass.  The upper-left part of
Figure~\ref{fig:SPS1aPrime_cut} shows the relative yield of the total
EW contributions to $\sigma(M_{\text{inv,min}})$, together with the
breakdown into the individual channels.  The lower cut
$M_{\text{inv,min}}$ excludes the region where the EW contributions
are positive. Therefore the EW contributions decreases as
$M_{\text{inv,min}}$ increases, while their relative impact
increases. For instance, in the region $M_{\text{inv,min}}\ge
1500~\GeV$ the relative yield of the EW contributions $\dNLO$ exceeds
$-5\%$.  In this region, $\dNLO$ is amplified by a factor of seven with
respect to its value in the case of the fully inclusive cross section,
\cf\ Table~\ref{tab:totalCS_14TeV}. The cross section is reduced by a
factor of five for $M_{\text{inv,min}} =1500~\GeV$ (upper-right panel
of Figure~\ref{fig:SPS1aPrime_cut}).

Figure~\ref{fig:b1b1_SPS1aPrime_dist}(c) shows the pseudo-rapidity
distribution, where always the squark with the higher absolute value
of the pseudo-rapidity $\eta$  is
considered. The gap for zero rapidity is a consequence  of this
definition. The NLO EW contributions peak at $|\eta|= 1$ and dominate
the EW contribution at this region. The contribution is negative for
small values of $|\eta|$ 
The total effect on the LO QCD cross section is up to $2\%$.
\medskip

Figure~\ref{fig:b2b2_SPS1aPrime_dist} shows the differential
distributions for $\by\by^\ast$ production in the transverse momentum
(a), in the invariant mass (b), and in the pseudo-rapidity (c). In
contrast to $\bx\bx^\ast$ production, the threshold behavior initiated
by $\ty$ is mild and hardly visible in both, the transverse momentum
and the invariant mass distributions. The EW contribution is small and
its relative yield stays below $5\%$, even in the high energy region.
This is expected since in the \SPA{} scenario $\by$ is mostly
right-handed. Interestingly, the contributions from the $q\bar{q}$
channel and the $g\gamma$ channel almost cancel in most parts of the
phase-space. Therefore the EW contributions are well approximated by
the $gg$ channel corrections.
The lower panels of Figure~\ref{fig:SPS1aPrime_cut} show
$\sigma(M_{\text{inv,min}})$ in the case of $\by\by^\ast$ production.
In this scenario the EW contributions to $\by\by^\ast$ production can
be safely neglected for each value of $M_{\text{inv,min}}$. Even in
the case $M_{\text{inv,min}}= 2~\TeV$ the EW contributions change the
Born cross section only by an amount of the order of $2\%$.
\medskip

In Figure~\ref{fig:b1b1_SPS8_dist} we consider the differential
transverse momentum distributions for $\bx\bx^\ast$ production in the SPS8
scenario for $14$~\TeV{} (upper panels) and $7$~\TeV{} (lower
panels). It is worth analyzing the EW corrections to the $gg$ channel
in this scenario.  The sbottom mass is heavier in the SPS8 scenario
than in the \SPA{} scenario. Therefore the gluons producing a
sbottom--anti-sbottom pair have a bigger typical momentum fraction $x$
in the SPS8 scenario than in the \SPA{} scenario. Since the gluon PDF
falls off rapidly at high $x$ values, the negative EW contributions of
the $gg$ channel in the high $p_T$ region are strongly suppressed in
the SPS8 scenario. In the $\sqrt{S} =7$~\TeV{} case the typical value
of $x$ it is even bigger and this phenomenon is enhanced. In the
SPS8 scenario the cancellation between the positive low $p_T$
corrections and the negative high $p_T$ ones is less effective.
Therefore, as mentioned in Section~\ref{sec:total_had_CS}, the overall
(positive) $gg$ channel contributions get relatively enhanced in the
$\sqrt{S} =7$~\TeV{} case.

\FIGURE{
  \includegraphics[width=.49\textwidth]
  {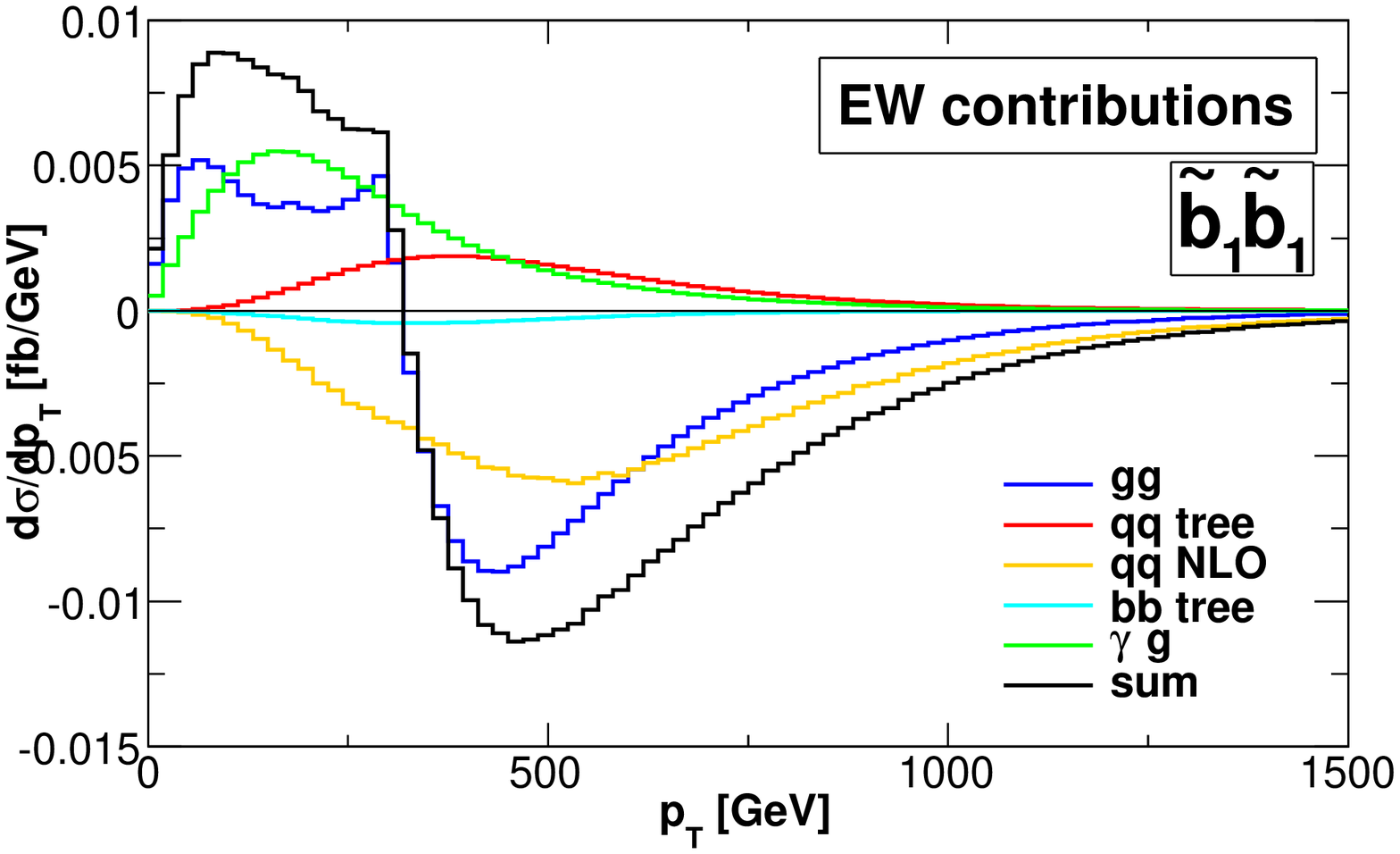}
  \includegraphics[width=.49\textwidth]
  {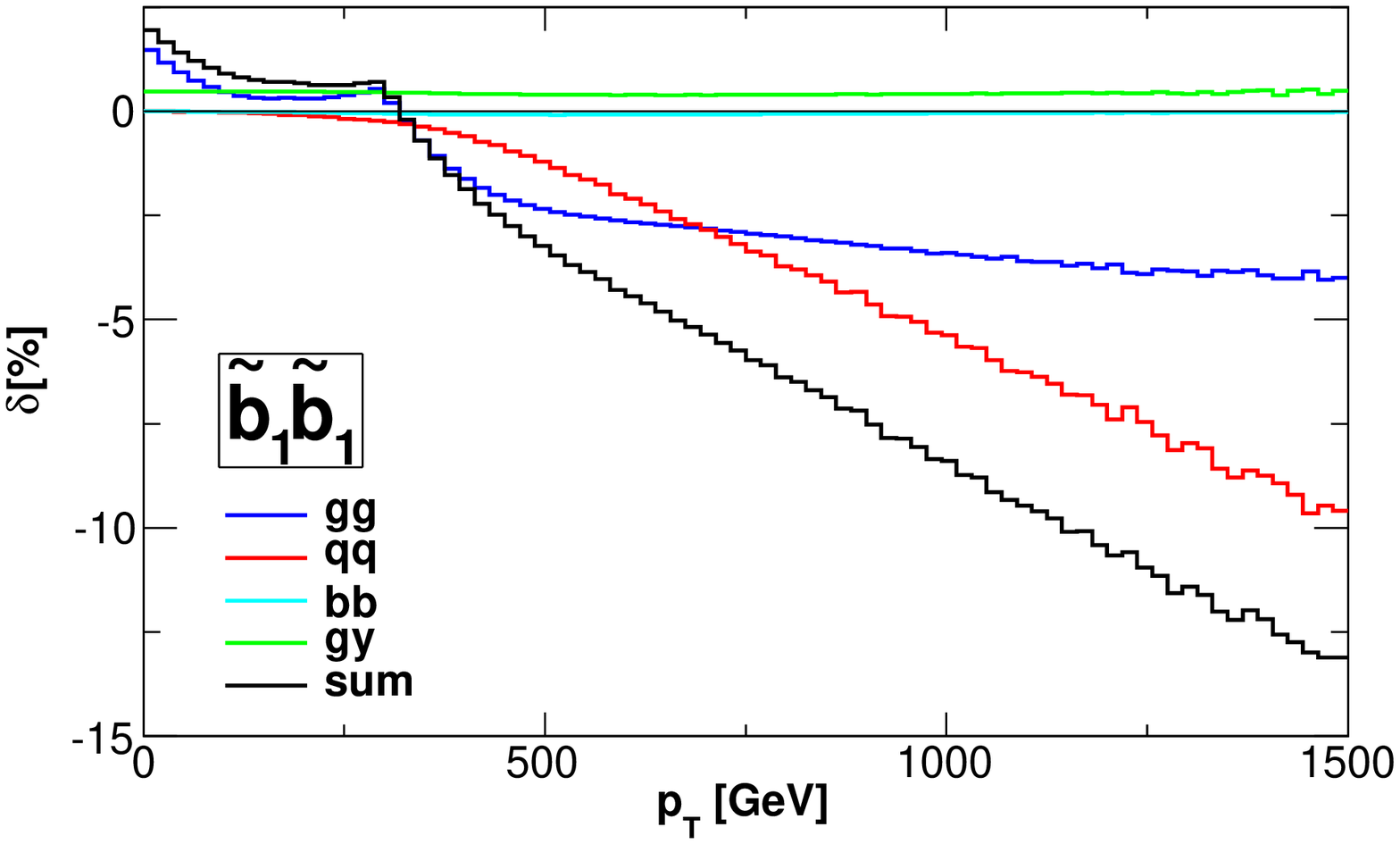}
  \\[-5pt]
  {\hspace*{85pt} (a) $p_T$ distribution}
  \\[5pt]
  \includegraphics[width=.49\textwidth]
  {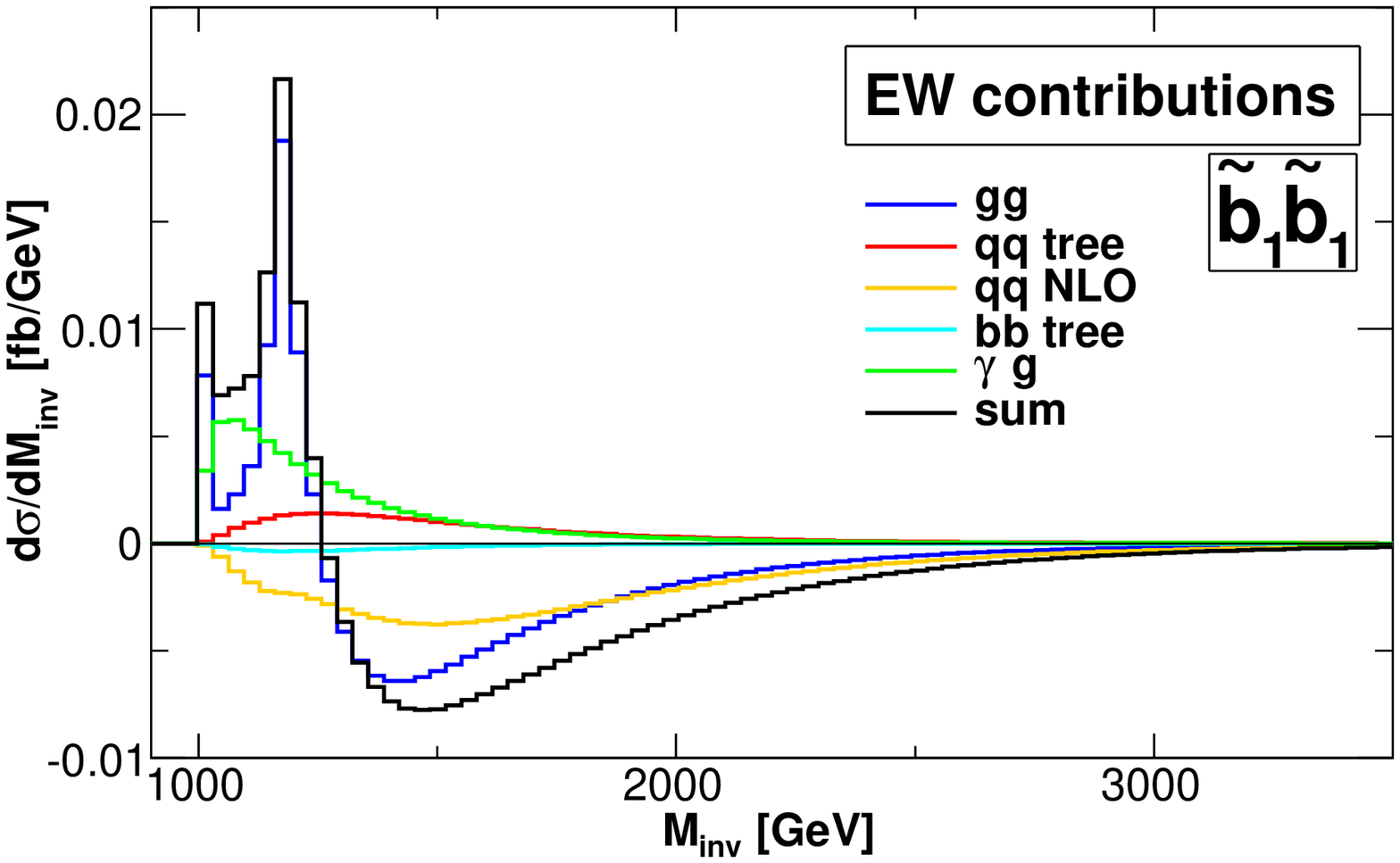}
  \includegraphics[width=.49\textwidth]
  {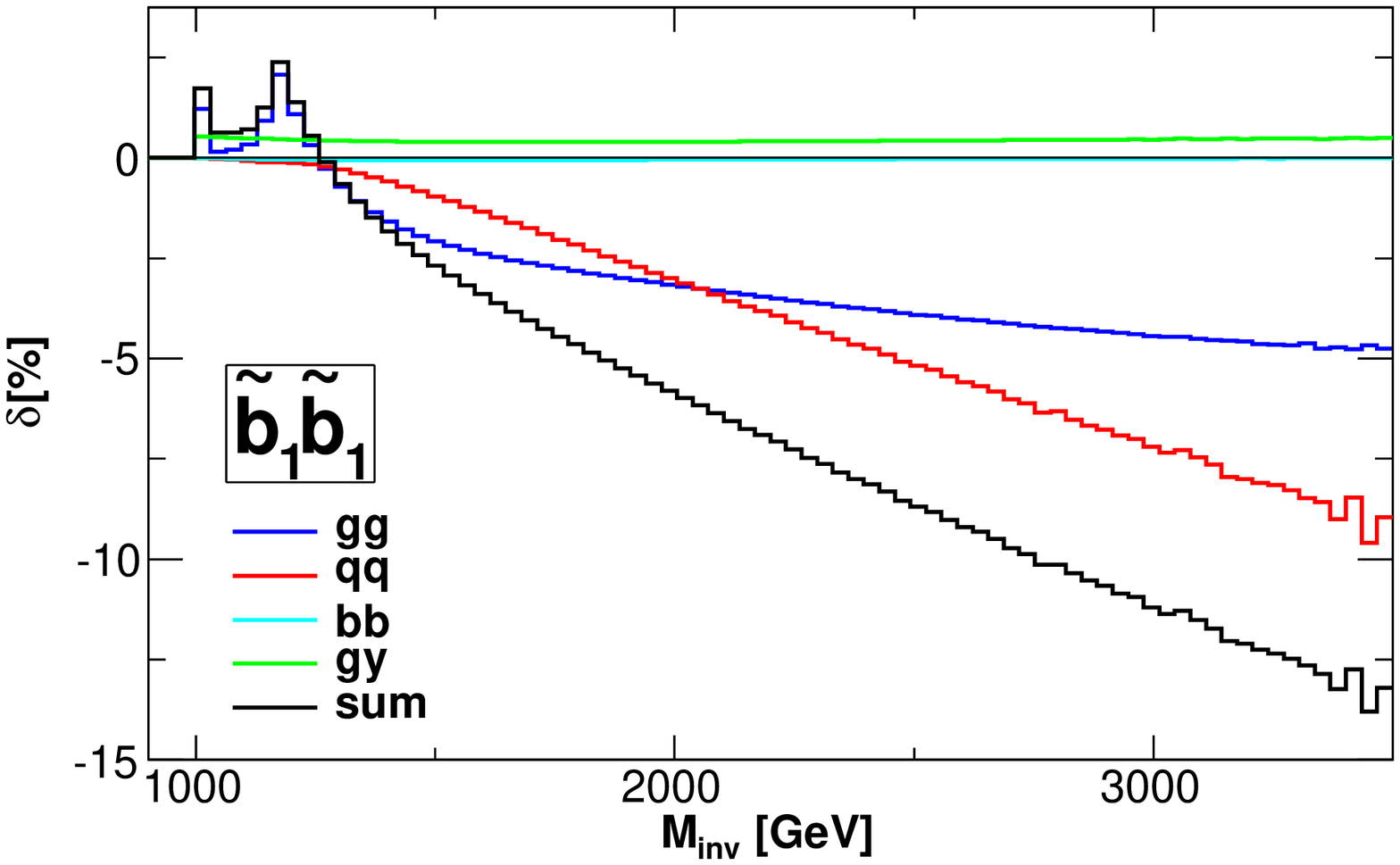}
  \\[-5pt]
  \hspace*{85pt}  (b) $M_{\text{inv}}$ distribution
  \\[5pt]
  \includegraphics[width=.49\textwidth]
  {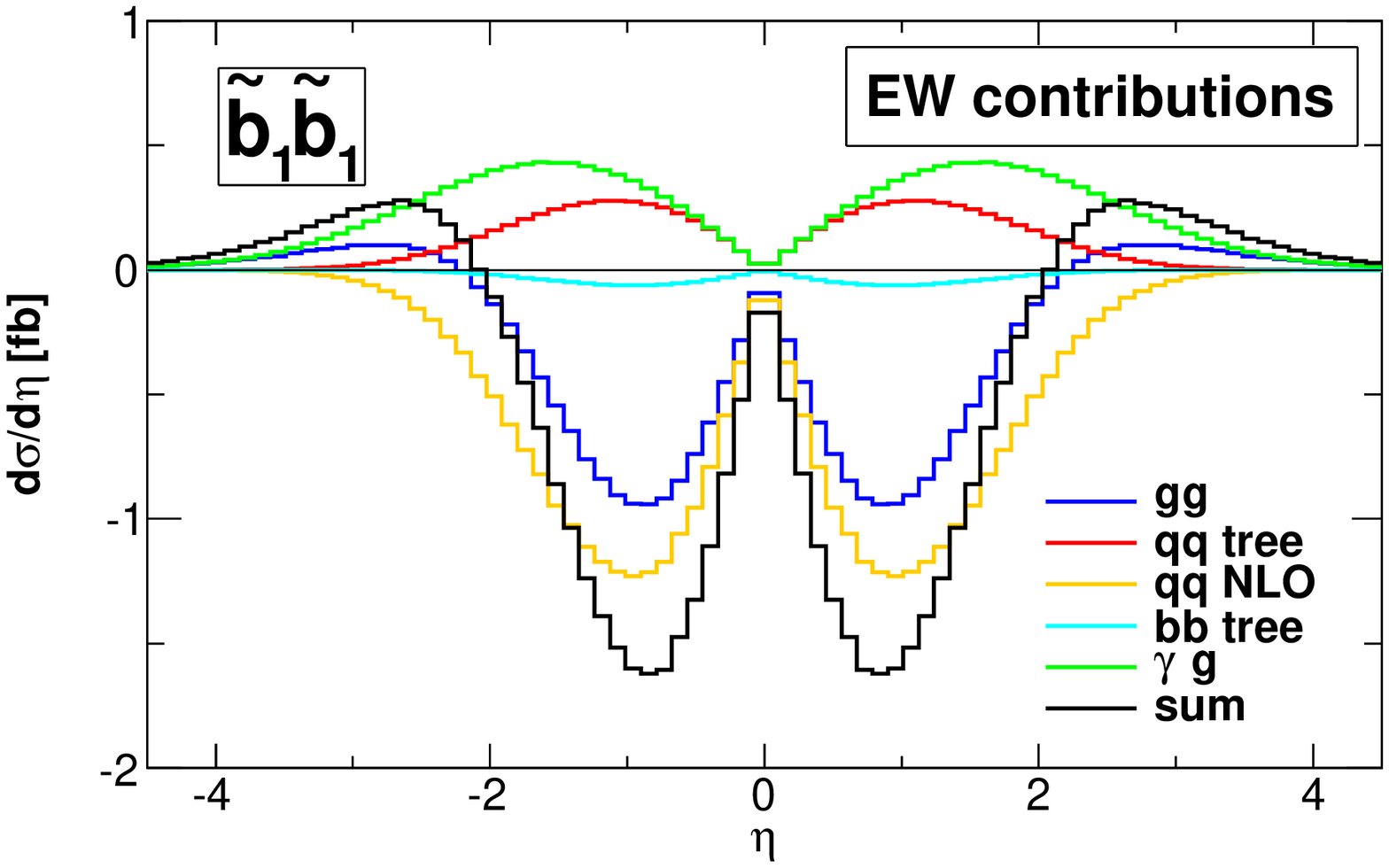}
  \includegraphics[width=.49\textwidth]
  {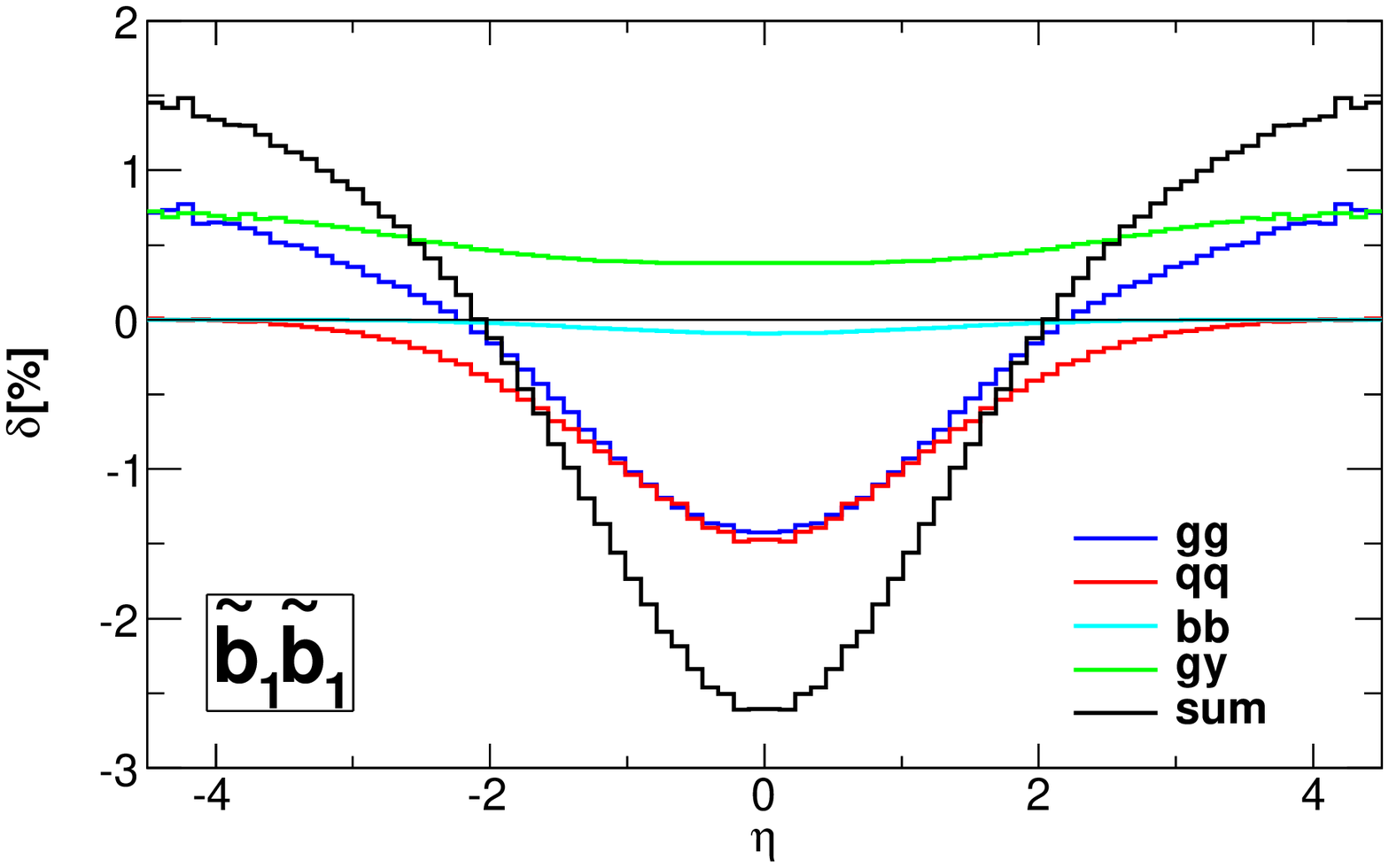}
  \\[-5pt]
  \hspace*{85pt}  (c) $\eta$ distribution
  \\[5pt]
  \caption{Differential distributions for the transverse momentum
    $p_T$, the invariant mass $M_{\text{inv}}$ and the pseudo-rapidity
    $\eta$ for $\bx\bx^\ast$ production at the 14~{\TeV} LHC within
    the \SPA{} scenario. Shown are the tree-level and NLO EW cross
    section contributions for the various production channels (left)
    and the impact of the NLO EW contributions relative to the LO QCD
    cross section (right).  In the left panels the 
    tree-level and NLO EW contributions for the $q\bar{q}$ channel
    are plotted separately. In the right panels they  are
    treated inclusively.}
  \label{fig:b1b1_SPS1aPrime_dist}
}

\FIGURE{
  \includegraphics[width=.49\textwidth]
  {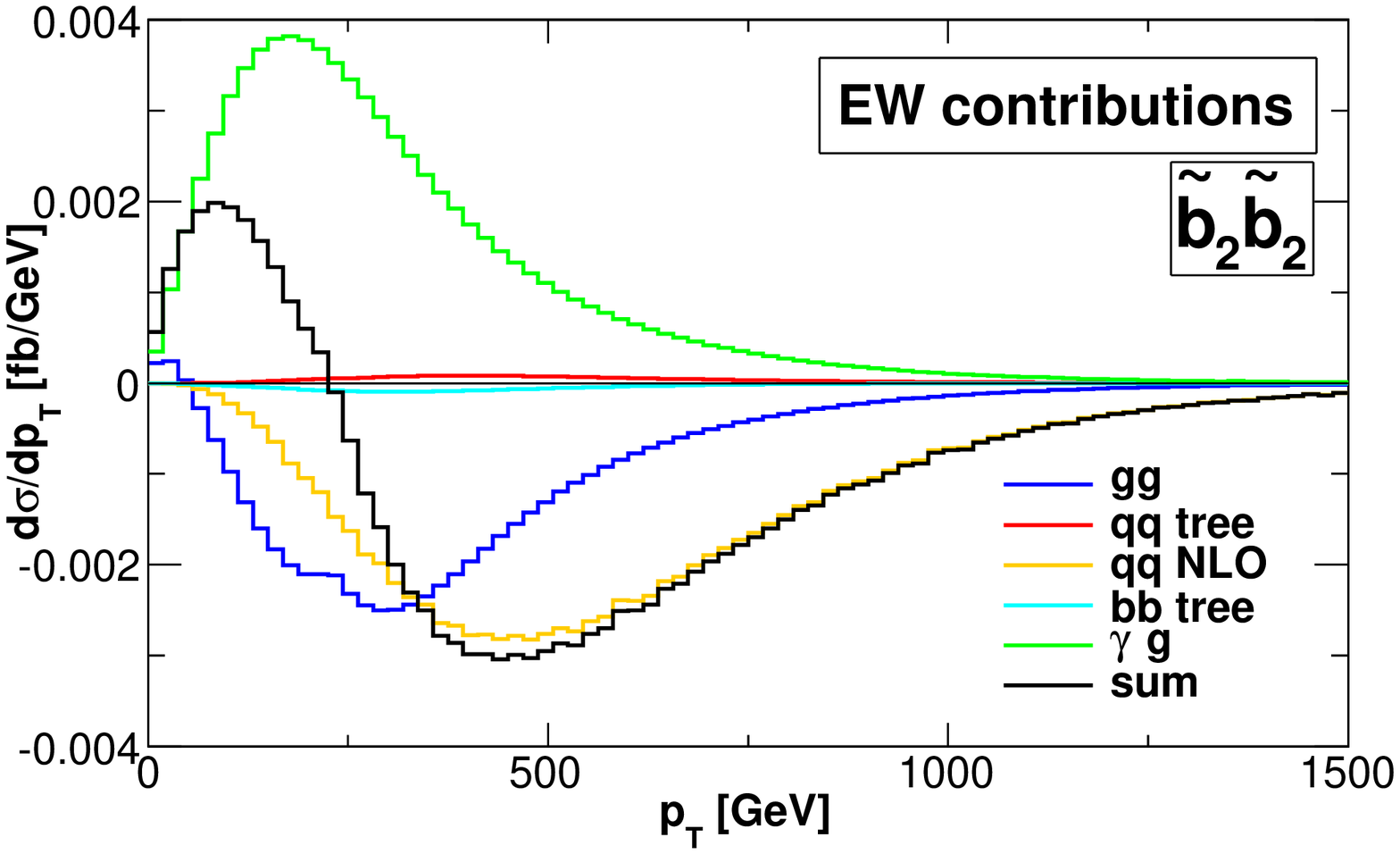}
  \includegraphics[width=.49\textwidth]
  {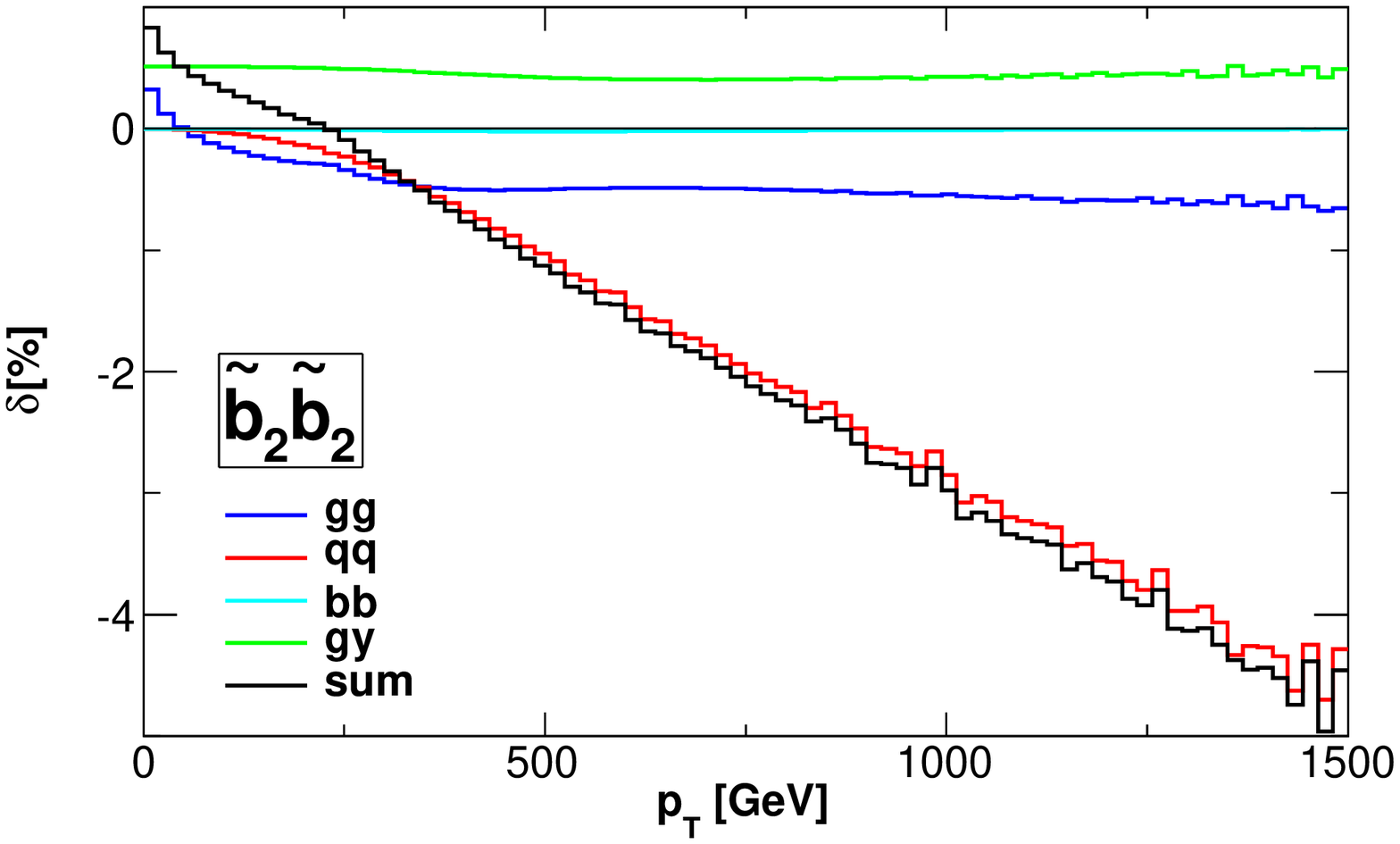}
  \\[-5pt]
  \hspace*{85pt}  (a) $p_T$ distribution
  \\[5pt]
  \includegraphics[width=.49\textwidth]
  {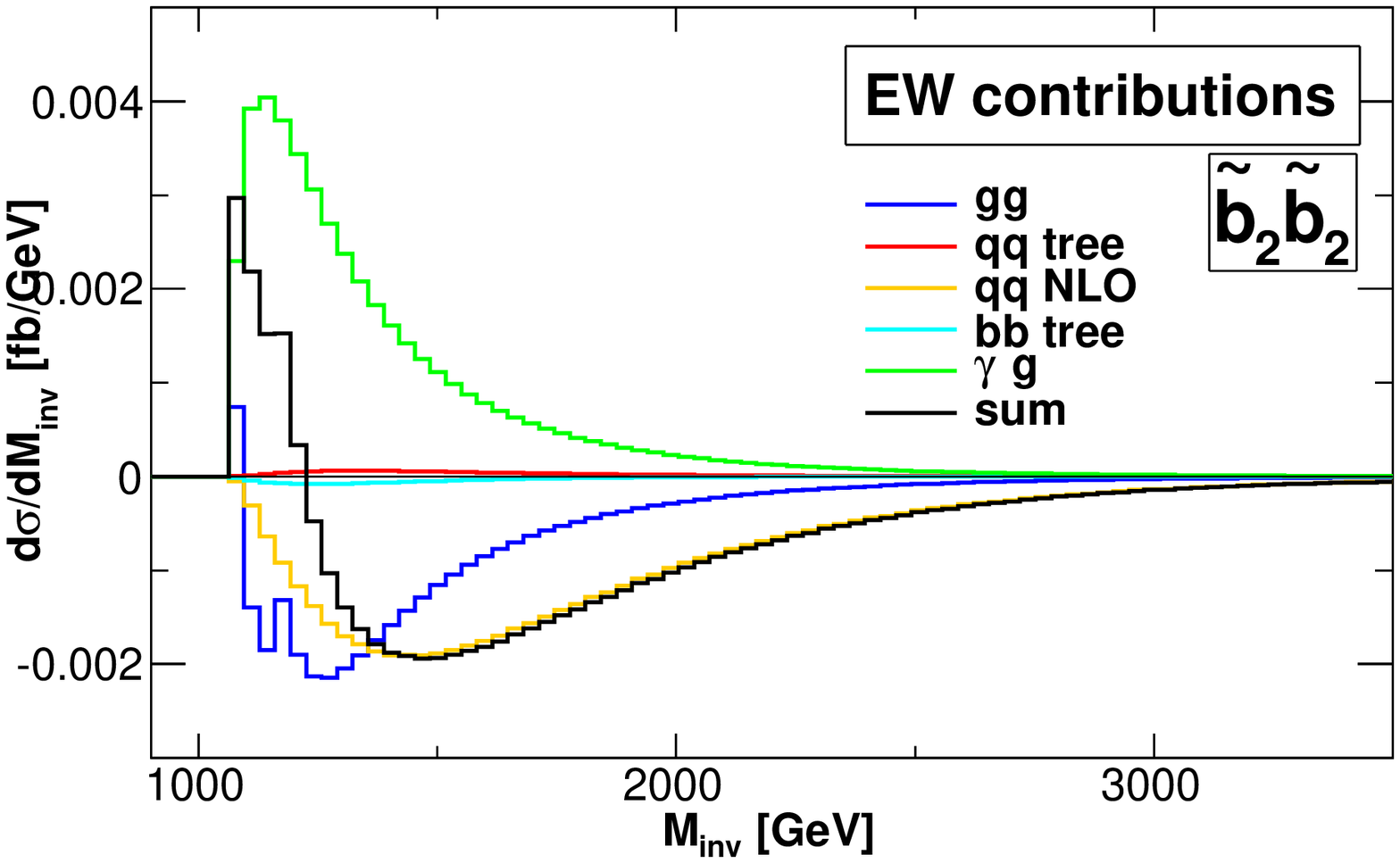}
  \includegraphics[width=.49\textwidth]
  {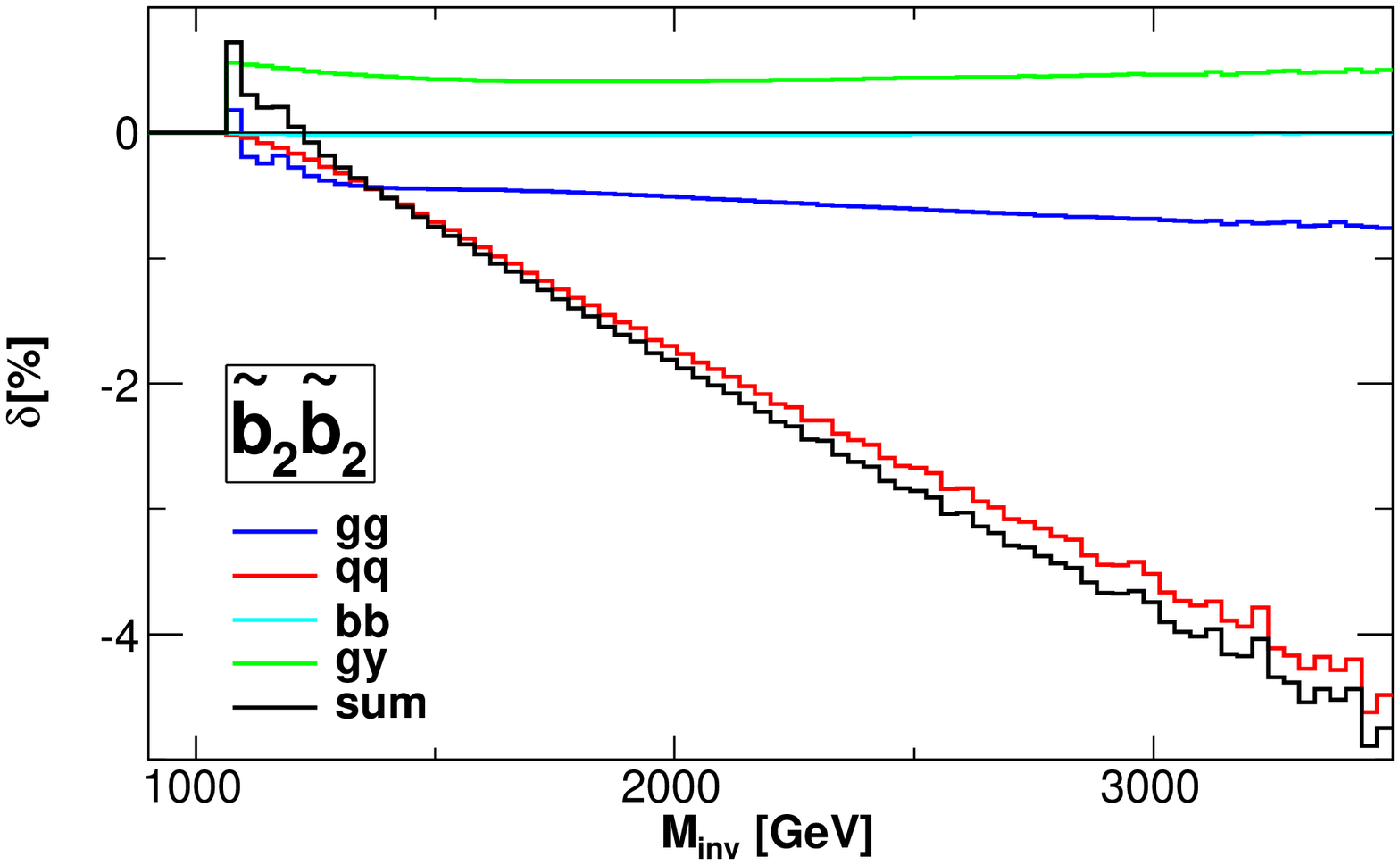}
  \\[-5pt]
  \hspace*{85pt}  (b) $M_{\text{inv}}$ distribution
  \\[5pt]
  \includegraphics[width=.49\textwidth]
  {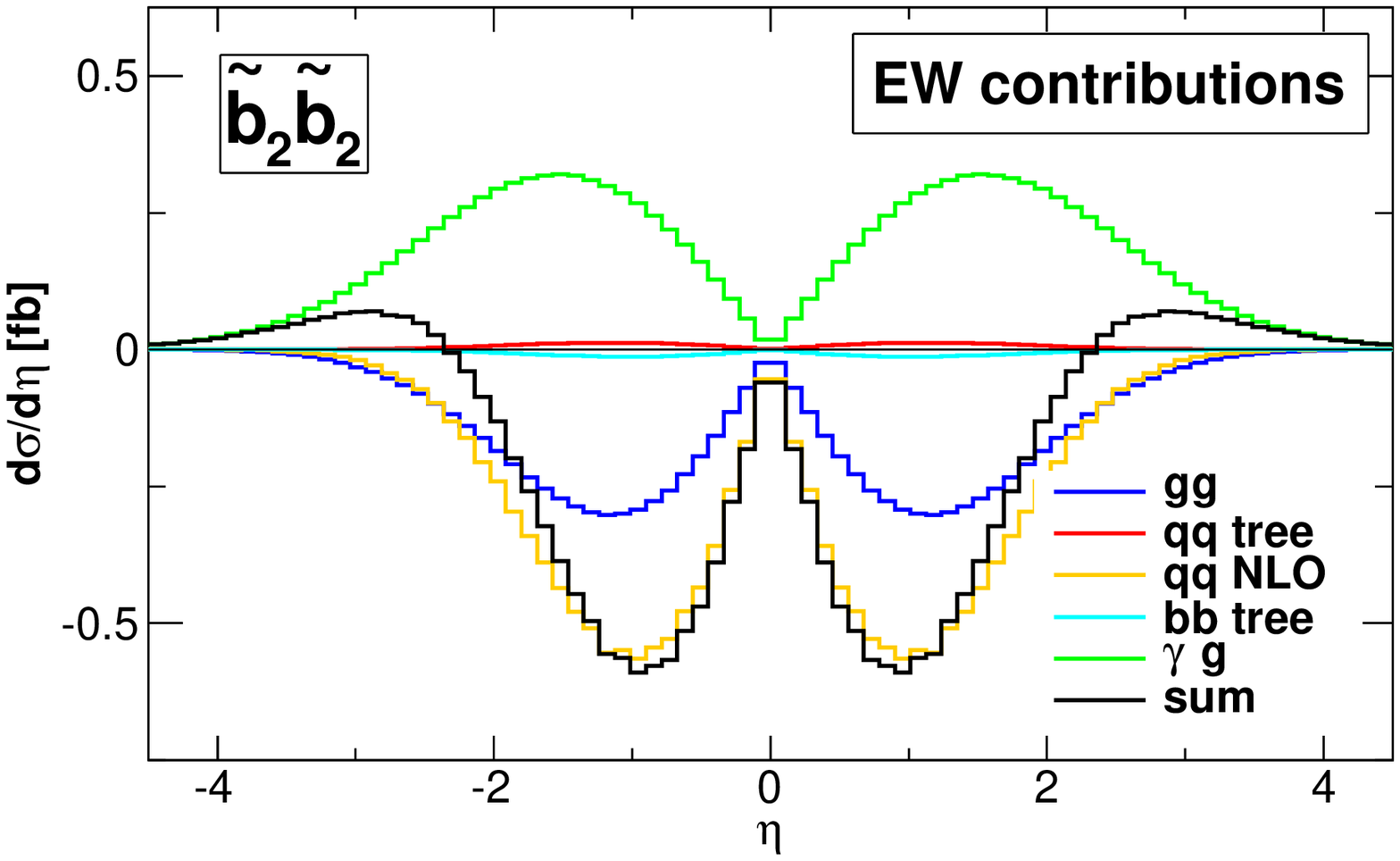}
  \includegraphics[width=.49\textwidth]
  {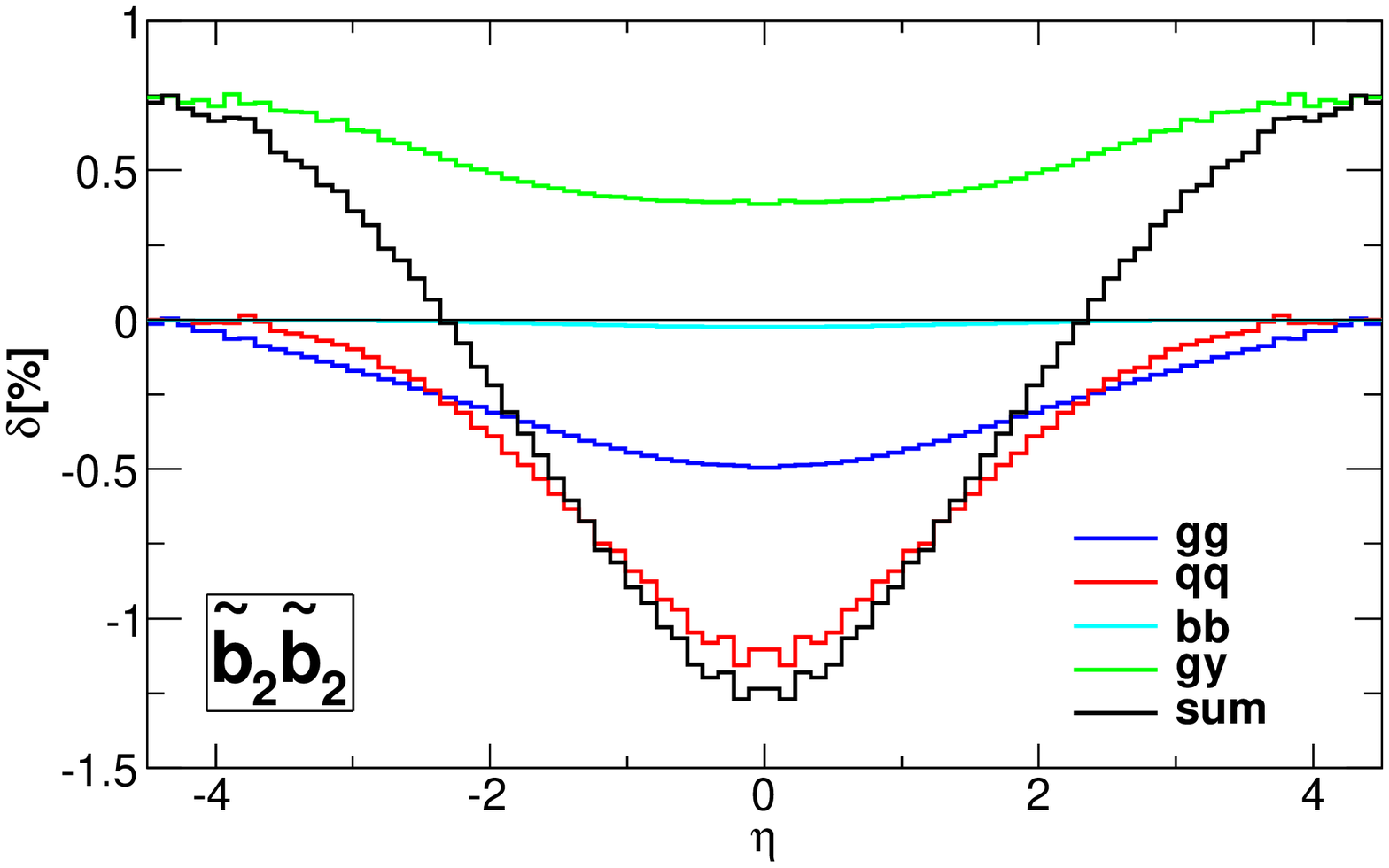}
  \\[-5pt]
  \hspace*{85pt}  (c) $\eta$ distribution
  \\[5pt]
  \caption{Same as Figure~\ref{fig:b1b1_SPS1aPrime_dist} but considering $\by\by^\ast$ production. 
}
  \label{fig:b2b2_SPS1aPrime_dist}
}

\FIGURE{
  \includegraphics[width=.49\textwidth]
  {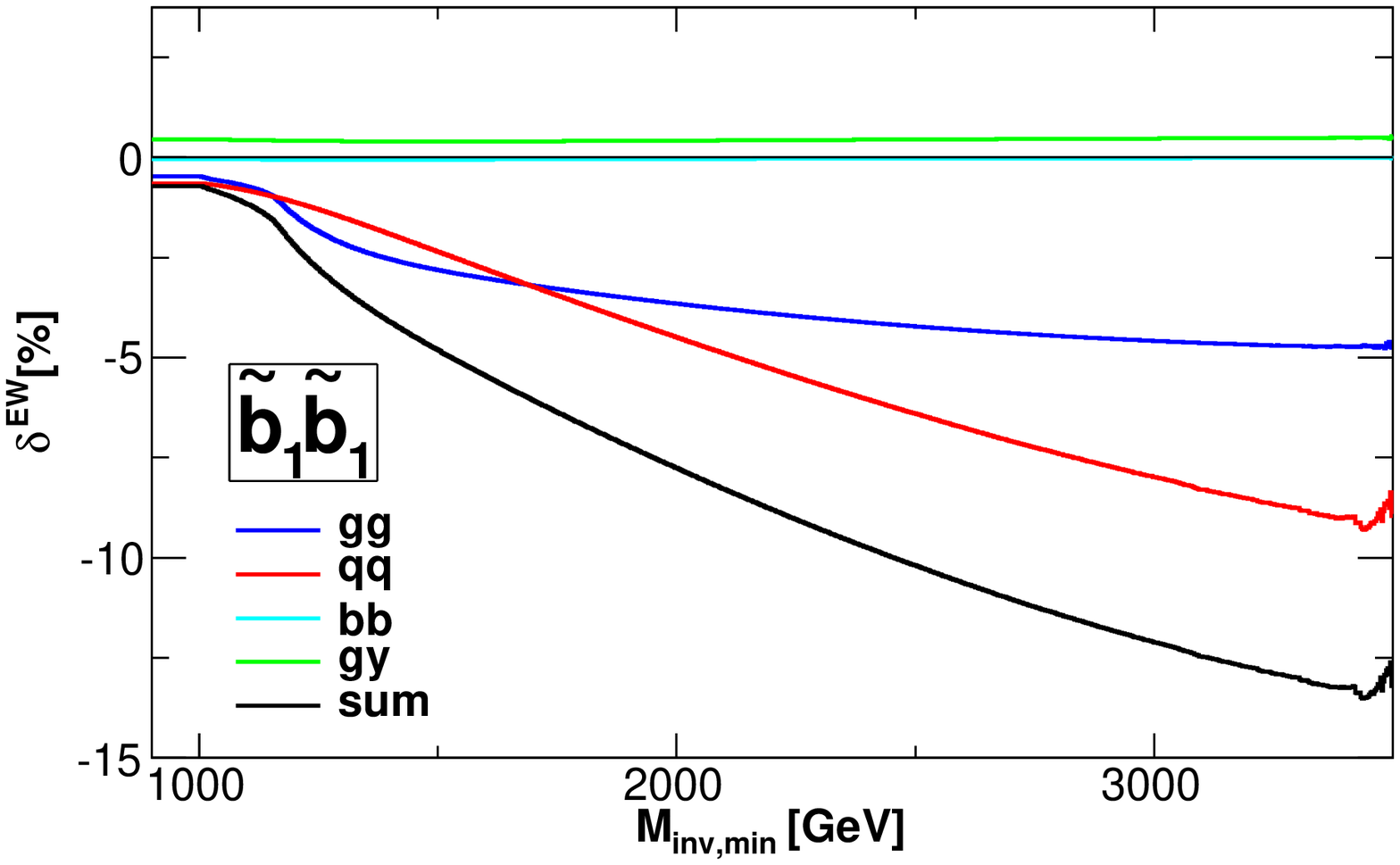}
  \includegraphics[width=.49\textwidth]
  {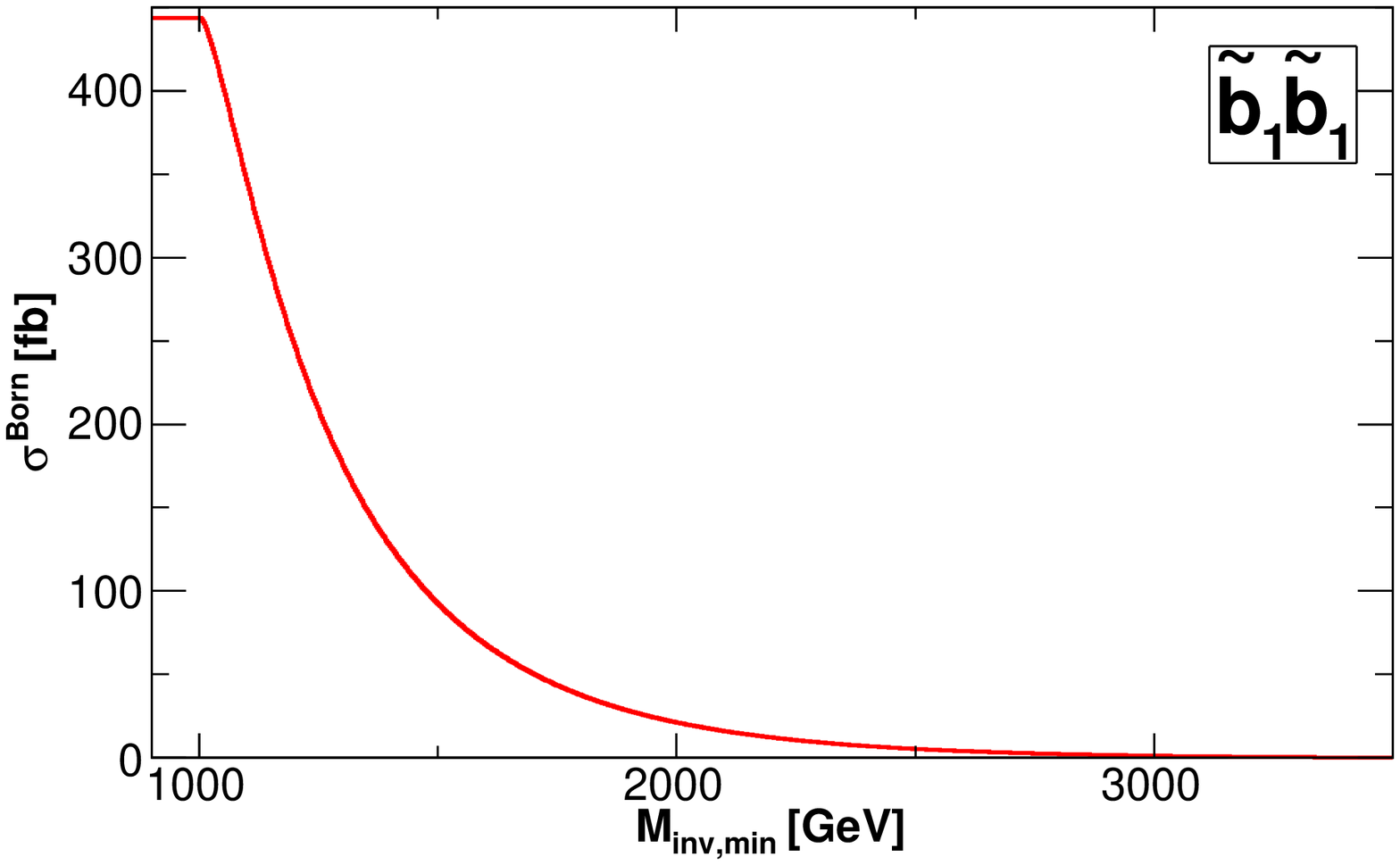}
  \includegraphics[width=.49\textwidth]
  {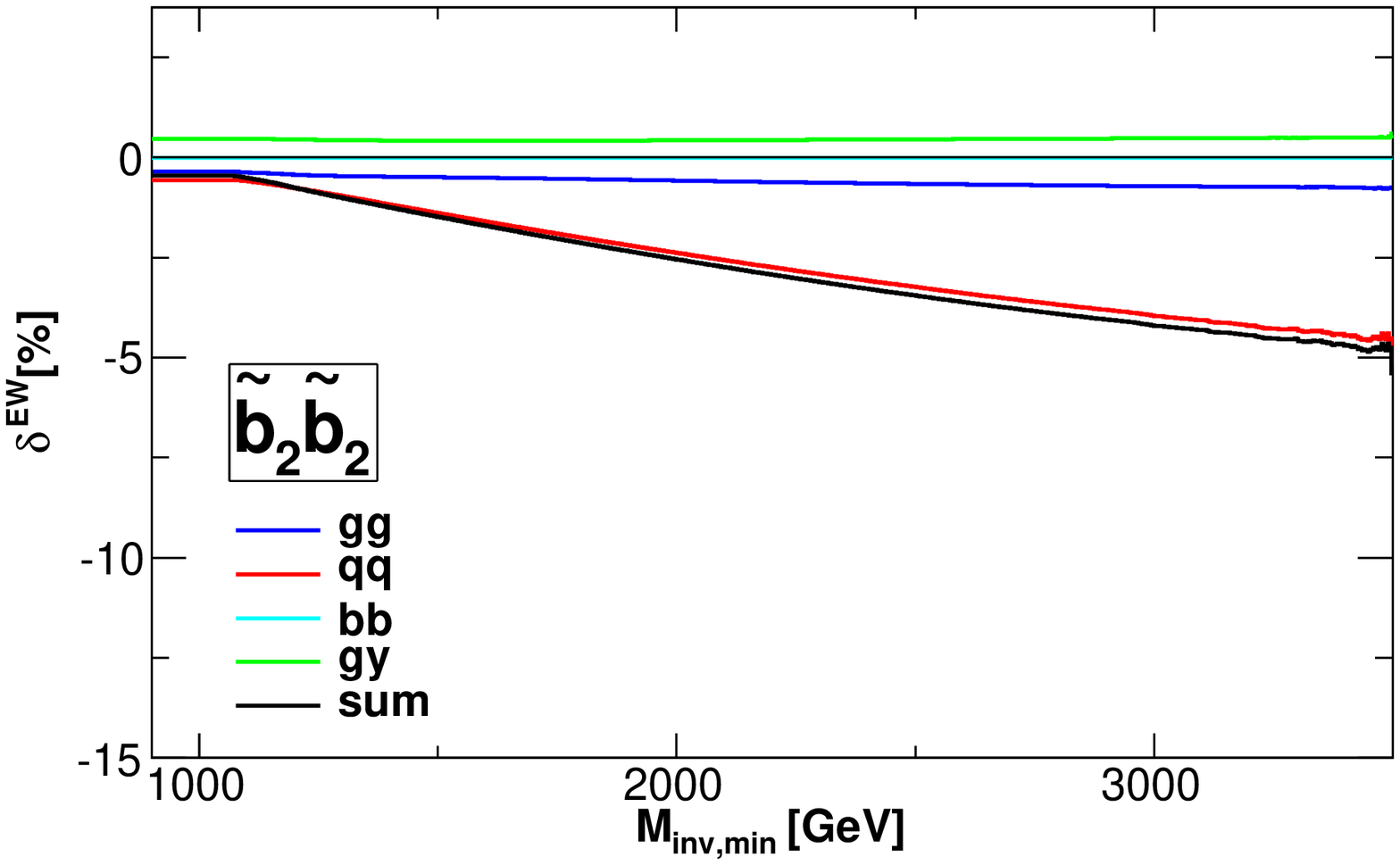}
  \includegraphics[width=.49\textwidth]
  {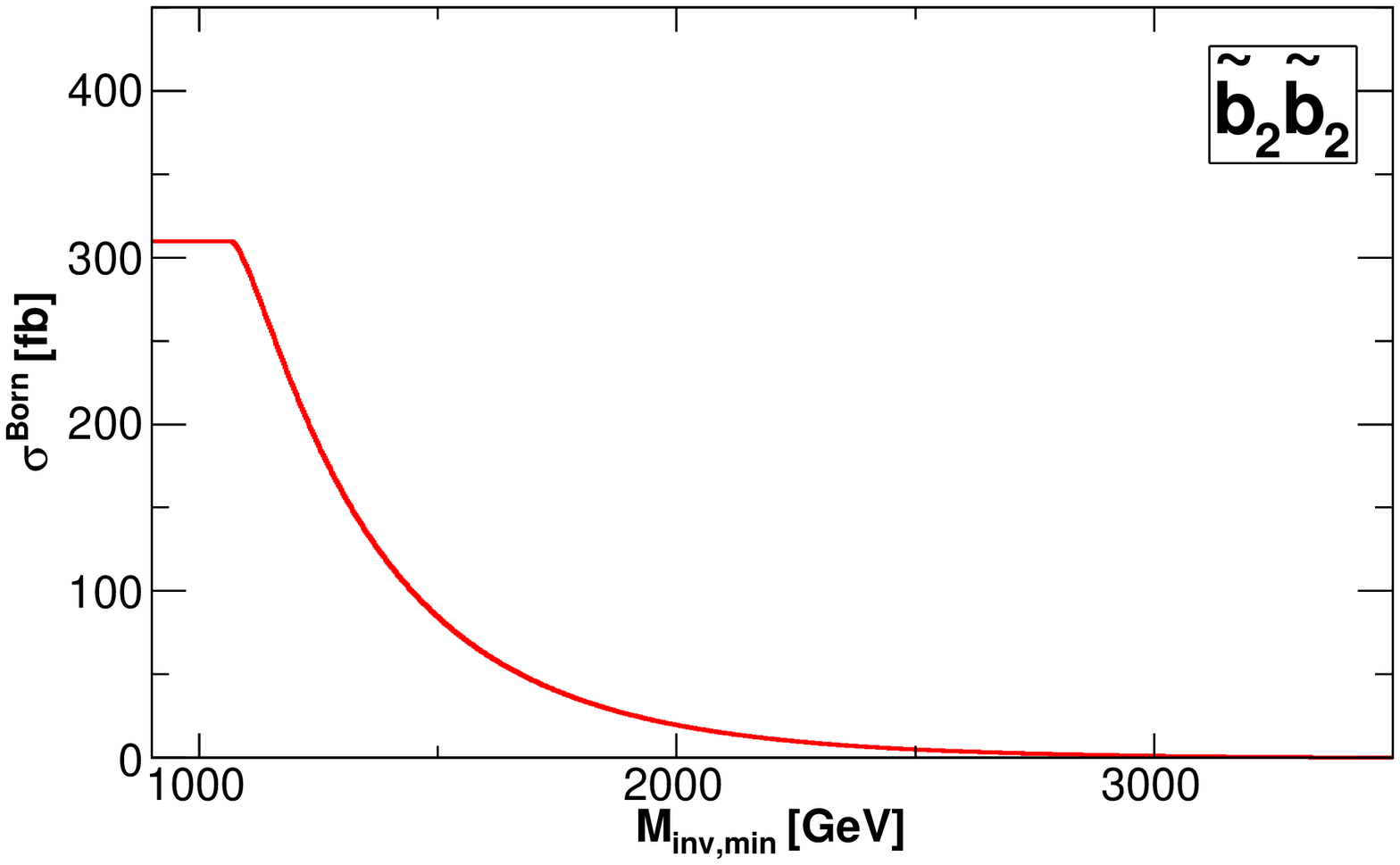}
  \vspace{-20pt}
  \caption{Left: Relative yield of the EW contributions of the
    different production channels in $\sigma(M_{\text{inv,min}})$.
    Right: LO QCD prediction for the same observable.}
  \label{fig:SPS1aPrime_cut}
}

\FIGURE{
  \includegraphics[width=.49\textwidth]
  {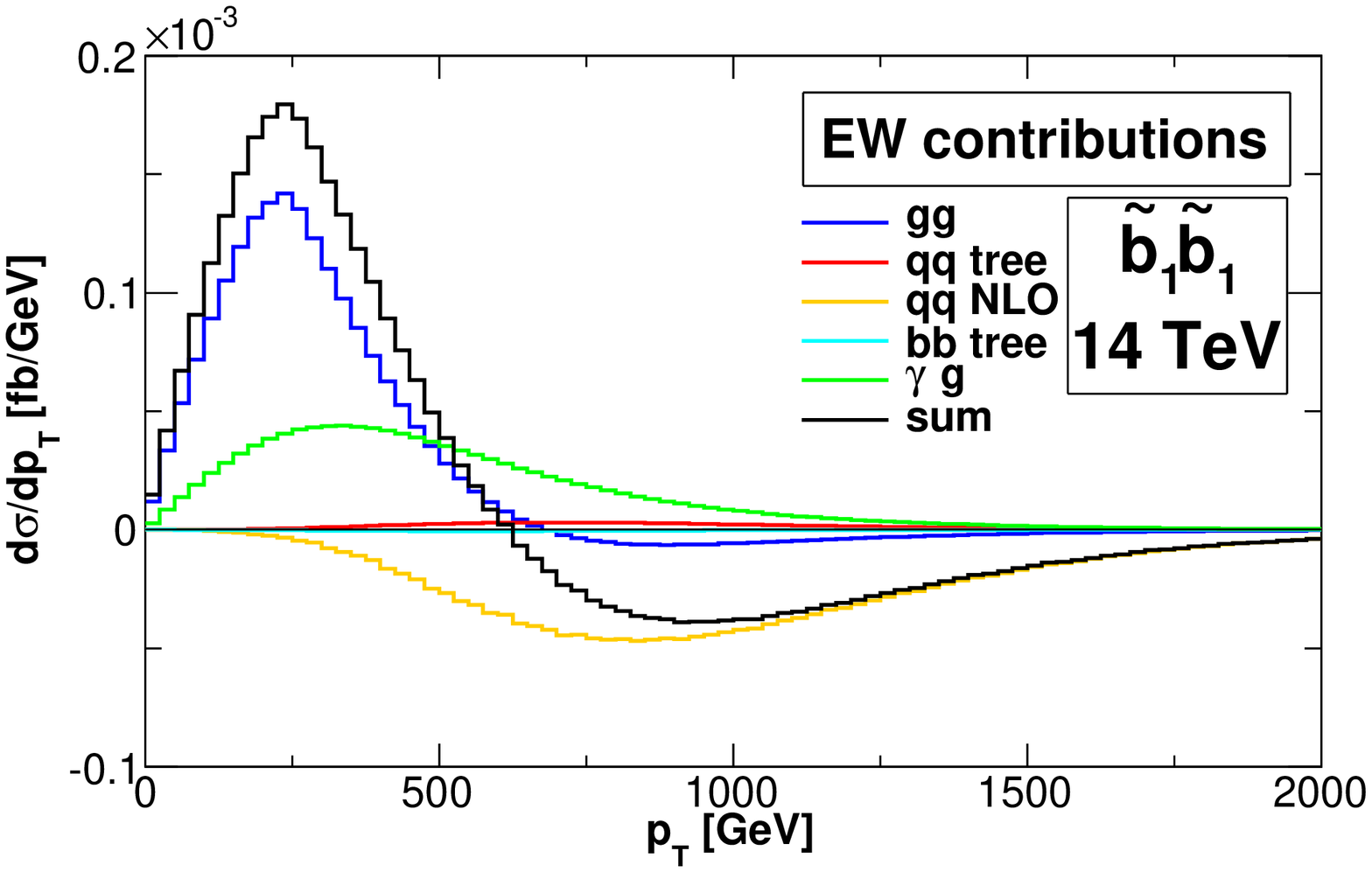}
  \includegraphics[width=.49\textwidth]
  {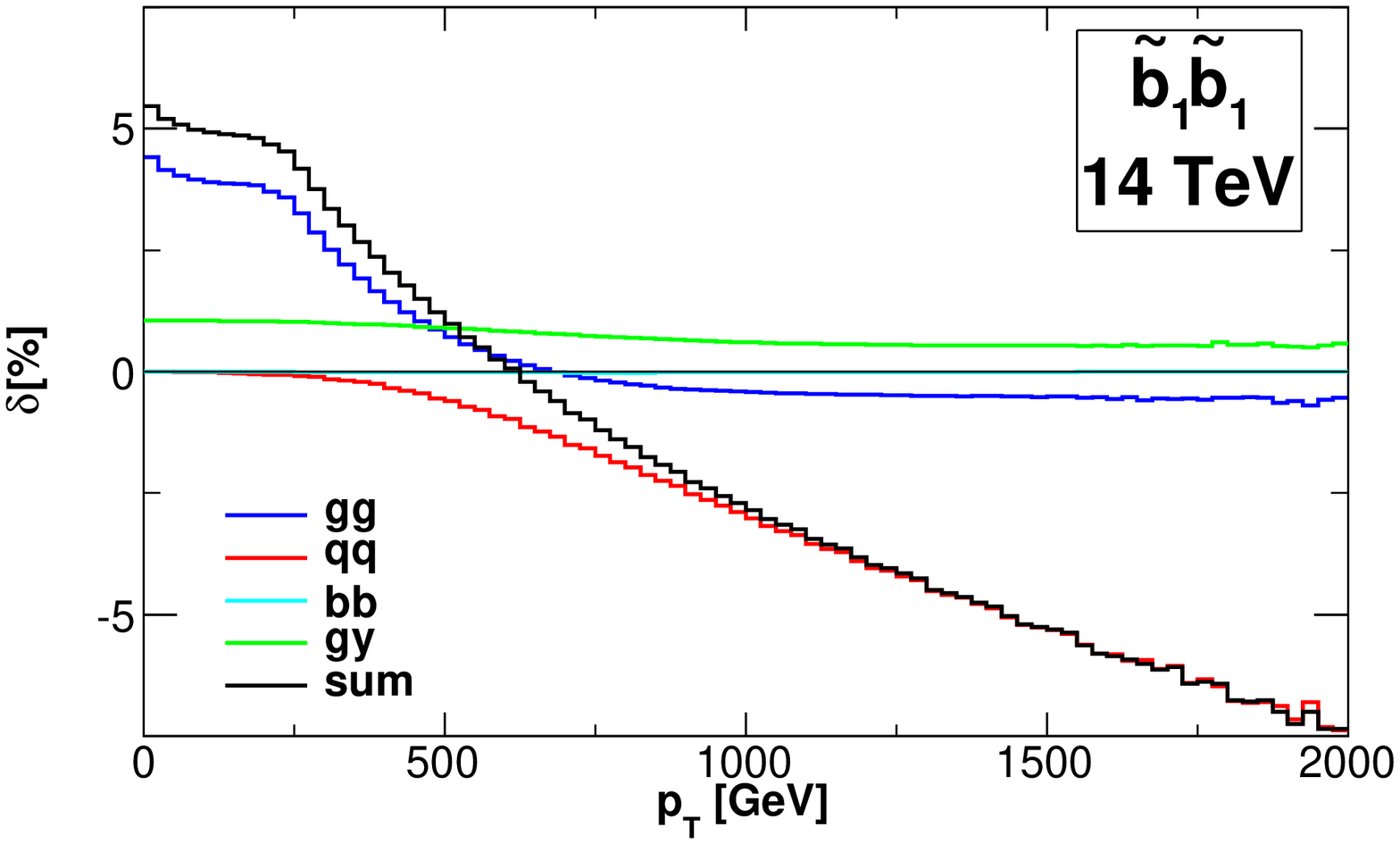}
    \includegraphics[width=.49\textwidth]
  {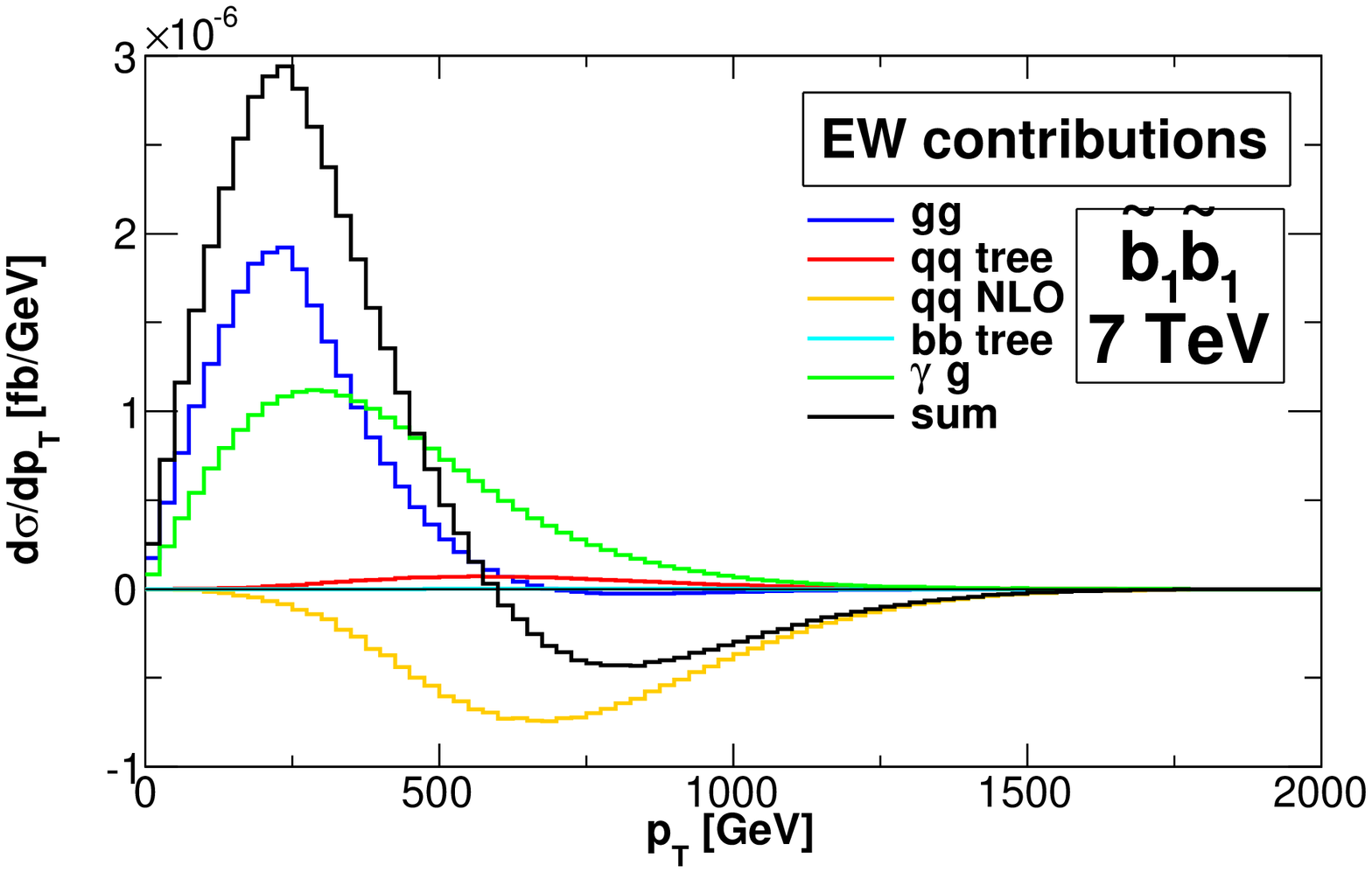}
  \includegraphics[width=.49\textwidth]
  {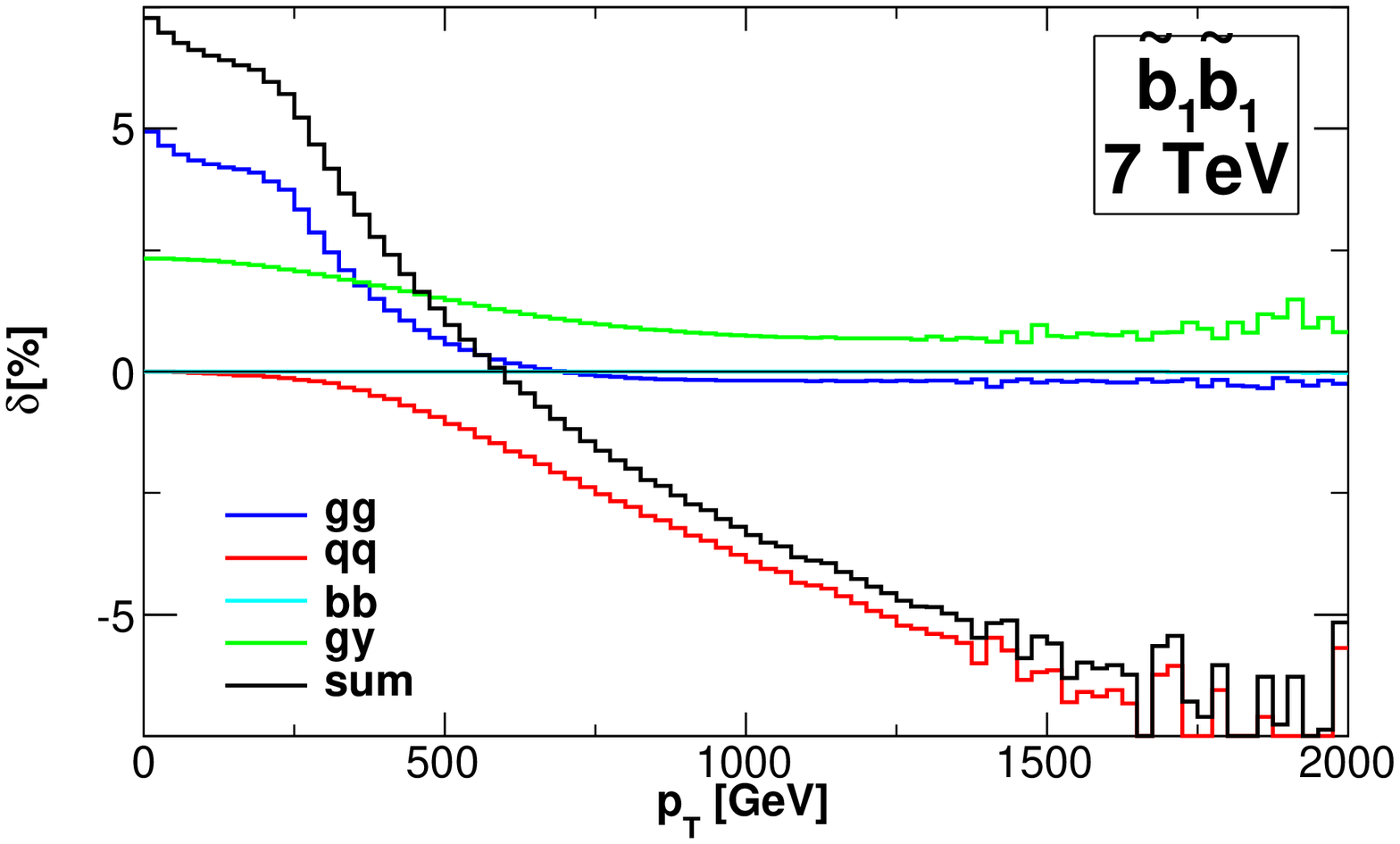}
  \vspace{-20pt}
  \caption{Differential transverse momentum distribution for
    $\bx\bx^\ast$ production within the SPS8 scenario for the
    14~\TeV{} (upper plots) and 7~\TeV{} (lower plots) LHC.}
  \label{fig:b1b1_SPS8_dist}
}
\clearpage


\section{Other processes leading to bottom-squark pair production}
\label{sect_other}
In this section we describe the remaining processes 
of (\ref{eq:tree})
leading to a pair of bottom-squarks in the final state.  The
processes are non-diagonal sbottom--anti-sbottom production and
(anti-)sbottom--(anti-)sbottom production,
\begin{equation}
PP \to  \tilde{b}_1 \tilde{b}^*_2, \;\tilde{b}_2\tilde{b}^\ast_1; \qquad
PP \to \ba \tilde{b}_{\beta},\; \ba^\ast \tilde{b}^\ast_{\beta}; \qquad \alpha,\beta \in \{1,2\}.
\label{eq:Other}
\end{equation}
\FIGURE[t]{
  \hspace{50pt}\includegraphics[width=.5\textwidth]
  {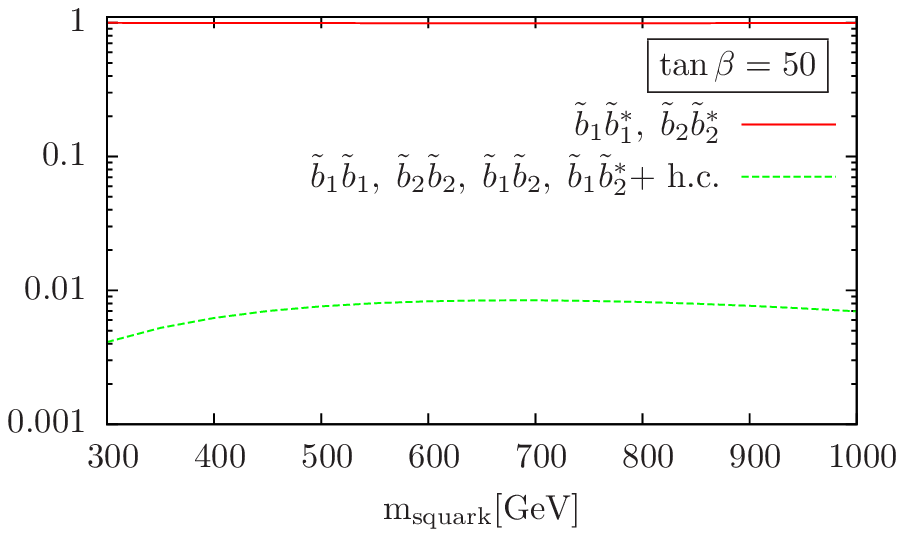}\hspace{50pt}
  \caption{Relative yield of the various hadronic processes with a
    bottom-squark pair in the final state as a function of the common
    squark mass breaking parameter $m_{\text{squark}}$. The parameter
    $m_{\text{squark}}$ is defined in Section~\ref{sec:parameterscan}.
  }
  \label{fig:channels-yield}
}
The leading-order cross sections, of the order $\Order(\alpha_s^2)$, are
given by
\begin{align}
\begin{split}
  {\rm d}\sigma^{\mbox{\tiny LO QCD}}_{PP \to  \tilde{b}_1\tilde{b}_2^\ast, \;   \tilde{b}_2\tilde{b}_1^\ast}(S) &=
   \int_{\tau_0(1,2)}^1 {\rm d}\tau \; \frac{{\rm d}L_{b \bar b}}{{\rm d}\tau}
  {\rm d} \hat  \sigma^{2,\, 0}_{b \bar b \to \tilde{b}_1 \tilde{b}_2^\ast}(\hat s) +
    \int_{\tau_0(1,2)}^1 {\rm d}\tau \; \frac{{\rm d}L_{b \bar b}}{{\rm d}\tau}
  {\rm d} \hat  \sigma^{2,\, 0}_{b \bar b \to \tilde{b}_2 \tilde{b}_1^\ast}(\hat s), \\
  {\rm d}\sigma^{\mbox{\tiny LO QCD}}_{PP \to  \tilde{b}_\alpha \tilde{b}_\beta, \;   \tilde{b}^\ast_\alpha \tilde{b}^\ast_\beta}(S)   &=
  \int_{\tau_0(\alpha,\beta)}^1 {\rm d}\tau \; \frac{{\rm d}L_{ b  b}}{{\rm d}\tau}
  {\rm d} \hat  \sigma^{2,\, 0}_{b b \to \tilde{b}_\alpha \tilde{b}_\beta}(\hat s) +
    \int_{\tau_0(\alpha,\beta)}^1 {\rm d}\tau \; \frac{{\rm d}L_{\bar b  \bar b}}{{\rm d}\tau}
  {\rm d} \hat  \sigma^{2,\, 0}_{\bar b \bar b \to \tilde{b}_\alpha^\ast \tilde{b}_\beta^\ast}(\hat s), \\
\end{split}
\end{align}
where $\tau_0(\alpha, \beta) \equiv (m_{\ba}+m_{\tilde{b}_\beta})^2/S$
is the production threshold. The parton luminosities ${\rm d}
L_{ij}/{\rm d}\tau$ are defined according to Eq.~(\ref{eq:Lumi}). The
tree-level EW contributions are of the order $\Order(\alpha_s \alpha)$
and $\Order(\alpha^2)$ and read as follows,
\begin{align}
\begin{split}
\label{eq:LOEWot}
 {\rm d}\sigma^{\mbox{\tiny LO EW}}_{PP \to  \tilde{b}_1\tilde{b}_2^\ast, \;   \tilde{b}_2\tilde{b}_1^\ast}(S) &=
     \int_{\tau_0(1,2)}^1 {\rm d}\tau \; \frac{{\rm d}L_{b \bar b}}{{\rm d}\tau}  
     \left [ 
  {\rm d} \hat  \sigma^{1,\, 1}_{b \bar b \to \tilde{b}_1 \tilde{b}_2^\ast}(\hat s)  +  {\rm d} \hat  \sigma^{0,\, 2}_{b \bar b \to \tilde{b}_1 \tilde{b}_2^\ast}(\hat s)   \right ]  \\
  &+
     \int_{\tau_0(1,2)}^1 {\rm d}\tau \; \frac{{\rm d}L_{b \bar b}}{{\rm d}\tau}  
  \left [
  {\rm d} \hat  \sigma^{1,\, 1}_{b \bar b \to \tilde{b}_2 \tilde{b}_1^\ast}(\hat s)  +   {\rm d} \hat  \sigma^{0,\, 2}_{b \bar b \to \tilde{b}_2 \tilde{b}_1^\ast}(\hat s)    \right ],\\
   {\rm d}\sigma^{\mbox{\tiny LO EW}}_{PP \to  \tilde{b}_\alpha \tilde{b}_\beta,  \;   \tilde{b}^\ast_\alpha \tilde{b}^\ast_\beta}(S)   &=
  \int_{\tau_0(\alpha,\beta)}^1 {\rm d}\tau \; \frac{{\rm d}L_{b  b}}{{\rm d}\tau} \left [
  {\rm d} \hat  \sigma^{1,\, 1}_{b b \to \tilde{b}_\alpha \tilde{b}_\beta}(\hat s) + {\rm d} \hat  \sigma^{0,\, 2}_{b b \to \tilde{b}_\alpha \tilde{b}_\beta}(\hat s) \right ] \\
  &+
    \int_{\tau_0(\alpha,\beta)}^1 {\rm d}\tau \; \frac{{\rm d}L_{\bar b \bar  b}}{{\rm d}\tau} \left [ 
  {\rm d} \hat  \sigma^{1,\, 1}_{\bar b \bar b \to \tilde{b}_\alpha^\ast \tilde{b}_\beta^\ast}(\hat s)+   {\rm d} \hat  \sigma^{0,\, 2}_{\bar b \bar b \to \tilde{b}_\alpha^\ast \tilde{b}_\beta^\ast}(\hat s)  \right ].\\
  \end{split}
\end{align}
The numerical impact of these processes on bottom-squark pair
production at tree-level is rather small. In
\figref{fig:channels-yield} we show the relative yield of diagonal
sbottom--anti-sbottom production, \eqref{eq:tree1}, and of all the
processes described in this section as a function of the soft breaking
parameter $m_{\mbox{\tiny squark}}$. Owing to the smallness of the
(anti-)bottom PDF, the tree-level contribution of the
processes~(\ref{eq:Other}) is below $1\%$.  We therefore do not
include NLO EW corrections; they are expected to be even smaller and
thus can be safely neglected.


\section{Conclusions}
We have studied the EW contributions to $b$-squark pair
production at the LHC within the MSSM.
The tree-level EW contributions from both the $q \bar
q$-annihilation and the gluon fusion channels have been supplemented 
by the contribution from the photon-induced channel, not included in 
previous discussions.
We have presented the first complete computation of the NLO EW
contributions to diagonal sbottom--anti-sbottom production;
together with the QCD corrections they complete the
NLO analysis of the class of squark--antisquark production
processes.

Renormalization of the $\tilde{b}$-sector is shown in detail, with a
check of the reliability of the adopted renormalization scheme in the
numerical analysis.  The potentially large corrections to the bottom
Yukawa couplings have been resummed, introducing effective couplings.
In the scenarios considered, the main effect is to change the value of
the masses of the bottom squarks.

The EW contributions to the total cross section are strongly scenario
dependent. However, in all the scenarios considered they are of the
order of few percents of the LO contribution in inclusive cross
sections. Their size is partly due to strong cancellations among
different channels. The EW contributions of different channels peak in
different regions of the phase space. Therefore, the EW contributions
are enhanced whenever kinematical cuts are applied. For similar
reasons the impact of the EW contributions is more important in
differential distributions, in particular in the high-energy region.
In the \SPA{} scenario, and in the case of $\tilde b_1 \tilde
b^\ast_1$ production, they can even exceed $10 \%$ of the LO
contributions for invariant mass and transverse momentum
distributions.

\begin{acknowledgments}
  E.M. is supported by the European Research Council under Advanced
  Investigator Grant ERC-AdG-228301. This work was supported in part
  by the European Communitys Marie-Curie Research Training Network
  under contract MRTN-CT-2006-035505 Tools and Precision Calculations
  for Physics Discoveries at Colliders (HEPTOOLS).
\end{acknowledgments}

\appendix
\numberwithin{figure}{section}
\newpage

\section{Renormalization of the stop and sbottom sector}
\label{sec:stopsbot}
In the MSSM the kinetic terms in the stop and sbottom sector read as
follows
\begin{equation}
\label{Eq:LagSquark2}
\mathcal{L}= \sum_{\tilde{q}=\tilde{t},\tilde{b}} \Bigg\{
\left(\partial_{\mu} \tilde{q}_L^*, \partial_{\mu} \tilde{q}_R^* \right)
\left(\begin{array}{cc} \partial^{\mu}\tilde{q}_L \\  \partial^{\mu}\tilde{q}_R  \end{array}\right)-
\left(\tilde{q}_L^*,\tilde{q}_R^* \right){\bf M}^2_{\tilde{q}}
\left(\begin{array}{cc} 
\tilde{q}_L \\  \tilde{q}_R  \end{array}\right) \Bigg\},
\qquad 
\end{equation}
with the squared-mass matrix 
\begin{equation}
  \label{Eq:MassMatrix2}
  {\bf M}^2_{\tilde{q}}= 
  \left( \begin{array}{cc}  M_{L}^2 + m_{q}^2 + m_Z^2 \cos 2
      \beta(T^3_q- e_q  s^2 _{\rm w})  &   
      m_{q}(A_{\tilde{q}}-\mu \lambda_{\tilq})
      \\ 
      m_{q}(A_{\tilde{q}}-\mu \lambda_{\tilq})
      &
      M_{\tilde{q},R}^2 + m_{q}^2 + e_q M_Z^2 \cos
      2\beta s^2 _{\rm w} \\ 
\end{array} \right).
\end{equation}
$M_L,~M_{\tilde{q},R},~A_{\tilde{q}}$ are soft breaking parameters,
while $\mu$ is the supersymmetric Higgs mass
parameter. $m_q,~e_q,~T^3_q$ are the mass, the charge and the isospin
of the quark $q$, respectively. $s_{\rm w}$ is the sine of the weak
mixing angle, while $\lambda_{\tilde t} =\cot \beta$ and
$\lambda_{\tilde b} = \tan \beta$. The matrix (\ref{Eq:MassMatrix2})
is symmetric and can be diagonalised by an orthogonal matrix ${\bf
  R}_{\tilde{q}}$ such that
\begin{displaymath}
\left(\begin{array}{cc} \tilde{q}_1 \\  \tilde{q}_2  \end{array}\right) = 
{\bf R}_{\tilde{q}}     \left(\begin{array}{cc} \tilde{q}_L \\  \tilde{q}_R  \end{array}\right), 
~~~~~~~
{\bf R}_{\tilde{q}}=\left( \begin{array}{cc}   R_{\tilde{q}1,1}     &   
                           R_{\tilde{q}1,2}                             \\
                           R_{\tilde{q}2,1}                         & 
                           R_{\tilde{q}2,2}                              \\ 
\end{array} \right).
\end{displaymath}
In the rotated basis the squared-mass matrix is diagonal
\begin{displaymath}
{\bf D}_{\tilde{q}}={\bf R}_{\tilde{q}}{\bf M}_{\tilde{q}}{\bf R}_{\tilde{q}}^\top = \left( \begin{array}{cc}  
                         m_{\tilde{q}_1}^2  &   
                         0                    \\
                         0                  & 
                         m_{\tilde{q}_2}^2    \\ 
\end{array} \right).
\end{displaymath}
The rotation matrix ${\bf R}_{\tilde{q}}$ can be parametrized in terms
of a mixing angle $\theta_{\tilde q}$
\begin{displaymath}
{\bf R}_{\tilde{q}}=\left( \begin{array}{cc}   \cos\theta_{\tilde{q}}     &   
                           \sin\theta_{\tilde{q}}                             \\
                          -\sin\theta_{\tilde{q}}                         & 
                           \cos\theta_{\tilde{q}}                              \\ 
\end{array} \right)
~~~ \mbox{or} ~~~
{\bf R}_{\tilde{q}}=\left( \begin{array}{cc}   -\sin\theta_{\tilde{q}}     &   
                           \cos\theta_{\tilde{q}}                             \\
                          \cos\theta_{\tilde{q}}                         & 
                           \sin\theta_{\tilde{q}}                              \\ 
\end{array} \right),
\end{displaymath}
depending on the sign of the determinant of ${\bf R}_{\tilde{q}}$.
The mixing angle $\theta_{\tilq}$ and the trilinear coupling
$A_{\tilq}$ are related via
\begin{equation}
\label{Eq:RelationTeta2}
\sin2\theta_{\tilde{q}} = \frac{2 m_q (A_{\tilde{q}}-\mu \lambda)}{m_{\tilde{q}_1}^2 -
 m_{\tilde{q}_2}^2} \; \xi_{\tilde q},~~~~~~~ \xi_{\tilq}  \equiv \mbox{det} \left [ {\bf R}_{\tilde{q}} \right ]. 
\end{equation}
Because of SU(2)-invariance the eigenvalues of ${\bf M}^2_{\tilde{t}}$
and ${\bf M}^2_{\tilde{b}}$ are connected. In particular they have to
satisfy the relation
\begin{equation}
\label{Eq:SU2constraint2}
R_{\tilde b 1,1}^2  m_{\tilde{b}_1}^2  + R_{\tilde b 2,1}^2   m_{\tilde{b}_2}^2 - m_{b}^2 =
R_{\tilde t 1,1}^2   m_{\tilde{t}_1}^2  + R_{\tilde t 2,1}^2  
m_{\tilde{t}_2}^2 - m_{t}^2 - m_W^2 \cos 2\beta.  
\end{equation}
\subsection*{Definition of the ``$\overline{\mbox{DR}}$ bottom-quark mass'' 
 renormalization scheme}

In this scheme the independent parameters are chosen to be
\begin{equation}
m^2_{\tilde t_1},~~m^2_{\tilde t_2} ,~~m^2_{\tilde b_2},~~m_t,~~m_b,~~A_{\tilde b},~~\theta_{\tilde t}.
\end{equation}
The squark mass squared and the top quark mass are defined in the
on-shell (OS) scheme, while the bottom quark mass and the trilinear
coupling $A_{\tilde b}$ are fixed using the $\overline{\mbox{DR}}$
prescription,
\begin{eqnarray}
\delta m^2_{\tilde q_a} &=& \Re \left \{  \Sigma_{\tilde q\, a,a}(m^2_{\tilde q_a}) \right \}, \hspace{0.2cm} \mbox{with } \tilde q_a =   \tilde t_1, \tilde t_2,\tilde b_2, \nonumber\\
\delta m_t &=& \frac{m_t}{2} \Re \left \{  \Sigma_{t L}(m_t) +   \Sigma_{t R}(m_t) +  2  \Sigma_{t S}(m_t)   
\right \}, \nonumber \\
\delta m_b &=& \frac{m_b}{2} \Re \left \{  \Sigma^{\text{div.}}_{b
    L}(m_b) +   \Sigma^{\text{div.}}_{b R}(m_b) +  2
  \Sigma^{\text{div.}}_{b S}(m_b)   \right \},
\label{eq:Ab}\\
 \delta A_{\tilde b} &=& \frac{1}{m_b} \Bigg [
  \frac{R_{\tilde b 1,1}  R_{\tilde b 2,2} +R_{\tilde b 1,2}  R_{\tilde b 2,1}}{2} \left (
\Re \{  \Sigma^{\text{div.}}_{\tilde b\, 1,2}(m^2_{\tilde b_2}) \}  +
  \Re \{  \Sigma^{\text{div.}}_{\tilde b\, 1,2}(m^2_{\tilde b_1}) \} \right ) \nonumber \\
&~&   + R_{\tilde b 1,1}  R_{\tilde b 1,2} \left (\Re \left \{  \Sigma^{\text{ div.}}_{\tilde b\, 1,1}(m^2_{\tilde b_1}) \right \} - \Re \left \{  \Sigma^{\text {div.}}_{\tilde b\, 2,2}(m^2_{\tilde b_2}) \right \} \right )  
 - \frac{A_{\tilde b} - \mu \tan\beta}{2}   \delta m_b \Bigg ] \nonumber \\
 &~& + \delta \mu  \tan \beta +  \mu \delta \tan \beta. \nonumber
\end{eqnarray}
$\delta \mu$ and $\delta \tan \beta$ appearing in \eqref{eq:Ab} are
defined in the $\overline{\mbox{DR}}$ scheme.  $\Sigma^{\text{div.}}$
is the divergent part of the scalar self energies defined according to
the following Lorentz decomposition,
\begin{equation}
\Sigma_q(p)  = \pslash  \omega_- \Sigma_{qL}(p) + \pslash \omega_+ \Sigma_{qR}(p) 
+  m_q \Sigma_{qS}(p). 
\label{eq:LorDec}
\end{equation}
The stop mixing angle is defined according to
 \begin{equation}
 \delta \theta_{\tilde t}  = \frac{ \xi_{\tilde t}  (\Re \{  \Sigma_{\tilde t\, 1,2}(m^2_{\tilde t_1}) \}  -   \Re \{\Sigma_{\tilde t\, 1,2}(m^2_{\tilde t_2}) \} ) }{2 (m^2_{\tilde t_1}   
- m^2_{\tilde t_2} )   }. 
\end{equation}
In the ``$\overline{\mbox{DR}}$ bottom-quark mass'' scheme $A_{\tilde
  t}$, $\theta_{\tilde{b}}$ and $m^2_{\tilde b_1}$ are dependent
quantities. Their counterterms read as follows
\begin{eqnarray}
\delta A_{\tilde t} &=&  \frac{1}{m_t} \Bigg [
R_{\tilde t 1,1} R_{\tilde t 1,2} 
 \left ( \delta m^2_{\tilde t_1} - \delta m^2_{\tilde t_2} \right ) + \xi_{\tilde t} 
 \left (R_{\tilde t 1,1} R_{\tilde t 2,2}- R_{\tilde t 1,2} R_{\tilde t 2,1}  \right) \, \delta \theta_{\tilde t}( m^2_{\tilde t_1} -  m^2_{\tilde t_2} ) \nonumber \\
&~& - (A_{\tilde t} - \mu \cot \beta) \delta m_t   
\Bigg ] + \delta \mu \cot \beta - \mu \cot^2\beta \delta \tan \beta, \\[0.5ex]
\label{eq:At}
\delta m^2_{\tilde{b}_1} &=& \frac{1}{R^2_{\tilde b1,1}} \Bigg [ 
(1-2R^2_{\tilde b1,2}) \Bigg ( 
R^2_{\tilde t 1,1}  \delta m^2_{\tilde t_1} +R^2_{\tilde t 1,2}  \delta m^2_{\tilde t_2} 
-2 \xi_{\tilde t}  R^2_{\tilde t 1,2}R^2_{\tilde t 2,2} 
( m^2_{\tilde t_1}- m^2_{\tilde t_2}) \delta \theta_{\tilde t}  \nonumber \\
&~& -2 m_t \delta m_t
-\delta m^2_W \cos 2\beta -m^2_W \delta \cos 2\beta 
\Bigg ) +R^2_{\tilde b1,2} \delta m^2_{\tilde b_2}  \nonumber \\
&~& +
2 R_{\tilde b1,1}R_{\tilde b1,2} \left ( \delta A_{\tilde b} - \delta \mu \tan\beta -
\mu \delta \tan \beta\right) \nonumber \\
&~&  + \delta m_b  \Big (
2 R_{\tilde b1,1}R_{\tilde b1,2} (A_{\tilde b} -\mu \tan \beta)  
+2(1-2 R^2_{\tilde b1,2} ) m_b
\Big )
\Bigg ],  \\
\delta \theta_{\tilde b} &=&
 \frac{\xi_{\tilde b} (m^2_{\tilde b_1} -  m^2_{\tilde b_2})^{-1}}{R_{\tilde b1,1}R_{\tilde b2,2}+R_{\tilde b1,2}R_{\tilde b2,1}} 
\Bigg [  R_{\tilde b1,1}R_{\tilde b1,2} \left (  \delta m^2_{\tilde b_2}   -  \delta m^2_{\tilde b_1}   \right ) + (A_{\tilde b} -\mu \tan \beta ) \delta m_b  \nonumber \\
&~& + m_b ( \delta A_{\tilde b} -\mu \delta \tan \beta - \delta \mu \tan \beta  )
\Bigg]. 
\label{eq:thetaA}
\end{eqnarray}
The mass of the $W$ boson, $m_W$, is renormalized on-shell.
In order to get finite Green functions, we also need the wavefunction
renormalization of the bottom and sbottom,
\begin{align}
\left(\begin{array}{cc}\tilde{q}_1^{\text{ bare}} \\
   \tilde{q}_2^{\text{ bare}}  \end{array}\right) &= 
\left({\bf 1}+\frac{\delta {\bf Z}_{\tilde{q}}}{2} \right)
\left(\begin{array}{cc} \tilde{q}_1^{\text{ ren}} \\
    \tilde{q}_2^{\text{ ren}}  \end{array}\right),
    ~~~~~~
\delta {\bf Z}_{\tilde{q}}  =     \left( \begin{array}{cc}  
                         \delta Z_{\tilq 1,1}    &   
                         \delta Z_{\tilq 1,2}    \\
                         \delta Z_{\tilq 2,1}    & 
                        \delta Z_{\tilq  2, 2}    \\ 
\end{array} \right), \\
\omega_\pm q^{\text{bare}} &= \omega_\pm  \left  (1+\frac{1}{2} \delta
Z_{qR/L} \right ) q^{\text{ren}} .
\end{align}
In the case of $\tilde b \tilde b^\ast$ production only the diagonal
entries of the matrix $\delta {\bf Z}_{\tilde{q}}$ are needed. They
are defined as
\begin{displaymath}
\delta Z_{\tilq \alpha,\alpha}=-\Re \left\{  
 \Sigma^{\prime}_{\tilq\alpha,\alpha }(m_b^2) 
\right \}, 
   ~~~~~~
\Sigma^{\prime}(m_b^2)  \equiv
\frac{\partial\Sigma(k^2)}
{\partial k^2}_{\big|{k^2=m^2_{\tilde{q}_\alpha}}}.
\end{displaymath}
The wavefunction renormalization constants of the bottom-quark read as follows
\begin{equation}
  \delta Z_{bL/R} = - \Re \left \{  \Sigma_{bL/R}(m_b^2) + m_b^2
    \Big( \Sigma^{\prime}_{bL}(m_b^2) +\Sigma^{\prime}_{bR}(m_b^2) +
      2 \Sigma^{\prime}_{bS}(m_b^2) \Big) \right \}.
\end{equation}

In the processes considered, $\tilde b_1$ is an external particle and
its mass has to be defined on-shell. Therefore we set the value of the
$\tilde b_1$ mass to its OS value, obtained using the following
relation
\begin{align}
\begin{split}
        m_{\tilde{b}_1,~\os}^{2}&=  m^2_{\tilde{b}_1}
                + \delta m_{\tilde{b}_1}^2 - \Re \left \{ \Sigma_{\tilde{b} 1,1}(m_{\tilde{b}_1}^2)\right \}.
\label{eq_app_4thmass}
\end{split}
\end{align}
The renormalization constant of $m^2_{\tilde{b}_1}$ enters the counter
terms in the last diagram of \figref{fig:feynman_ggvirt} and it is
fixed in accordance with our choice of the $\tilde b_1$ mass,
\begin{equation}
 \delta m_{\tilde{b}_1,~\os}^2 = \Re \left \{ \Sigma_{\tilde{b} 1,1} (m_{\tilde{b}_1,~\os}^2)\right \}.
\end{equation}

The renormalization constants have to be evaluated at $\Order(\alpha)$
and enter the calculation via the counterters in
\figref{fig:feynman_ggvirt} and \figref{fig:feynman_qqvirtEW}. The
explicit expressions for the counterterms are given by

\bgroup
\unitlength=0.6bp%
\hfill
\begin{tabular}{cl}
  \begin{feynartspicture}(100,100)(1,1)
    \FADiagram{}
    \FAProp(0.,10.)(10.,10.)(0.,){/ScalarDash}{1}
    \FALabel(0.,15.18)[t]{$\ba$}
    \FAProp(10.,10.)(20.,10.)(0.,){/ScalarDash}{1}
    \FALabel(20.,15.18)[t]{$\ba$}
    \FAVert(10.,10.){1}
  \end{feynartspicture}
  &
  \parbox{0.6\textwidth}{\vspace{-55pt}\qquad$=$\qquad
    $\I\left[\left(p^2-m_{\ba}^2\right) \delta
      Z_{\Sb\alpha,\alpha} - \delta m_{\ba}^2\right]$,
  }\\
  \begin{feynartspicture}(100,100)(1,1)
    \FADiagram{}
    \FAProp(-0.,17.)(10.,10.)(0.,){/Cycles}{0}
    \FALabel(0.,16.)[tr]{g}
    \FAProp(0.,3.)(10.,10.)(0.,){/Cycles}{0}
    \FALabel(0.,6.)[tr]{g}
    \FAProp(10.,10.)(20.,17.)(0.,){/ScalarDash}{1}
    \FALabel(20.,16.)[tl]{$\ba$}
    \FAProp(10.,10.)(20.,3.)(0.,){/ScalarDash}{-1}
    \FALabel(20.,6.)[tl]{$\ba$}
    \FAVert(10.,10.){1}
  \end{feynartspicture}
  &
  \parbox{0.6\textwidth}{\vspace{-55pt}\qquad$=$\qquad
    $\I g_s^2\left(T^{c_1}T^{c_2}+T^{c_2}T^{c_1}\right) \delta
    Z_{\Sb\alpha,\alpha} g_{\mu\nu}$,
  }\\
  \begin{feynartspicture}(100,100)(1,1)
    \FADiagram{}
    \FAProp(0.,10.)(10.,10.)(0.,){/Cycles}{0}
    \FALabel(0.,14.)[t]{$g$}
    \FAProp(10.,10.)(20.,15.)(0.,){/ScalarDash}{1}
    \FALabel(20.1013,19.)[tl]{$\ba$}
    \FAProp(10.,10.)(20.,5.)(0.,){/ScalarDash}{-1}
    \FALabel(20.,3.8172)[tl]{$\ba$}
    \FAVert(10.,10.){1}
  \end{feynartspicture}
  &
  \parbox{0.6\textwidth}{\vspace{-55pt}\qquad$=$\qquad
    $-\I g_s \, T^c \, \delta Z_{\Sb\alpha,\alpha} \left( k + k^\prime
    \right)_{\mu}$,
  }\\
  \begin{feynartspicture}(100,100)(1,1)
    \FADiagram{}
    \FAProp(0.,10.)(10.,10.)(0.,){/Cycles}{0}
    \FALabel(0.,14.)[t]{$g$}
    \FAProp(10.,10.)(20.,15.)(0.,){/Straight}{1}
    \FALabel(20.1013,19.)[tl]{$q$}
    \FAProp(10.,10.)(20.,5.)(0.,){/Straight}{-1}
    \FALabel(20.,3.8172)[tl]{$\bar{q}$}
    \FAVert(10.,10.){1}
  \end{feynartspicture}
  &
  \parbox{0.6\textwidth}{\vspace{-55pt}\qquad$=$\qquad
      $-\I g_s \, T^c \, \left( \delta Z_{qL} \gamma_{\mu} \omega_-
        + \delta Z_{qR} \gamma_{\mu} \omega_+
      \right)$.
    }
\end{tabular}
\egroup


\newpage
\section{Feynman diagrams}
\label{sec:diagrams}

In this Appendix we list the Feynman diagrams relevant for
$\ba\ba^\ast$ production at tree-level and at next-to leading order
electroweak. $S$ ($S^{\pm}$) denotes the neutral (charged) scalar
Higgs and Goldstone bosons. $V=\gamma,Z$.

\FIGURE{
\footnotesize\unitlength=0.26bp%
\begin{feynartspicture}(300,300)(1,1)
\FADiagram{}
\FAProp(0.,15.)(10.,10.)(0.,){/Cycles}{0}
\FALabel(4.78682,11.5936)[tr]{$g$}
\FAProp(0.,5.)(10.,10.)(0.,){/Cycles}{0}
\FALabel(5.21318,6.59364)[tl]{$g$}
\FAProp(20.,15.)(10.,10.)(0.,){/ScalarDash}{-1}
\FALabel(14.7868,13.4064)[br]{$\ba$}
\FAProp(20.,5.)(10.,10.)(0.,){/ScalarDash}{1}
\FALabel(15.2132,8.40636)[bl]{$\ba$}
\FAVert(10.,10.){0}
\end{feynartspicture}
\begin{feynartspicture}(300,300)(1,1)
\FADiagram{}
\FAProp(0.,15.)(6.,10.)(0.,){/Cycles}{0}
\FALabel(2.48771,11.7893)[tr]{$g$}
\FAProp(0.,5.)(6.,10.)(0.,){/Cycles}{0}
\FALabel(3.51229,6.78926)[tl]{$g$}
\FAProp(20.,15.)(14.,10.)(0.,){/ScalarDash}{-1}
\FALabel(16.4877,13.2107)[br]{$\ba$}
\FAProp(20.,5.)(14.,10.)(0.,){/ScalarDash}{1}
\FALabel(17.5123,8.21074)[bl]{$\ba$}
\FAProp(6.,10.)(14.,10.)(0.,){/Cycles}{0}
\FALabel(10.,8.93)[t]{$g$}
\FAVert(6.,10.){0}
\FAVert(14.,10.){0}
\end{feynartspicture}
\begin{feynartspicture}(300,300)(1,1)
\FADiagram{}
\FAProp(0.,15.)(10.,14.)(0.,){/Cycles}{0}
\FALabel(4.84577,13.4377)[t]{$g$}
\FAProp(0.,5.)(10.,6.)(0.,){/Cycles}{0}
\FALabel(5.15423,4.43769)[t]{$g$}
\FAProp(20.,15.)(10.,14.)(0.,){/ScalarDash}{-1}
\FALabel(14.8458,15.5623)[b]{$\ba$}
\FAProp(20.,5.)(10.,6.)(0.,){/ScalarDash}{1}
\FALabel(15.1542,6.56231)[b]{$\ba$}
\FAProp(10.,14.)(10.,6.)(0.,){/ScalarDash}{-1}
\FALabel(8.93,10.)[r]{$\ba$}
\FAVert(10.,14.){0}
\FAVert(10.,6.){0}
\end{feynartspicture}
\begin{feynartspicture}(300,300)(1,1)
\FADiagram{}
\FAProp(0.,15.)(10.,14.)(0.,){/Cycles}{0}
\FALabel(4.84577,13.4377)[t]{$g$}
\FAProp(0.,5.)(10.,6.)(0.,){/Cycles}{0}
\FALabel(5.15423,4.43769)[t]{$g$}
\FAProp(20.,15.)(10.,6.)(0.,){/ScalarDash}{-1}
\FALabel(16.8128,13.2058)[br]{$\ba$}
\FAProp(20.,5.)(10.,14.)(0.,){/ScalarDash}{1}
\FALabel(17.6872,8.20582)[bl]{$\ba$}
\FAProp(10.,14.)(10.,6.)(0.,){/ScalarDash}{1}
\FALabel(9.03,10.)[r]{$\ba$}
\FAVert(10.,14.){0}
\FAVert(10.,6.){0}
\end{feynartspicture}
\\[-10pt]
\hspace*{98pt}(a)
\\
\begin{feynartspicture}(300,300)(1,1)
\FADiagram{}
\FAProp(0.,15.)(6.,10.)(0.,){/Straight}{1}
\FALabel(2.48771,11.7893)[tr]{$q$}
\FAProp(0.,5.)(6.,10.)(0.,){/Straight}{-1}
\FALabel(3.51229,6.78926)[tl]{$q$}
\FAProp(20.,15.)(14.,10.)(0.,){/ScalarDash}{-1}
\FALabel(16.4877,13.2107)[br]{$\ba$}
\FAProp(20.,5.)(14.,10.)(0.,){/ScalarDash}{1}
\FALabel(17.5123,8.21074)[bl]{$\ba$}
\FAProp(6.,10.)(14.,10.)(0.,){/Cycles}{0}
\FALabel(10.,8.93)[t]{$g$}
\FAVert(6.,10.){0}
\FAVert(14.,10.){0}
\end{feynartspicture}
\begin{feynartspicture}(300,300)(1,1)
\FADiagram{}
\FAProp(0.,15.)(6.,10.)(0.,){/Straight}{1}
\FALabel(2.48771,11.7893)[tr]{$b$}
\FAProp(0.,5.)(6.,10.)(0.,){/Straight}{-1}
\FALabel(3.51229,6.78926)[tl]{$b$}
\FAProp(20.,15.)(14.,10.)(0.,){/ScalarDash}{-1}
\FALabel(16.4877,13.2107)[br]{$\ba$}
\FAProp(20.,5.)(14.,10.)(0.,){/ScalarDash}{1}
\FALabel(17.5123,8.21074)[bl]{$\ba$}
\FAProp(6.,10.)(14.,10.)(0.,){/Cycles}{0}
\FALabel(10.,8.93)[t]{$g$}
\FAVert(6.,10.){0}
\FAVert(14.,10.){0}
\end{feynartspicture}
\begin{feynartspicture}(300,300)(1,1)
\FADiagram{}
\FAProp(0.,15.)(10.,14.)(0.,){/Straight}{1}
\FALabel(4.84577,13.4377)[t]{$b$}
\FAProp(0.,5.)(10.,6.)(0.,){/Straight}{-1}
\FALabel(5.15423,4.43769)[t]{$b$}
\FAProp(20.,15.)(10.,14.)(0.,){/ScalarDash}{-1}
\FALabel(14.8458,15.5623)[b]{$\ba$}
\FAProp(20.,5.)(10.,6.)(0.,){/ScalarDash}{1}
\FALabel(15.1542,6.56231)[b]{$\ba$}
\FAProp(10.,14.)(10.,6.)(0.,){/Straight}{0}
\FAProp(10.,14.)(10.,6.)(0.,){/Cycles}{0}
\FALabel(9.18,10.)[r]{$\tilde g$}
\FAVert(10.,14.){0}
\FAVert(10.,6.){0}
\end{feynartspicture}
\begin{feynartspicture}(300,300)(1,1)
\end{feynartspicture}
\\[-10pt]
\hspace*{98pt}(b)
\\
\begin{feynartspicture}(300,300)(1,1)
\FADiagram{}
\FAProp(0.,15.)(6.,10.)(0.,){/Straight}{1}
\FALabel(2.48771,11.7893)[tr]{$q$}
\FAProp(0.,5.)(6.,10.)(0.,){/Straight}{-1}
\FALabel(3.51229,6.78926)[tl]{$q$}
\FAProp(20.,15.)(14.,10.)(0.,){/ScalarDash}{-1}
\FALabel(16.4877,13.2107)[br]{$\ba$}
\FAProp(20.,5.)(14.,10.)(0.,){/ScalarDash}{1}
\FALabel(17.5123,8.21074)[bl]{$\ba$}
\FAProp(6.,10.)(14.,10.)(0.,){/Sine}{0}
\FALabel(10.,8.93)[t]{$V$}
\FAVert(6.,10.){0}
\FAVert(14.,10.){0}
\end{feynartspicture}
\begin{feynartspicture}(300,300)(1,1)
\FADiagram{}
\FAProp(0.,15.)(6.,10.)(0.,){/Straight}{1}
\FALabel(2.48771,11.7893)[tr]{$b$}
\FAProp(0.,5.)(6.,10.)(0.,){/Straight}{-1}
\FALabel(3.51229,6.78926)[tl]{$b$}
\FAProp(20.,15.)(14.,10.)(0.,){/ScalarDash}{-1}
\FALabel(16.4877,13.2107)[br]{$\ba$}
\FAProp(20.,5.)(14.,10.)(0.,){/ScalarDash}{1}
\FALabel(17.5123,8.21074)[bl]{$\ba$}
\FAProp(6.,10.)(14.,10.)(0.,){/Sine}{0}
\FALabel(10.,8.93)[t]{$V$}
\FAVert(6.,10.){0}
\FAVert(14.,10.){0}
\end{feynartspicture}
\begin{feynartspicture}(300,300)(1,1)
\FADiagram{}
\FAProp(0.,15.)(6.,10.)(0.,){/Straight}{1}
\FALabel(2.48771,11.7893)[tr]{$b$}
\FAProp(0.,5.)(6.,10.)(0.,){/Straight}{-1}
\FALabel(3.51229,6.78926)[tl]{$b$}
\FAProp(20.,15.)(14.,10.)(0.,){/ScalarDash}{-1}
\FALabel(16.4877,13.2107)[br]{$\ba$}
\FAProp(20.,5.)(14.,10.)(0.,){/ScalarDash}{1}
\FALabel(17.5123,8.21074)[bl]{$\ba$}
\FAProp(6.,10.)(14.,10.)(0.,){/ScalarDash}{0}
\FALabel(10.,9.18)[t]{$S$}
\FAVert(6.,10.){0}
\FAVert(14.,10.){0}
\end{feynartspicture}
\begin{feynartspicture}(300,300)(1,1)
\FADiagram{}
\FAProp(0.,15.)(10.,14.)(0.,){/Straight}{1}
\FALabel(4.84577,13.4377)[t]{$b$}
\FAProp(0.,5.)(10.,6.)(0.,){/Straight}{-1}
\FALabel(5.15423,4.43769)[t]{$b$}
\FAProp(20.,15.)(10.,14.)(0.,){/ScalarDash}{-1}
\FALabel(14.8458,15.5623)[b]{$\ba$}
\FAProp(20.,5.)(10.,6.)(0.,){/ScalarDash}{1}
\FALabel(15.1542,6.56231)[b]{$\ba$}
\FAProp(10.,14.)(10.,6.)(0.,){/Straight}{0}
\FAProp(10.,14.)(10.,6.)(0.,){/Sine}{0}
\FALabel(9.18,10.)[r]{$\tilde \chi_m^0$}
\FAVert(10.,14.){0}
\FAVert(10.,6.){0}
\end{feynartspicture}\\[-10pt]
\hspace*{98pt}(c)
\\
\hspace*{78pt}
\begin{feynartspicture}(300,300)(1,1)
\FADiagram{}
\FAProp(0.,15.)(10.,14.)(0.,){/Straight}{1}
\FALabel(4.84577,13.4377)[t]{$u/c$}
\FAProp(0.,5.)(10.,6.)(0.,){/Straight}{-1}
\FALabel(5.15423,4.43769)[t]{$u/c$}
\FAProp(20.,15.)(10.,14.)(0.,){/ScalarDash}{-1}
\FALabel(14.8458,15.5623)[b]{$\ba$}
\FAProp(20.,5.)(10.,6.)(0.,){/ScalarDash}{1}
\FALabel(15.1542,6.56231)[b]{$\ba$}
\FAProp(10.,14.)(10.,6.)(0.,){/Sine}{0}
\FAProp(10.,14.)(10.,6.)(0.,){/Straight}{0}
\FALabel(9.18,10.)[r]{$\tilde \chi_m^\pm$}
\FAVert(10.,14.){0}
\FAVert(10.,6.){0}
\end{feynartspicture}\\[-10pt]
(d)
\caption{Tree-level Feynman diagrams for $\ba\ba^\ast$ production. (a)
  and (b) show the QCD diagrams for the $gg$, $q\bar{q}$ and
  $b\bar{b}$ channels. (c) are the EW diagrams. They are
  not present in the $gg$ channel at tree-level. (d) is the  EW
  tree-level diagram involving CKM matrix. It is only present
  for initial $u\bar{u}$ or $c\bar{c}$ quarks.
  \label{fig:feynman_tree}}
}

\FIGURE{
\footnotesize\unitlength=0.26bp%
\hspace*{20pt}
\begin{feynartspicture}(300,300)(1,1)
\FADiagram{}
\FAProp(0.,15.)(10.,10.)(0.,){/Cycles}{0}
\FALabel(4.78682,11.5936)[tr]{$g$}
\FAProp(0.,5.)(10.,10.)(0.,){/Sine}{0}
\FALabel(5.21318,6.59364)[tl]{$\gamma$}
\FAProp(20.,15.)(10.,10.)(0.,){/ScalarDash}{-1}
\FALabel(14.7868,13.4064)[br]{$\ba$}
\FAProp(20.,5.)(10.,10.)(0.,){/ScalarDash}{1}
\FALabel(15.2132,8.40636)[bl]{$\ba$}
\FAVert(10.,10.){0}
\end{feynartspicture}
\begin{feynartspicture}(300,300)(1,1)
\FADiagram{}
\FAProp(0.,15.)(10.,14.)(0.,){/Cycles}{0}
\FALabel(4.84577,13.4377)[t]{$g$}
\FAProp(0.,5.)(10.,6.)(0.,){/Sine}{0}
\FALabel(5.15423,4.43769)[t]{$\gamma$}
\FAProp(20.,15.)(10.,14.)(0.,){/ScalarDash}{-1}
\FALabel(14.8458,15.5623)[b]{$\ba$}
\FAProp(20.,5.)(10.,6.)(0.,){/ScalarDash}{1}
\FALabel(15.1542,6.56231)[b]{$\ba$}
\FAProp(10.,14.)(10.,6.)(0.,){/ScalarDash}{-1}
\FALabel(8.93,10.)[r]{$\ba$}
\FAVert(10.,14.){0}
\FAVert(10.,6.){0}
\end{feynartspicture}
\begin{feynartspicture}(300,300)(1,1)
\FADiagram{}
\FAProp(0.,15.)(10.,14.)(0.,){/Cycles}{0}
\FALabel(4.84577,13.4377)[t]{$g$}
\FAProp(0.,5.)(10.,6.)(0.,){/Sine}{0}
\FALabel(5.15423,4.43769)[t]{$\gamma$}
\FAProp(20.,15.)(10.,6.)(0.,){/ScalarDash}{-1}
\FALabel(16.8128,13.2058)[br]{$\ba$}
\FAProp(20.,5.)(10.,14.)(0.,){/ScalarDash}{1}
\FALabel(17.6872,8.20582)[bl]{$\ba$}
\FAProp(10.,14.)(10.,6.)(0.,){/ScalarDash}{1}
\FALabel(9.03,10.)[r]{$\ba$}
\FAVert(10.,14.){0}
\FAVert(10.,6.){0}
\end{feynartspicture}
\hspace*{20pt}
\caption{Lowest-order  Feynman diagrams for the gluon-photon fusion process  $g \gamma \to \ba\ba^\ast$.\label{fig:feynman_ggamma}}
} 

\FIGURE{
\footnotesize\unitlength=0.22bp%
\begin{feynartspicture}(300,300)(1,1)
\FADiagram{}
\FAProp(0.,15.)(4.,10.)(0.,){/Cycles}{0}
\FALabel(1.26965,12.0117)[tr]{$g$}
\FAProp(0.,5.)(4.,10.)(0.,){/Cycles}{0}
\FALabel(2.73035,7.01172)[tl]{$g$}
\FAProp(20.,15.)(16.,13.5)(0.,){/ScalarDash}{-1}
\FALabel(17.4558,15.2213)[b]{$\ba$}
\FAProp(20.,5.)(16.,6.5)(0.,){/ScalarDash}{1}
\FALabel(17.4558,4.77869)[t]{$\ba$}
\FAProp(4.,10.)(10.,10.)(0.,){/Cycles}{0}
\FALabel(7.,11.77)[b]{$g$}
\FAProp(16.,13.5)(16.,6.5)(0.,){/Straight}{0}
\FALabel(16.82,10.)[l]{$\neu_i/\cha_i$}
\FAProp(16.,13.5)(10.,10.)(0.,){/Straight}{-1}
\FALabel(12.699,12.6089)[br]{$b/t$}
\FAProp(16.,6.5)(10.,10.)(0.,){/Straight}{1}
\FALabel(12.699,7.39114)[tr]{$b/t$}
\FAVert(4.,10.){0}
\FAVert(16.,13.5){0}
\FAVert(16.,6.5){0}
\FAVert(10.,10.){0}
\end{feynartspicture}
\begin{feynartspicture}(300,300)(1,1)
\FADiagram{}
\FAProp(0.,15.)(4.,10.)(0.,){/Cycles}{0}
\FALabel(1.26965,12.0117)[br]{$g$}
\FAProp(0.,5.)(4.,10.)(0.,){/Cycles}{0}
\FALabel(2.73035,7.01172)[tl]{$g$}
\FAProp(20.,15.)(16.,13.5)(0.,){/ScalarDash}{-1}
\FALabel(17.4558,15.2213)[b]{$\ba$}
\FAProp(20.,5.)(16.,6.5)(0.,){/ScalarDash}{1}
\FALabel(17.4558,4.77869)[t]{$\ba$}
\FAProp(4.,10.)(10.,10.)(0.,){/Cycles}{0}
\FALabel(7.,11.77)[b]{$g$}
\FAProp(16.,13.5)(16.,6.5)(0.,){/ScalarDash}{-1}
\FALabel(17.07,10.)[l]{$S/S^\pm$}
\FAProp(16.,13.5)(10.,10.)(0.,){/ScalarDash}{-1}
\FALabel(14.,13.6089)[br]{$\bb/\tb$}
\FAProp(16.,6.5)(10.,10.)(0.,){/ScalarDash}{1}
\FALabel(14.,7.39114)[tr]{$\bb/\tb$}
\FAVert(4.,10.){0}
\FAVert(16.,13.5){0}
\FAVert(16.,6.5){0}
\FAVert(10.,10.){0}
\end{feynartspicture}
\begin{feynartspicture}(300,300)(1,1)
\FADiagram{}
\FAProp(0.,15.)(4.,10.)(0.,){/Cycles}{0}
\FALabel(1.26965,12.0117)[br]{$g$}
\FAProp(0.,5.)(4.,10.)(0.,){/Cycles}{0}
\FALabel(2.73035,7.01172)[tl]{$g$}
\FAProp(20.,15.)(16.,13.5)(0.,){/ScalarDash}{-1}
\FALabel(17.4558,15.2213)[b]{$\ba$}
\FAProp(20.,5.)(16.,6.5)(0.,){/ScalarDash}{1}
\FALabel(17.4558,4.77869)[t]{$\ba$}
\FAProp(4.,10.)(10.,10.)(0.,){/Cycles}{0}
\FALabel(7.,11.77)[b]{$g$}
\FAProp(16.,13.5)(16.,6.5)(0.,){/Sine}{0}
\FALabel(17.07,10.)[l]{$V/W$}
\FAProp(16.,13.5)(10.,10.)(0.,){/ScalarDash}{-1}
\FALabel(14.,13.6089)[br]{$\bb/\tb$}
\FAProp(16.,6.5)(10.,10.)(0.,){/ScalarDash}{1}
\FALabel(14.,7.39114)[tr]{$\bb/\tb$}
\FAVert(4.,10.){0}
\FAVert(16.,13.5){0}
\FAVert(16.,6.5){0}
\FAVert(10.,10.){0}
\end{feynartspicture}
\begin{feynartspicture}(300,300)(1,1)
\FADiagram{}
\FAProp(0.,15.)(10.,14.5)(0.,){/Cycles}{0}
\FALabel(5.11236,16.5172)[b]{$g$}
\FAProp(0.,5.)(6.5,5.5)(0.,){/Cycles}{0}
\FALabel(3.36888,4.18457)[t]{$g$}
\FAProp(20.,15.)(10.,14.5)(0.,){/ScalarDash}{-1}
\FALabel(14.9226,15.8181)[b]{$\ba$}
\FAProp(20.,5.)(13.5,5.5)(0.,){/ScalarDash}{1}
\FALabel(16.6311,4.18457)[t]{$\ba$}
\FAProp(10.,14.5)(10.,11.)(0.,){/ScalarDash}{-1}
\FALabel(11.07,12.75)[l]{$\ba$}
\FAProp(6.5,5.5)(13.5,5.5)(0.,){/Straight}{-1}
\FALabel(10.,4.43)[t]{$b/t$}
\FAProp(6.5,5.5)(10.,11.)(0.,){/Straight}{1}
\FALabel(7.42232,8.60216)[br]{$b/t$}
\FAProp(13.5,5.5)(10.,11.)(0.,){/Straight}{0}
\FALabel(12.3668,8.46794)[bl]{$\neu_i/\cha_i$}
\FAVert(10.,14.5){0}
\FAVert(6.5,5.5){0}
\FAVert(13.5,5.5){0}
\FAVert(10.,11.){0}
\end{feynartspicture}
\begin{feynartspicture}(300,300)(1,1)
\FADiagram{}
\FAProp(0.,15.)(10.,14.5)(0.,){/Cycles}{0}
\FALabel(5.11236,16.5172)[b]{$g$}
\FAProp(0.,5.)(6.5,5.5)(0.,){/Cycles}{0}
\FALabel(3.36888,4.18457)[t]{$g$}
\FAProp(20.,15.)(10.,14.5)(0.,){/ScalarDash}{-1}
\FALabel(14.9226,15.8181)[b]{$\ba$}
\FAProp(20.,5.)(13.5,5.5)(0.,){/ScalarDash}{1}
\FALabel(16.6311,4.18457)[t]{$\ba$}
\FAProp(10.,14.5)(10.,11.)(0.,){/ScalarDash}{-1}
\FALabel(11.07,12.75)[l]{$\ba$}
\FAProp(6.5,5.5)(13.5,5.5)(0.,){/ScalarDash}{-1}
\FALabel(10.,4.43)[t]{$\bb/\tb$}
\FAProp(6.5,5.5)(10.,11.)(0.,){/ScalarDash}{1}
\FALabel(7.42232,8.60216)[br]{$\bb/\tb$}
\FAProp(13.5,5.5)(10.,11.)(0.,){/ScalarDash}{1}
\FALabel(12.5777,8.60216)[bl]{$S/S^\pm$}
\FAVert(10.,14.5){0}
\FAVert(6.5,5.5){0}
\FAVert(13.5,5.5){0}
\FAVert(10.,11.){0}
\end{feynartspicture}
\begin{feynartspicture}(300,300)(1,1)
\FADiagram{}
\FAProp(0.,15.)(10.,14.5)(0.,){/Cycles}{0}
\FALabel(5.11236,16.5172)[b]{$g$}
\FAProp(0.,5.)(6.5,5.5)(0.,){/Cycles}{0}
\FALabel(3.36888,4.18457)[t]{$g$}
\FAProp(20.,15.)(10.,14.5)(0.,){/ScalarDash}{-1}
\FALabel(14.9226,15.8181)[b]{$\ba$}
\FAProp(20.,5.)(13.5,5.5)(0.,){/ScalarDash}{1}
\FALabel(16.6311,4.18457)[t]{$\ba$}
\FAProp(10.,14.5)(10.,11.)(0.,){/ScalarDash}{-1}
\FALabel(11.07,12.75)[l]{$\ba$}
\FAProp(6.5,5.5)(13.5,5.5)(0.,){/ScalarDash}{-1}
\FALabel(10.,4.43)[t]{$\bb/\tb$}
\FAProp(6.5,5.5)(10.,11.)(0.,){/ScalarDash}{1}
\FALabel(7.42232,8.60216)[br]{$\bb/\tb$}
\FAProp(13.5,5.5)(10.,11.)(0.,){/Sine}{0}
\FALabel(12.5777,8.60216)[bl]{$V/W$}
\FAVert(10.,14.5){0}
\FAVert(6.5,5.5){0}
\FAVert(13.5,5.5){0}
\FAVert(10.,11.){0}
\end{feynartspicture}
\begin{feynartspicture}(300,300)(1,1)
\FADiagram{}
\FAProp(0.,15.)(6.5,14.5)(0.,){/Cycles}{0}
\FALabel(3.42257,16.5134)[b]{$g$}
\FAProp(0.,5.)(10.,5.5)(0.,){/Cycles}{0}
\FALabel(5.0774,4.18193)[t]{$g$}
\FAProp(20.,15.)(13.5,14.5)(0.,){/ScalarDash}{-1}
\FALabel(16.6311,15.8154)[b]{$\ba$}
\FAProp(20.,5.)(10.,5.5)(0.,){/ScalarDash}{1}
\FALabel(14.9226,4.18193)[t]{$\ba$}
\FAProp(10.,5.5)(10.,8.5)(0.,){/ScalarDash}{1}
\FALabel(11.,7.)[l]{$\ba$}
\FAProp(6.5,14.5)(13.5,14.5)(0.,){/Straight}{1}
\FALabel(10.,15.57)[b]{$b/t$}
\FAProp(6.5,14.5)(10.,8.5)(0.,){/Straight}{-1}
\FALabel(7.39114,11.199)[tr]{$b/t$}
\FAProp(13.5,14.5)(10.,8.5)(0.,){/Straight}{0}
\FALabel(12.3929,11.325)[tl]{$\neu_i/\cha_i$}
\FAVert(6.5,14.5){0}
\FAVert(10.,5.5){0}
\FAVert(13.5,14.5){0}
\FAVert(10.,8.5){0}
\end{feynartspicture}

\begin{feynartspicture}(300,300)(1,1)
\FADiagram{}
\FAProp(0.,15.)(6.5,14.5)(0.,){/Cycles}{0}
\FALabel(3.42257,16.5134)[b]{$g$}
\FAProp(0.,5.)(10.,5.5)(0.,){/Cycles}{0}
\FALabel(5.0774,4.18193)[t]{$g$}
\FAProp(20.,15.)(13.5,14.5)(0.,){/ScalarDash}{-1}
\FALabel(16.6311,15.8154)[b]{$\ba$}
\FAProp(20.,5.)(10.,5.5)(0.,){/ScalarDash}{1}
\FALabel(14.9226,4.18193)[t]{$\ba$}
\FAProp(10.,5.5)(10.,8.5)(0.,){/ScalarDash}{1}
\FALabel(11.,7.)[l]{$\ba$}
\FAProp(6.5,14.5)(13.5,14.5)(0.,){/ScalarDash}{1}
\FALabel(10.,15.57)[b]{$\bb/\tb$}
\FAProp(6.5,14.5)(10.,8.5)(0.,){/ScalarDash}{-1}
\FALabel(7.39114,11.199)[tr]{$\bb/\tb$}
\FAProp(13.5,14.5)(10.,8.5)(0.,){/ScalarDash}{-1}
\FALabel(12.6089,11.199)[tl]{$S/S^\pm$}
\FAVert(6.5,14.5){0}
\FAVert(10.,5.5){0}
\FAVert(13.5,14.5){0}
\FAVert(10.,8.5){0}
\end{feynartspicture}

\begin{feynartspicture}(300,300)(1,1)
\FADiagram{}
\FAProp(0.,15.)(6.5,14.5)(0.,){/Cycles}{0}
\FALabel(3.42257,16.5134)[b]{$g$}
\FAProp(0.,5.)(10.,5.5)(0.,){/Cycles}{0}
\FALabel(5.0774,4.18193)[t]{$g$}
\FAProp(20.,15.)(13.5,14.5)(0.,){/ScalarDash}{-1}
\FALabel(16.6311,15.8154)[b]{$\ba$}
\FAProp(20.,5.)(10.,5.5)(0.,){/ScalarDash}{1}
\FALabel(14.9226,4.18193)[t]{$\ba$}
\FAProp(10.,5.5)(10.,8.5)(0.,){/ScalarDash}{1}
\FALabel(11.,7.)[l]{$\ba$}
\FAProp(6.5,14.5)(13.5,14.5)(0.,){/ScalarDash}{1}
\FALabel(10.,15.57)[b]{$\bb/\tb$}
\FAProp(6.5,14.5)(10.,8.5)(0.,){/ScalarDash}{-1}
\FALabel(7.39114,11.199)[tr]{$\bb/\tb$}
\FAProp(13.5,14.5)(10.,8.5)(0.,){/Sine}{0}
\FALabel(12.6089,11.199)[tl]{$V/W$}
\FAVert(6.5,14.5){0}
\FAVert(10.,5.5){0}
\FAVert(13.5,14.5){0}
\FAVert(10.,8.5){0}
\end{feynartspicture}

\begin{feynartspicture}(300,300)(1,1)
\FADiagram{}
\FAProp(0.,15.)(4.,13.5)(0.,){/Cycles}{0}
\FALabel(2.79003,15.8767)[b]{$g$}
\FAProp(0.,5.)(4.,6.5)(0.,){/Cycles}{0}
\FALabel(2.54424,4.77869)[t]{$g$}
\FAProp(20.,15.)(16.,10.)(0.,){/ScalarDash}{-1}
\FALabel(17.2697,12.9883)[br]{$\ba$}
\FAProp(20.,5.)(16.,10.)(0.,){/ScalarDash}{1}
\FALabel(18.7303,7.98828)[tr]{$\ba$}
\FAProp(16.,10.)(10.,10.)(0.,){/ScalarDash}{0}
\FALabel(13.,9.18)[t]{$S$}
\FAProp(4.,13.5)(4.,6.5)(0.,){/Straight}{-1}
\FALabel(2.93,10.)[r]{$b/t$}
\FAProp(4.,13.5)(10.,10.)(0.,){/Straight}{1}
\FALabel(7.301,12.6089)[bl]{$b/t$}
\FAProp(4.,6.5)(10.,10.)(0.,){/Straight}{-1}
\FALabel(7.301,7.39114)[tl]{$b/t$}
\FAVert(4.,13.5){0}
\FAVert(4.,6.5){0}
\FAVert(16.,10.){0}
\FAVert(10.,10.){0}
\end{feynartspicture}

\begin{feynartspicture}(300,300)(1,1)
\FADiagram{}
\FAProp(0.,15.)(4.,13.5)(0.,){/Cycles}{0}
\FALabel(2.79003,15.8767)[b]{$g$}
\FAProp(0.,5.)(4.,6.5)(0.,){/Cycles}{0}
\FALabel(2.54424,4.77869)[t]{$g$}
\FAProp(20.,15.)(16.,10.)(0.,){/ScalarDash}{-1}
\FALabel(17.2697,12.9883)[br]{$\ba$}
\FAProp(20.,5.)(16.,10.)(0.,){/ScalarDash}{1}
\FALabel(18.7303,7.98828)[bl]{$\ba$}
\FAProp(16.,10.)(10.,10.)(0.,){/ScalarDash}{0}
\FALabel(13.,9.18)[t]{$S$}
\FAProp(4.,13.5)(4.,6.5)(0.,){/ScalarDash}{-1}
\FALabel(2.93,10.)[r]{$\bb/\tb$}
\FAProp(4.,13.5)(10.,10.)(0.,){/ScalarDash}{1}
\FALabel(7.301,12.6089)[bl]{$\bb/\tb$}
\FAProp(4.,6.5)(10.,10.)(0.,){/ScalarDash}{-1}
\FALabel(7.301,7.39114)[tl]{$\bb/\tb$}
\FAVert(4.,13.5){0}
\FAVert(4.,6.5){0}
\FAVert(16.,10.){0}
\FAVert(10.,10.){0}
\end{feynartspicture}

\begin{feynartspicture}(300,300)(1,1)
\FADiagram{}
\FAProp(0.,15.)(5.,10.)(0.,){/Cycles}{0}
\FALabel(1.88398,11.884)[tr]{$g$}
\FAProp(0.,5.)(5.,10.)(0.,){/Cycles}{0}
\FALabel(3.11602,6.88398)[tl]{$g$}
\FAProp(20.,15.)(14.5,13.5)(0.,){/ScalarDash}{-1}
\FALabel(16.8422,15.2654)[b]{$\ba$}
\FAProp(20.,5.)(12.5,8.)(0.,){/ScalarDash}{1}
\FALabel(15.6743,5.54086)[t]{$\ba$}
\FAProp(5.,10.)(12.5,8.)(0.,){/Cycles}{0}
\FALabel(8.35062,7.98234)[t]{$g$}
\FAProp(14.5,13.5)(12.5,8.)(0.8,){/ScalarDash}{-1}
\FALabel(11.3233,14.0797)[r]{$\bb/\tb$}
\FAProp(14.5,13.5)(12.5,8.)(-0.8,){/Sine}{0}
\FALabel(16.6767,9.4203)[l]{$V/W$}
\FAVert(5.,10.){0}
\FAVert(14.5,13.5){0}
\FAVert(12.5,8.){0}
\end{feynartspicture}

\begin{feynartspicture}(300,300)(1,1)
\FADiagram{}
\FAProp(0.,15.)(5.,10.)(0.,){/Cycles}{0}
\FALabel(1.88398,11.884)[tr]{$g$}
\FAProp(0.,5.)(5.,10.)(0.,){/Cycles}{0}
\FALabel(3.11602,6.88398)[tl]{$g$}
\FAProp(20.,15.)(12.5,12.)(0.,){/ScalarDash}{-1}
\FALabel(15.6743,14.4591)[b]{$\ba$}
\FAProp(20.,5.)(14.5,6.5)(0.,){/ScalarDash}{1}
\FALabel(16.8422,4.73462)[t]{$\ba$}
\FAProp(5.,10.)(12.5,12.)(0.,){/Cycles}{0}
\FALabel(8.17026,12.694)[b]{$g$}
\FAProp(14.5,6.5)(12.5,12.)(0.8,){/ScalarDash}{1}
\FALabel(16.6767,10.5797)[l]{$\bb/\tb$}
\FAProp(14.5,6.5)(12.5,12.)(-0.8,){/Sine}{0}
\FALabel(10.3233,7.9203)[r]{$V/W$}
\FAVert(5.,10.){0}
\FAVert(14.5,6.5){0}
\FAVert(12.5,12.){0}
\end{feynartspicture}

\begin{feynartspicture}(300,300)(1,1)
\FADiagram{}
\FAProp(0.,15.)(10.,14.)(0.,){/Cycles}{0}
\FALabel(5.22388,16.2588)[b]{$g$}
\FAProp(0.,5.)(10.,6.)(0.,){/Cycles}{0}
\FALabel(5.15423,4.43769)[t]{$g$}
\FAProp(20.,15.)(10.,14.)(0.,){/ScalarDash}{-1}
\FALabel(14.8458,15.5623)[b]{$\ba$}
\FAProp(20.,5.)(15.55,5.4)(0.,){/ScalarDash}{1}
\FALabel(17.9138,6.26378)[b]{$\ba$}
\FAProp(10.,14.)(10.,6.)(0.,){/ScalarDash}{-1}
\FALabel(8.93,10.)[r]{$\ba$}
\FAProp(15.55,5.4)(10.,6.)(0.8,){/ScalarDash}{1}
\FALabel(14.1816,8.98102)[b]{$\bb/\tb$}
\FAProp(15.55,5.4)(10.,6.)(-0.8,){/Sine}{0}
\FALabel(12.3684,2.41898)[t]{$V/W$}
\FAVert(10.,14.){0}
\FAVert(15.55,5.4){0}
\FAVert(10.,6.){0}
\end{feynartspicture}

\begin{feynartspicture}(300,300)(1,1)
\FADiagram{}
\FAProp(0.,15.)(10.,14.)(0.,){/Cycles}{0}
\FALabel(5.22388,16.2588)[b]{$g$}
\FAProp(0.,5.)(10.,6.)(0.,){/Cycles}{0}
\FALabel(5.15423,4.43769)[t]{$g$}
\FAProp(20.,15.)(15.6,14.55)(0.,){/ScalarDash}{-1}
\FALabel(17.6423,15.837)[b]{$\ba$}
\FAProp(20.,5.)(10.,6.)(0.,){/ScalarDash}{1}
\FALabel(14.8458,4.43769)[t]{$\ba$}
\FAProp(10.,6.)(10.,14.)(0.,){/ScalarDash}{1}
\FALabel(8.93,10.)[r]{$\ba$}
\FAProp(15.6,14.55)(10.,14.)(0.8,){/ScalarDash}{-1}
\FALabel(12.4285,17.5776)[b]{$\bb/\tb$}
\FAProp(15.6,14.55)(10.,14.)(-0.8,){/Sine}{0}
\FALabel(13.1715,10.9724)[t]{$V/W$}
\FAVert(10.,6.){0}
\FAVert(15.6,14.55){0}
\FAVert(10.,14.){0}
\end{feynartspicture}

\begin{feynartspicture}(300,300)(1,1)
\FADiagram{}
\FAProp(0.,15.)(10.,14.5)(0.,){/Cycles}{0}
\FALabel(5.11236,16.5172)[b]{$g$}
\FAProp(0.,5.)(10.,5.5)(0.,){/Cycles}{0}
\FALabel(5.0774,4.18193)[t]{$g$}
\FAProp(20.,15.)(10.,14.5)(0.,){/ScalarDash}{-1}
\FALabel(14.9226,15.8181)[b]{$\ba$}
\FAProp(20.,5.)(10.,5.5)(0.,){/ScalarDash}{1}
\FALabel(14.9226,4.18193)[t]{$\ba$}
\FAProp(10.,14.5)(10.,11.)(0.,){/ScalarDash}{-1}
\FALabel(11.07,12.75)[l]{$\ba$}
\FAProp(10.,11.)(10.,5.5)(0.8,){/ScalarDash}{-1}
\FALabel(6.73,8.25)[r]{$\bb/\tb$}
\FAProp(10.,11.)(10.,5.5)(-0.8,){/Sine}{0}
\FALabel(13.27,8.25)[l]{$V/W$}
\FAVert(10.,14.5){0}
\FAVert(10.,11.){0}
\FAVert(10.,5.5){0}
\end{feynartspicture}

\begin{feynartspicture}(300,300)(1,1)
\FADiagram{}
\FAProp(0.,15.)(10.,14.5)(0.,){/Cycles}{0}
\FALabel(5.11236,16.5172)[b]{$g$}
\FAProp(0.,5.)(10.,5.5)(0.,){/Cycles}{0}
\FALabel(5.0774,4.18193)[t]{$g$}
\FAProp(20.,15.)(10.,14.5)(0.,){/ScalarDash}{-1}
\FALabel(14.9226,15.8181)[b]{$\ba$}
\FAProp(20.,5.)(10.,5.5)(0.,){/ScalarDash}{1}
\FALabel(14.9226,4.18193)[t]{$\ba$}
\FAProp(10.,5.5)(10.,9.)(0.,){/ScalarDash}{1}
\FALabel(11.,7.25)[l]{$\ba$}
\FAProp(10.,9.)(10.,14.5)(0.8,){/ScalarDash}{1}
\FALabel(13.27,11.75)[l]{$\bb/\tb$}
\FAProp(10.,9.)(10.,14.5)(-0.8,){/Sine}{0}
\FALabel(6.73,11.75)[r]{$V/W$}
\FAVert(10.,5.5){0}
\FAVert(10.,9.){0}
\FAVert(10.,14.5){0}
\end{feynartspicture}

\begin{feynartspicture}(300,300)(1,1)
\FADiagram{}
\FAProp(0.,15.)(4.,10.)(0.,){/Cycles}{0}
\FALabel(2.36018,14.621)[bl]{$g$}
\FAProp(0.,5.)(4.,10.)(0.,){/Cycles}{0}
\FALabel(0.723045,8.42556)[br]{$g$}
\FAProp(20.,15.)(16.5,10.)(0.,){/ScalarDash}{-1}
\FALabel(17.4602,12.9089)[br]{$\ba$}
\FAProp(20.,5.)(16.5,10.)(0.,){/ScalarDash}{1}
\FALabel(19.0398,7.90887)[bl]{$\ba$}
\FAProp(16.5,10.)(9.5,10.)(0.,){/ScalarDash}{0}
\FALabel(13.,9.18)[t]{$S$}
\FAProp(9.5,10.)(4.,10.)(0.8,){/ScalarDash}{-1}
\FALabel(6.75,13.27)[b]{$\bb/\tb$}
\FAProp(9.5,10.)(4.,10.)(-0.8,){/ScalarDash}{1}
\FALabel(6.75,6.73)[t]{$\bb/\tb$}
\FAVert(16.5,10.){0}
\FAVert(9.5,10.){0}
\FAVert(4.,10.){0}
\end{feynartspicture}
\begin{feynartspicture}(300,300)(1,1)
\FADiagram{}
\FAProp(0.,15.)(6.,13.5)(0.,){/Cycles}{0}
\FALabel(3.54571,15.9528)[b]{$g$}
\FAProp(0.,5.)(6.,6.5)(0.,){/Cycles}{0}
\FALabel(3.37593,4.72628)[t]{$g$}
\FAProp(20.,15.)(12.,10.)(0.,){/ScalarDash}{-1}
\FALabel(15.6585,13.3344)[br]{$\ba$}
\FAProp(20.,5.)(12.,10.)(0.,){/ScalarDash}{1}
\FALabel(16.3415,8.3344)[bl]{$\ba$}
\FAProp(6.,13.5)(6.,6.5)(0.,){/ScalarDash}{-1}
\FALabel(4.93,10.)[r]{$\qa$}
\FAProp(6.,13.5)(12.,10.)(0.,){/ScalarDash}{1}
\FALabel(9.301,12.6089)[bl]{$\qa$}
\FAProp(6.,6.5)(12.,10.)(0.,){/ScalarDash}{-1}
\FALabel(9.301,7.39114)[tl]{$\qa$}
\FAVert(6.,13.5){0}
\FAVert(6.,6.5){0}
\FAVert(12.,10.){0}
\end{feynartspicture}

\begin{feynartspicture}(300,300)(1,1)
\FADiagram{}
\FAProp(0.,15.)(6.5,13.)(0.,){/Cycles}{0}
\FALabel(3.91169,15.6705)[b]{$g$}
\FAProp(0.,5.)(10.,7.)(0.,){/Cycles}{0}
\FALabel(5.30398,4.9601)[t]{$g$}
\FAProp(20.,15.)(13.5,13.)(0.,){/ScalarDash}{-1}
\FALabel(16.2942,15.0015)[b]{$\ba$}
\FAProp(20.,5.)(10.,7.)(0.,){/ScalarDash}{1}
\FALabel(14.696,4.9601)[t]{$\ba$}
\FAProp(6.5,13.)(13.5,13.)(0.,){/ScalarDash}{1}
\FALabel(10.,14.07)[b]{$\bb/\tb$}
\FAProp(6.5,13.)(10.,7.)(0.,){/ScalarDash}{-1}
\FALabel(7.39114,10.699)[tr]{$\bb/\tb$}
\FAProp(13.5,13.)(10.,7.)(0.,){/Sine}{0}
\FALabel(12.6089,9.699)[tl]{$V/W$}
\FAVert(6.5,13.){0}
\FAVert(13.5,13.){0}
\FAVert(10.,7.){0}
\end{feynartspicture}

\begin{feynartspicture}(300,300)(1,1)
\FADiagram{}
\FAProp(0.,15.)(10.,13.)(0.,){/Cycles}{0}
\FALabel(5.44126,15.7263)[b]{$g$}
\FAProp(0.,5.)(6.5,7.)(0.,){/Cycles}{0}
\FALabel(3.70583,4.99854)[t]{$g$}
\FAProp(20.,15.)(10.,13.)(0.,){/ScalarDash}{-1}
\FALabel(14.696,15.0399)[b]{$\ba$}
\FAProp(20.,5.)(13.5,7.)(0.,){/ScalarDash}{1}
\FALabel(16.2942,4.99854)[t]{$\ba$}
\FAProp(6.5,7.)(13.5,7.)(0.,){/ScalarDash}{-1}
\FALabel(10.,5.93)[t]{$\bb/\tb$}
\FAProp(6.5,7.)(10.,13.)(0.,){/ScalarDash}{1}
\FALabel(7.39114,10.301)[br]{$\tilde b^1$}
\FAProp(13.5,7.)(10.,13.)(0.,){/Sine}{0}
\FALabel(12.6089,10.301)[bl]{$V/W$}
\FAVert(6.5,7.){0}
\FAVert(13.5,7.){0}
\FAVert(10.,13.){0}
\end{feynartspicture}

\begin{feynartspicture}(300,300)(1,1)
\FADiagram{}
\FAProp(0.,15.)(8.,10.)(0.,){/Cycles}{0}
\FALabel(3.6585,11.6656)[tr]{$g$}
\FAProp(0.,5.)(8.,10.)(0.,){/Cycles}{0}
\FALabel(4.3415,6.6656)[tl]{$g$}
\FAProp(20.,15.)(14.,13.5)(0.,){/ScalarDash}{-1}
\FALabel(16.6241,15.2737)[b]{$\ba$}
\FAProp(20.,5.)(14.,6.5)(0.,){/ScalarDash}{1}
\FALabel(16.6241,4.72628)[t]{$\ba$}
\FAProp(14.,13.5)(14.,6.5)(0.,){/ScalarDash}{-1}
\FALabel(15.07,10.)[l]{$S/S^\pm$}
\FAProp(14.,13.5)(8.,10.)(0.,){/ScalarDash}{-1}
\FALabel(12.699,12.6089)[br]{$\bb/\tb$}
\FAProp(14.,6.5)(8.,10.)(0.,){/ScalarDash}{1}
\FALabel(12.699,7.39114)[tr]{$\bb/\tb$}
\FAVert(14.,13.5){0}
\FAVert(14.,6.5){0}
\FAVert(8.,10.){0}
\end{feynartspicture}

\begin{feynartspicture}(300,300)(1,1)
\FADiagram{}
\FAProp(0.,15.)(8.,10.)(0.,){/Cycles}{0}
\FALabel(3.6585,11.6656)[tr]{$g$}
\FAProp(0.,5.)(8.,10.)(0.,){/Cycles}{0}
\FALabel(4.3415,6.6656)[tl]{$g$}
\FAProp(20.,15.)(14.,13.5)(0.,){/ScalarDash}{-1}
\FALabel(16.6241,15.2737)[b]{$\ba$}
\FAProp(20.,5.)(14.,6.5)(0.,){/ScalarDash}{1}
\FALabel(16.6241,4.72628)[t]{$\ba$}
\FAProp(14.,13.5)(14.,6.5)(0.,){/Sine}{0}
\FALabel(15.07,10.)[l]{$V/W$}
\FAProp(14.,13.5)(8.,10.)(0.,){/ScalarDash}{-1}
\FALabel(12.699,12.6089)[br]{$\bb/\tb$}
\FAProp(14.,6.5)(8.,10.)(0.,){/ScalarDash}{1}
\FALabel(12.699,7.39114)[tr]{$\bb/\tb$}
\FAVert(14.,13.5){0}
\FAVert(14.,6.5){0}
\FAVert(8.,10.){0}
\end{feynartspicture}

\begin{feynartspicture}(300,300)(1,1)
\FADiagram{}
\FAProp(0.,15.)(6.5,13.5)(0.,){/Cycles}{0}
\FALabel(3.75593,15.9624)[b]{$g$}
\FAProp(0.,5.)(6.5,6.5)(0.,){/Cycles}{0}
\FALabel(3.59853,4.71969)[t]{$g$}
\FAProp(20.,15.)(13.5,13.5)(0.,){/ScalarDash}{-1}
\FALabel(16.4015,15.2803)[b]{$\ba$}
\FAProp(20.,5.)(13.5,6.5)(0.,){/ScalarDash}{1}
\FALabel(16.4015,4.71969)[t]{$\ba$}
\FAProp(6.5,13.5)(6.5,6.5)(0.,){/Straight}{-1}
\FALabel(5.43,10.)[r]{$b/t$}
\FAProp(6.5,13.5)(13.5,13.5)(0.,){/Straight}{1}
\FALabel(10.,14.57)[b]{$b/t$}
\FAProp(6.5,6.5)(13.5,6.5)(0.,){/Straight}{-1}
\FALabel(10.,5.43)[t]{$b/t$}
\FAProp(13.5,13.5)(13.5,6.5)(0.,){/Straight}{0}
\FALabel(14.32,10.)[l]{$\neu_i/\cha_i$}
\FAVert(6.5,13.5){0}
\FAVert(6.5,6.5){0}
\FAVert(13.5,13.5){0}
\FAVert(13.5,6.5){0}
\end{feynartspicture}

\begin{feynartspicture}(300,300)(1,1)
\FADiagram{}
\FAProp(0.,15.)(6.5,13.5)(0.,){/Cycles}{0}
\FALabel(3.75593,15.9624)[b]{$g$}
\FAProp(0.,5.)(6.5,6.5)(0.,){/Cycles}{0}
\FALabel(3.59853,4.71969)[t]{$g$}
\FAProp(20.,15.)(13.5,13.5)(0.,){/ScalarDash}{-1}
\FALabel(16.4015,15.2803)[b]{$\ba$}
\FAProp(20.,5.)(13.5,6.5)(0.,){/ScalarDash}{1}
\FALabel(16.4015,4.71969)[t]{$\ba$}
\FAProp(6.5,13.5)(6.5,6.5)(0.,){/ScalarDash}{-1}
\FALabel(5.43,10.)[r]{$\bb/\tb$}
\FAProp(6.5,13.5)(13.5,13.5)(0.,){/ScalarDash}{1}
\FALabel(10.,14.57)[b]{$\bb/\tb$}
\FAProp(6.5,6.5)(13.5,6.5)(0.,){/ScalarDash}{-1}
\FALabel(10.,5.43)[t]{$\bb/\tb$}
\FAProp(13.5,13.5)(13.5,6.5)(0.,){/ScalarDash}{-1}
\FALabel(14.57,10.)[l]{$S/S^\pm$}
\FAVert(6.5,13.5){0}
\FAVert(6.5,6.5){0}
\FAVert(13.5,13.5){0}
\FAVert(13.5,6.5){0}
\end{feynartspicture}

\begin{feynartspicture}(300,300)(1,1)
\FADiagram{}
\FAProp(0.,15.)(6.5,13.5)(0.,){/Cycles}{0}
\FALabel(3.75593,15.9624)[b]{$g$}
\FAProp(0.,5.)(6.5,6.5)(0.,){/Cycles}{0}
\FALabel(3.59853,4.71969)[t]{$g$}
\FAProp(20.,15.)(13.5,13.5)(0.,){/ScalarDash}{-1}
\FALabel(16.4015,15.2803)[b]{$\ba$}
\FAProp(20.,5.)(13.5,6.5)(0.,){/ScalarDash}{1}
\FALabel(16.4015,4.71969)[t]{$\ba$}
\FAProp(6.5,13.5)(6.5,6.5)(0.,){/ScalarDash}{-1}
\FALabel(5.43,10.)[r]{$\bb/\tb$}
\FAProp(6.5,13.5)(13.5,13.5)(0.,){/ScalarDash}{1}
\FALabel(10.,14.57)[b]{$\bb/\tb$}
\FAProp(6.5,6.5)(13.5,6.5)(0.,){/ScalarDash}{-1}
\FALabel(10.,5.43)[t]{$\bb/\tb$}
\FAProp(13.5,13.5)(13.5,6.5)(0.,){/Sine}{0}
\FALabel(14.57,10.)[l]{$V/W$}
\FAVert(6.5,13.5){0}
\FAVert(6.5,6.5){0}
\FAVert(13.5,13.5){0}
\FAVert(13.5,6.5){0}
\end{feynartspicture}

\begin{feynartspicture}(300,300)(1,1)
\FADiagram{}
\FAProp(0.,15.)(6.,10.)(0.,){/Cycles}{0}
\FALabel(2.48771,11.7893)[tr]{$g$}
\FAProp(0.,5.)(6.,10.)(0.,){/Cycles}{0}
\FALabel(3.51229,6.78926)[tl]{$g$}
\FAProp(20.,15.)(14.,10.)(0.,){/ScalarDash}{-1}
\FALabel(16.4877,13.2107)[br]{$\ba$}
\FAProp(20.,5.)(14.,10.)(0.,){/ScalarDash}{1}
\FALabel(17.5123,8.21074)[bl]{$\ba$}
\FAProp(6.,10.)(14.,10.)(0.8,){/ScalarDash}{-1}
\FALabel(10.,5.73)[t]{$\qa$}
\FAProp(6.,10.)(14.,10.)(-0.8,){/ScalarDash}{1}
\FALabel(10.,14.27)[b]{$\qa$}
\FAVert(6.,10.){0}
\FAVert(14.,10.){0}
\end{feynartspicture}

\begin{feynartspicture}(300,300)(1,1)
\FADiagram{}
\FAProp(0.,15.)(10.,14.)(0.,){/Cycles}{0}
\FALabel(5.22388,16.2588)[b]{$g$}
\FAProp(0.,5.)(10.,6.)(0.,){/Cycles}{0}
\FALabel(5.15423,4.43769)[t]{$g$}
\FAProp(20.,15.)(10.,14.)(0.,){/ScalarDash}{-1}
\FALabel(14.8458,15.5623)[b]{$\ba$}
\FAProp(20.,5.)(10.,6.)(0.,){/ScalarDash}{1}
\FALabel(14.8458,4.43769)[t]{$\ba$}
\FAProp(10.,14.)(10.,6.)(0.8,){/ScalarDash}{-1}
\FALabel(5.73,10.)[r]{$\bb/\tb$}
\FAProp(10.,14.)(10.,6.)(-0.8,){/Sine}{0}
\FALabel(14.27,10.)[l]{$V/W$}
\FAVert(10.,14.){0}
\FAVert(10.,6.){0}
\end{feynartspicture}
\begin{feynartspicture}(300,300)(1,1)
\FADiagram{}
\end{feynartspicture}
\begin{feynartspicture}(300,300)(1,1)
\FADiagram{}
\end{feynartspicture}
\begin{feynartspicture}(300,300)(1,1)
\FADiagram{}
\FAProp(0.,15.)(10.,14.5)(0.,){/Cycles}{0}
\FALabel(5.11236,16.5172)[b]{$g$}
\FAProp(0.,5.)(10.,5.5)(0.,){/Cycles}{0}
\FALabel(5.0774,4.18193)[t]{$g$}
\FAProp(20.,15.)(10.,14.5)(0.,){/ScalarDash}{-1}
\FALabel(14.9226,15.8181)[b]{$\ba$}
\FAProp(20.,5.)(10.,5.5)(0.,){/ScalarDash}{1}
\FALabel(14.9226,4.18193)[t]{$\ba$}
\FAProp(10.,14.5)(10.,10.)(0.,){/ScalarDash}{-1}
\FALabel(8.93,12.25)[r]{$\ba$}
\FAProp(10.,5.5)(10.,10.)(0.,){/ScalarDash}{1}
\FALabel(8.93,7.75)[r]{$\ba$}
\FAProp(10.,10.)(10.,10.)(15.,10.){/ScalarDash}{0}
\FALabel(15.82,10.)[lb]{$S,S^\pm,$}
\FALabel(15.82,10.)[lt]{$\qa,\tilde{l}_a$}
\FAVert(10.,14.5){0}
\FAVert(10.,5.5){0}
\FAVert(10.,10.){0}
\end{feynartspicture}

\begin{feynartspicture}(300,300)(1,1)
\FADiagram{}
\FAProp(0.,15.)(10.,14.5)(0.,){/Cycles}{0}
\FALabel(5.11236,16.5172)[b]{$g$}
\FAProp(0.,5.)(10.,5.5)(0.,){/Cycles}{0}
\FALabel(5.0774,4.18193)[t]{$g$}
\FAProp(20.,15.)(10.,14.5)(0.,){/ScalarDash}{-1}
\FALabel(14.9226,15.8181)[b]{$\ba$}
\FAProp(20.,5.)(10.,5.5)(0.,){/ScalarDash}{1}
\FALabel(14.9226,4.18193)[t]{$\ba$}
\FAProp(10.,14.5)(10.,10.)(0.,){/ScalarDash}{-1}
\FALabel(8.93,12.25)[r]{$\ba$}
\FAProp(10.,5.5)(10.,10.)(0.,){/ScalarDash}{1}
\FALabel(8.93,7.75)[r]{$\ba$}
\FAProp(10.,10.)(10.,10.)(15.,10.){/Sine}{0}
\FALabel(16.07,10.)[l]{$V$}
\FAVert(10.,14.5){0}
\FAVert(10.,5.5){0}
\FAVert(10.,10.){0}
\end{feynartspicture}

\begin{feynartspicture}(300,300)(1,1)
\FADiagram{}
\FAProp(0.,15.)(10.,14.5)(0.,){/Cycles}{0}
\FALabel(5.11236,16.5172)[b]{$g$}
\FAProp(0.,5.)(10.,5.5)(0.,){/Cycles}{0}
\FALabel(5.0774,4.18193)[t]{$g$}
\FAProp(20.,15.)(10.,14.5)(0.,){/ScalarDash}{-1}
\FALabel(14.9226,15.8181)[b]{$\ba$}
\FAProp(20.,5.)(10.,5.5)(0.,){/ScalarDash}{1}
\FALabel(14.9226,4.18193)[t]{$\ba$}
\FAProp(10.,14.5)(10.,12.)(0.,){/ScalarDash}{-1}
\FALabel(11.07,13.25)[l]{$\ba$}
\FAProp(10.,5.5)(10.,8.)(0.,){/ScalarDash}{1}
\FALabel(11.07,6.75)[l]{$\ba$}
\FAProp(10.,12.)(10.,8.)(1.,){/Straight}{-1}
\FALabel(6.93,10.)[r]{$b/t$}
\FAProp(10.,12.)(10.,8.)(-1.,){/Straight}{0}
\FALabel(12.82,10.)[l]{$\neu_i/\cha_i$}
\FAVert(10.,14.5){0}
\FAVert(10.,5.5){0}
\FAVert(10.,12.){0}
\FAVert(10.,8.){0}
\end{feynartspicture}

\begin{feynartspicture}(300,300)(1,1)
\FADiagram{}
\FAProp(0.,15.)(10.,14.5)(0.,){/Cycles}{0}
\FALabel(5.11236,16.5172)[b]{$g$}
\FAProp(0.,5.)(10.,5.5)(0.,){/Cycles}{0}
\FALabel(5.0774,4.18193)[t]{$g$}
\FAProp(20.,15.)(10.,14.5)(0.,){/ScalarDash}{-1}
\FALabel(14.9226,15.8181)[b]{$\ba$}
\FAProp(20.,5.)(10.,5.5)(0.,){/ScalarDash}{1}
\FALabel(14.9226,4.18193)[t]{$\ba$}
\FAProp(10.,14.5)(10.,12.)(0.,){/ScalarDash}{-1}
\FALabel(11.07,13.25)[l]{$\ba$}
\FAProp(10.,5.5)(10.,8.)(0.,){/ScalarDash}{1}
\FALabel(11.07,6.75)[l]{$\ba$}
\FAProp(10.,12.)(10.,8.)(1.,){/ScalarDash}{-1}
\FALabel(6.93,10.)[r]{$S/S^\pm$}
\FAProp(10.,12.)(10.,8.)(-1.,){/ScalarDash}{-1}
\FALabel(13.07,10.)[l]{$\bb/\tb$}
\FAVert(10.,14.5){0}
\FAVert(10.,5.5){0}
\FAVert(10.,12.){0}
\FAVert(10.,8.){0}
\end{feynartspicture}

\begin{feynartspicture}(300,300)(1,1)
\FADiagram{}
\FAProp(0.,15.)(10.,14.5)(0.,){/Cycles}{0}
\FALabel(5.11236,16.5172)[b]{$g$}
\FAProp(0.,5.)(10.,5.5)(0.,){/Cycles}{0}
\FALabel(5.0774,4.18193)[t]{$g$}
\FAProp(20.,15.)(10.,14.5)(0.,){/ScalarDash}{-1}
\FALabel(14.9226,15.8181)[b]{$\ba$}
\FAProp(20.,5.)(10.,5.5)(0.,){/ScalarDash}{1}
\FALabel(14.9226,4.18193)[t]{$\ba$}
\FAProp(10.,14.5)(10.,12.)(0.,){/ScalarDash}{-1}
\FALabel(11.07,13.25)[l]{$\ba$}
\FAProp(10.,5.5)(10.,8.)(0.,){/ScalarDash}{1}
\FALabel(11.07,6.75)[l]{$\ba$}
\FAProp(10.,12.)(10.,8.)(1.,){/ScalarDash}{-1}
\FALabel(6.93,10.)[r]{$\bb/\tb$}
\FAProp(10.,12.)(10.,8.)(-1.,){/Sine}{0}
\FALabel(13.07,10.)[l]{$V/W$}
\FAVert(10.,14.5){0}
\FAVert(10.,5.5){0}
\FAVert(10.,12.){0}
\FAVert(10.,8.){0}
\end{feynartspicture}
\begin{feynartspicture}(300,300)(1,1)
\FADiagram{}
\end{feynartspicture}

\begin{feynartspicture}(300,300)(1,1)
\FADiagram{}
\FAProp(0.,15.)(10.,10.)(0.,){/Cycles}{0}
\FALabel(4.78682,11.5936)[tr]{$g$}
\FAProp(0.,5.)(10.,10.)(0.,){/Cycles}{0}
\FALabel(5.21318,6.59364)[tl]{$g$}
\FAProp(20.,15.)(10.,10.)(0.,){/ScalarDash}{-1}
\FALabel(14.7868,13.4064)[br]{$\ba$}
\FAProp(20.,5.)(10.,10.)(0.,){/ScalarDash}{1}
\FALabel(15.2132,8.40636)[bl]{$\ba$}
\FAVert(10.,10.){1}
\end{feynartspicture}

\begin{feynartspicture}(300,300)(1,1)
\FADiagram{}
\FAProp(0.,15.)(6.,10.)(0.,){/Cycles}{0}
\FALabel(2.48771,11.7893)[tr]{$g$}
\FAProp(0.,5.)(6.,10.)(0.,){/Cycles}{0}
\FALabel(3.51229,6.78926)[tl]{$g$}
\FAProp(20.,15.)(14.,10.)(0.,){/ScalarDash}{-1}
\FALabel(16.4877,13.2107)[br]{$\ba$}
\FAProp(20.,5.)(14.,10.)(0.,){/ScalarDash}{1}
\FALabel(17.5123,8.21074)[bl]{$\ba$}
\FAProp(6.,10.)(14.,10.)(0.,){/Cycles}{0}
\FALabel(10.,8.93)[t]{$g$}
\FAVert(6.,10.){0}
\FAVert(14.,10.){1}
\end{feynartspicture}

\begin{feynartspicture}(300,300)(1,1)
\FADiagram{}
\FAProp(0.,15.)(10.,14.)(0.,){/Cycles}{0}
\FALabel(4.84577,13.4377)[t]{$g$}
\FAProp(0.,5.)(10.,6.)(0.,){/Cycles}{0}
\FALabel(5.15423,4.43769)[t]{$g$}
\FAProp(20.,15.)(10.,14.)(0.,){/ScalarDash}{-1}
\FALabel(14.8458,15.5623)[b]{$\ba$}
\FAProp(20.,5.)(10.,6.)(0.,){/ScalarDash}{1}
\FALabel(15.1542,6.56231)[b]{$\ba$}
\FAProp(10.,14.)(10.,6.)(0.,){/ScalarDash}{-1}
\FALabel(8.93,10.)[r]{$\ba$}
\FAVert(10.,14.){0}
\FAVert(10.,6.){1}
\end{feynartspicture}

\begin{feynartspicture}(300,300)(1,1)
\FADiagram{}
\FAProp(0.,15.)(10.,14.)(0.,){/Cycles}{0}
\FALabel(4.84577,13.4377)[t]{$g$}
\FAProp(0.,5.)(10.,6.)(0.,){/Cycles}{0}
\FALabel(5.15423,4.43769)[t]{$g$}
\FAProp(20.,15.)(10.,14.)(0.,){/ScalarDash}{-1}
\FALabel(14.8458,15.5623)[b]{$\ba$}
\FAProp(20.,5.)(10.,6.)(0.,){/ScalarDash}{1}
\FALabel(15.1542,6.56231)[b]{$\ba$}
\FAProp(10.,14.)(10.,6.)(0.,){/ScalarDash}{-1}
\FALabel(8.93,10.)[r]{$\ba$}
\FAVert(10.,6.){0}
\FAVert(10.,14.){1}
\end{feynartspicture}

\begin{feynartspicture}(300,300)(1,1)
\FADiagram{}
\FAProp(0.,15.)(10.,14.)(0.,){/Cycles}{0}
\FALabel(4.84577,13.4377)[t]{$g$}
\FAProp(0.,5.)(10.,6.)(0.,){/Cycles}{0}
\FALabel(5.15423,4.43769)[t]{$g$}
\FAProp(20.,15.)(10.,14.)(0.,){/ScalarDash}{-1}
\FALabel(14.8458,15.5623)[b]{$\ba$}
\FAProp(20.,5.)(10.,6.)(0.,){/ScalarDash}{1}
\FALabel(15.1542,6.56231)[b]{$\ba$}
\FAProp(10.,10.)(10.,14.)(0.,){/ScalarDash}{1}
\FALabel(11.07,12.)[l]{$\ba$}
\FAProp(10.,10.)(10.,6.)(0.,){/ScalarDash}{-1}
\FALabel(8.93,8.)[r]{$\ba$}
\FAVert(10.,14.){0}
\FAVert(10.,6.){0}
\FAVert(10.,10.){1}
\end{feynartspicture}

\caption{Virtual corrections to the process $gg \to\ba\ba^\ast$. A
  common label $V$ is used for the neutral gauge bosons
  $\gamma,Z^0$, while $S$ denotes any of the neutral Higgs bosons
  or the neural Goldstone boson $h^0,H^0,G^0$, and $S^\pm$ denotes the
  charged ones $H^\pm,G^\pm$. Crossed diagrams are not shown
  explicitly. The diagrams containing the counterterms  are depicted in the last line.  The
  counterterms have to be evaluated at $\Order(\alpha)$.
  \label{fig:feynman_ggvirt}}
}

\FIGURE{
\footnotesize\unitlength=0.26bp%
\begin{feynartspicture}(300,300)(1,1)
\FADiagram{}
\FAProp(0.,15.)(4.,10.)(0.,){/Straight}{1}
\FAProp(0.,5.)(4.,10.)(0.,){/Straight}{-1}
\FALabel(2.,14.2213)[b]{$q$}
\FALabel(2.,5.77869)[t]{$q$}
\FAProp(20.,15.)(16.,13.5)(0.,){/ScalarDash}{-1}
\FALabel(17.4558,15.2213)[b]{$\ba$}
\FAProp(20.,5.)(16.,6.5)(0.,){/ScalarDash}{1}
\FALabel(17.4558,4.77869)[t]{$\ba$}
\FAProp(4.,10.)(10.,10.)(0.,){/Cycles}{0}
\FALabel(7.,11.77)[b]{$g$}
\FAProp(16.,13.5)(16.,6.5)(0.,){/Straight}{0}
\FALabel(16.82,10.)[l]{$\neu_i/\cha_i$}
\FAProp(16.,13.5)(10.,10.)(0.,){/Straight}{-1}
\FALabel(12.699,12.6089)[br]{$b/t$}
\FAProp(16.,6.5)(10.,10.)(0.,){/Straight}{1}
\FALabel(12.699,7.39114)[tr]{$b/t$}
\FAVert(4.,10.){0}
\FAVert(16.,13.5){0}
\FAVert(16.,6.5){0}
\FAVert(10.,10.){0}
\end{feynartspicture}
\begin{feynartspicture}(300,300)(1,1)
\FADiagram{}
\FAProp(0.,15.)(4.,10.)(0.,){/Straight}{1}
\FAProp(0.,5.)(4.,10.)(0.,){/Straight}{-1}
\FALabel(2.,14.2213)[b]{$q$}
\FALabel(2.,5.77869)[t]{$q$}
\FAProp(20.,15.)(16.,13.5)(0.,){/ScalarDash}{-1}
\FALabel(17.4558,15.2213)[b]{$\ba$}
\FAProp(20.,5.)(16.,6.5)(0.,){/ScalarDash}{1}
\FALabel(17.4558,4.77869)[t]{$\ba$}
\FAProp(4.,10.)(10.,10.)(0.,){/Cycles}{0}
\FALabel(7.,10.77)[b]{$g$}
\FAProp(16.,13.5)(16.,6.5)(0.,){/ScalarDash}{-1}
\FALabel(17.07,10.)[l]{$S/S^\pm$}
\FAProp(16.,13.5)(10.,10.)(0.,){/ScalarDash}{-1}
\FALabel(13.699,12.6089)[br]{$\bb/\tb$}
\FAProp(16.,6.5)(10.,10.)(0.,){/ScalarDash}{1}
\FALabel(13.699,7.39114)[tr]{$\bb/\tb$}
\FAVert(4.,10.){0}
\FAVert(16.,13.5){0}
\FAVert(16.,6.5){0}
\FAVert(10.,10.){0}
\end{feynartspicture}
\begin{feynartspicture}(300,300)(1,1)
\FADiagram{}
\FAProp(0.,15.)(4.,10.)(0.,){/Straight}{1}
\FAProp(0.,5.)(4.,10.)(0.,){/Straight}{-1}
\FALabel(2.,14.2213)[b]{$q$}
\FALabel(2.,5.77869)[t]{$q$}
\FAProp(20.,15.)(16.,13.5)(0.,){/ScalarDash}{-1}
\FALabel(17.4558,15.2213)[b]{$\ba$}
\FAProp(20.,5.)(16.,6.5)(0.,){/ScalarDash}{1}
\FALabel(17.4558,4.77869)[t]{$\ba$}
\FAProp(4.,10.)(10.,10.)(0.,){/Cycles}{0}
\FALabel(7.,11.)[b]{$g$}
\FAProp(16.,13.5)(16.,6.5)(0.,){/Sine}{0}
\FALabel(17.07,10.)[l]{$V$}
\FAProp(16.,13.5)(10.,10.)(0.,){/ScalarDash}{-1}
\FALabel(13.699,12.6089)[br]{$\bb/\tb$}
\FAProp(16.,6.5)(10.,10.)(0.,){/ScalarDash}{1}
\FALabel(13.699,7.39114)[tr]{$\bb/\tb$}
\FAVert(4.,10.){0}
\FAVert(16.,13.5){0}
\FAVert(16.,6.5){0}
\FAVert(10.,10.){0}
\end{feynartspicture}
\begin{feynartspicture}(300,300)(1,1)
\FADiagram{}
\FAProp(0.,15.)(4.,13.5)(0.,){/Straight}{1}
\FALabel(2.54424,15.2213)[b]{$q$}
\FAProp(0.,5.)(4.,6.5)(0.,){/Straight}{-1}
\FALabel(2.54424,4.77869)[t]{$q$}
\FAProp(20.,15.)(16.,10.)(0.,){/ScalarDash}{-1}
\FALabel(17.2697,12.9883)[br]{$\ba$}
\FAProp(20.,5.)(16.,10.)(0.,){/ScalarDash}{1}
\FALabel(18.7303,7.98828)[tr]{$\ba$}
\FAProp(16.,10.)(10.,10.)(0.,){/Cycles}{0}
\FALabel(13.,8.23)[t]{$g$}
\FAProp(4.,13.5)(4.,6.5)(0.,){/Sine}{0}
\FALabel(2.93,10.)[r]{$V$}
\FAProp(4.,13.5)(10.,10.)(0.,){/Straight}{1}
\FALabel(7.301,12.6089)[bl]{$q$}
\FAProp(4.,6.5)(10.,10.)(0.,){/Straight}{-1}
\FALabel(7.301,7.39114)[tl]{$q$}
\FAVert(4.,13.5){0}
\FAVert(4.,6.5){0}
\FAVert(16.,10.){0}
\FAVert(10.,10.){0}
\end{feynartspicture}
\begin{feynartspicture}(300,300)(1,1)
\FADiagram{}
\FAProp(0.,15.)(4.,13.5)(0.,){/Straight}{1}
\FALabel(2.54424,15.2213)[b]{$q$}
\FAProp(0.,5.)(4.,6.5)(0.,){/Straight}{-1}
\FALabel(2.54424,4.77869)[t]{$q$}
\FAProp(20.,15.)(16.,10.)(0.,){/ScalarDash}{-1}
\FALabel(17.2697,12.9883)[br]{$\ba$}
\FAProp(20.,5.)(16.,10.)(0.,){/ScalarDash}{1}
\FALabel(18.7303,7.98828)[bl]{$\ba$}
\FAProp(16.,10.)(10.,10.)(0.,){/Cycles}{0}
\FALabel(13.,8.23)[t]{$g$}
\FAProp(4.,13.5)(4.,6.5)(0.,){/Straight}{0}
\FALabel(4.18,10.)[r]{$\neu_i/\cha_i$}
\FAProp(4.,13.5)(10.,10.)(0.,){/ScalarDash}{1}
\FALabel(7.301,12.6089)[bl]{$\qa$}
\FAProp(4.,6.5)(10.,10.)(0.,){/ScalarDash}{-1}
\FALabel(7.301,7.39114)[tl]{$\qa$}
\FAVert(4.,13.5){0}
\FAVert(4.,6.5){0}
\FAVert(16.,10.){0}
\FAVert(10.,10.){0}
\end{feynartspicture}
\begin{feynartspicture}(300,300)(1,1)
\FADiagram{}
\FAProp(0.,15.)(5.,10.)(0.,){/Straight}{1}
\FAProp(0.,5.)(5.,10.)(0.,){/Straight}{-1}
\FALabel(2.,14.2213)[b]{$q$}
\FALabel(2.,5.77869)[t]{$q$}
\FAProp(20.,15.)(12.5,12.)(0.,){/ScalarDash}{-1}
\FALabel(15.6743,14.4591)[b]{$\ba$}
\FAProp(20.,5.)(14.5,6.5)(0.,){/ScalarDash}{1}
\FALabel(16.8422,4.73462)[t]{$\ba$}
\FAProp(5.,10.)(12.5,12.)(0.,){/Cycles}{0}
\FALabel(8.17026,12.694)[b]{$g$}
\FAProp(14.5,6.5)(12.5,12.)(0.8,){/ScalarDash}{1}
\FALabel(16.6767,10.5797)[l]{$\bb/\tb$}
\FAProp(14.5,6.5)(12.5,12.)(-0.8,){/Sine}{0}
\FALabel(10.3233,7.9203)[r]{$V/W$}
\FAVert(5.,10.){0}
\FAVert(14.5,6.5){0}
\FAVert(12.5,12.){0}
\end{feynartspicture}

\begin{feynartspicture}(300,300)(1,1)
\FADiagram{}
\FAProp(0.,15.)(5.,10.)(0.,){/Straight}{1}
\FAProp(0.,5.)(5.,10.)(0.,){/Straight}{-1}
\FALabel(2.,14.2213)[b]{$q$}
\FALabel(2.,5.77869)[t]{$q$}
\FAProp(20.,15.)(14.5,13.5)(0.,){/ScalarDash}{-1}
\FALabel(16.8422,15.2654)[b]{$\ba$}
\FAProp(20.,5.)(12.5,8.)(0.,){/ScalarDash}{1}
\FALabel(15.6743,5.54086)[t]{$\ba$}
\FAProp(5.,10.)(12.5,8.)(0.,){/Cycles}{0}
\FALabel(8.35062,7.98234)[t]{$g$}
\FAProp(14.5,13.5)(12.5,8.)(0.8,){/ScalarDash}{-1}
\FALabel(10.3233,12.0797)[r]{$\bb/\tb$}
\FAProp(14.5,13.5)(12.5,8.)(-0.8,){/Sine}{0}
\FALabel(16.6767,9.4203)[l]{$V/W$}
\FAVert(5.,10.){0}
\FAVert(14.5,13.5){0}
\FAVert(12.5,8.){0}
\end{feynartspicture}
\begin{feynartspicture}(300,300)(1,1)
\FADiagram{}
\FAProp(0.,15.)(4.,10.)(0.,){/Straight}{1}
\FAProp(0.,5.)(4.,10.)(0.,){/Straight}{-1}
\FALabel(2.,14.2213)[b]{$q$}
\FALabel(2.,5.77869)[t]{$q$}
\FAProp(20.,15.)(15.5,10.)(0.,){/ScalarDash}{-1}
\FALabel(17.,13.9431)[bl]{$\ba$}
\FAProp(20.,5.)(15.5,10.)(0.,){/ScalarDash}{1}
\FALabel(18.4221,7.0569)[tr]{$\ba$}
\FAProp(4.,10.)(10.,10.)(0.,){/Cycles}{0}
\FALabel(7.,11.77)[b]{$g$}
\FAProp(10.,10.)(15.5,10.)(0.8,){/ScalarDash}{-1}
\FALabel(12.75,6.73)[t]{$\qa$}
\FAProp(10.,10.)(15.5,10.)(-0.8,){/ScalarDash}{1}
\FALabel(12.75,13.27)[b]{$\qa$}
\FAVert(4.,10.){0}
\FAVert(10.,10.){0}
\FAVert(15.5,10.){0}
\end{feynartspicture}

\begin{feynartspicture}(300,300)(1,1)
\FADiagram{}
\FAProp(0.,15.)(6.5,13.5)(0.,){/Straight}{1}
\FALabel(3.59853,15.2803)[b]{$q$}
\FAProp(0.,5.)(6.5,6.5)(0.,){/Straight}{-1}
\FALabel(3.59853,4.71969)[t]{$q$}
\FAProp(20.,15.)(13.5,13.5)(0.,){/ScalarDash}{-1}
\FALabel(16.4015,15.2803)[b]{$\ba$}
\FAProp(20.,5.)(13.5,6.5)(0.,){/ScalarDash}{1}
\FALabel(16.4015,4.71969)[t]{$\ba$}
\FAProp(6.5,13.5)(6.5,6.5)(0.,){/ScalarDash}{1}
\FALabel(5.43,10.)[r]{$\qa$}
\FAProp(6.5,13.5)(13.5,13.5)(0.,){/Straight}{0}
\FALabel(10.,14.32)[b]{$\neu_i$}
\FAProp(6.5,6.5)(13.5,6.5)(0.,){/Straight}{0}
\FALabel(10.,5.68)[t]{$\tilde g$}
\FAProp(13.5,13.5)(13.5,6.5)(0.,){/Straight}{-1}
\FALabel(14.57,10.)[l]{$b$}
\FAVert(6.5,13.5){0}
\FAVert(6.5,6.5){0}
\FAVert(13.5,13.5){0}
\FAVert(13.5,6.5){0}
\end{feynartspicture}
\begin{feynartspicture}(300,300)(1,1)
\FADiagram{}
\FAProp(0.,15.)(6.5,13.5)(0.,){/Straight}{1}
\FALabel(3.59853,15.2803)[b]{$q$}
\FAProp(0.,5.)(6.5,6.5)(0.,){/Straight}{-1}
\FALabel(3.59853,4.71969)[t]{$q$}
\FAProp(20.,15.)(13.5,13.5)(0.,){/ScalarDash}{-1}
\FALabel(16.4015,15.2803)[b]{$\ba$}
\FAProp(20.,5.)(13.5,6.5)(0.,){/ScalarDash}{1}
\FALabel(16.4015,4.71969)[t]{$\ba$}
\FAProp(6.5,13.5)(6.5,6.5)(0.,){/ScalarDash}{1}
\FALabel(5.43,10.)[r]{$\qa$}
\FAProp(6.5,13.5)(13.5,13.5)(0.,){/Straight}{0}
\FALabel(10.,14.32)[b]{$\tilde g$}
\FAProp(6.5,6.5)(13.5,6.5)(0.,){/Straight}{0}
\FALabel(10.,5.68)[t]{$\neu_i$}
\FAProp(13.5,13.5)(13.5,6.5)(0.,){/Straight}{-1}
\FALabel(14.57,10.)[l]{$b$}
\FAVert(6.5,13.5){0}
\FAVert(6.5,6.5){0}
\FAVert(13.5,13.5){0}
\FAVert(13.5,6.5){0}
\end{feynartspicture}
\begin{feynartspicture}(300,300)(1,1)
\FADiagram{}
\FAProp(0.,15.)(6.5,13.5)(0.,){/Straight}{1}
\FALabel(3.59853,15.2803)[b]{$q$}
\FAProp(0.,5.)(6.5,6.5)(0.,){/Straight}{-1}
\FALabel(3.59853,4.71969)[t]{$q$}
\FAProp(20.,15.)(13.5,13.5)(0.,){/ScalarDash}{-1}
\FALabel(16.4015,15.2803)[b]{$\ba$}
\FAProp(20.,5.)(13.5,6.5)(0.,){/ScalarDash}{1}
\FALabel(16.4015,4.71969)[t]{$\ba$}
\FAProp(6.5,13.5)(6.5,6.5)(0.,){/Straight}{1}
\FALabel(5.43,10.)[r]{$q$}
\FAProp(6.5,13.5)(13.5,13.5)(0.,){/Sine}{0}
\FALabel(10.,14.57)[b]{$V$}
\FAProp(6.5,6.5)(13.5,6.5)(0.,){/Cycles}{0}
\FALabel(10.,5.43)[t]{$g$}
\FAProp(13.5,13.5)(13.5,6.5)(0.,){/ScalarDash}{-1}
\FALabel(14.57,10.)[l]{$\ba$}
\FAVert(6.5,13.5){0}
\FAVert(6.5,6.5){0}
\FAVert(13.5,13.5){0}
\FAVert(13.5,6.5){0}
\end{feynartspicture}
\begin{feynartspicture}(300,300)(1,1)
\FADiagram{}
\FAProp(0.,15.)(6.5,13.5)(0.,){/Straight}{1}
\FALabel(3.59853,15.2803)[b]{$q$}
\FAProp(0.,5.)(6.5,6.5)(0.,){/Straight}{-1}
\FALabel(3.59853,4.71969)[t]{$q$}
\FAProp(20.,15.)(13.5,13.5)(0.,){/ScalarDash}{-1}
\FALabel(16.4015,15.2803)[b]{$\ba$}
\FAProp(20.,5.)(13.5,6.5)(0.,){/ScalarDash}{1}
\FALabel(16.4015,4.71969)[t]{$\ba$}
\FAProp(6.5,13.5)(6.5,6.5)(0.,){/Straight}{1}
\FALabel(5.43,10.)[r]{$q$}
\FAProp(6.5,13.5)(13.5,13.5)(0.,){/Cycles}{0}
\FALabel(10.,14.57)[b]{$g$}
\FAProp(6.5,6.5)(13.5,6.5)(0.,){/Sine}{0}
\FALabel(10.,5.43)[t]{$V$}
\FAProp(13.5,13.5)(13.5,6.5)(0.,){/ScalarDash}{-1}
\FALabel(14.57,10.)[l]{$\ba$}
\FAVert(6.5,13.5){0}
\FAVert(6.5,6.5){0}
\FAVert(13.5,13.5){0}
\FAVert(13.5,6.5){0}
\end{feynartspicture}
\begin{feynartspicture}(300,300)(1,1)
\FADiagram{}
\FAProp(0.,15.)(6.,10.)(0.,){/Straight}{1}
\FAProp(0.,5.)(6.,10.)(0.,){/Straight}{-1}
\FALabel(2.,14.2213)[b]{$q$}
\FALabel(2.,5.77869)[t]{$q$}
\FAProp(20.,15.)(14.,10.)(0.,){/ScalarDash}{-1}
\FALabel(16.4877,13.2107)[br]{$\ba$}
\FAProp(20.,5.)(14.,10.)(0.,){/ScalarDash}{1}
\FALabel(17.5123,8.21074)[bl]{$\ba$}
\FAProp(6.,10.)(14.,10.)(0.,){/Cycles}{0}
\FALabel(10.,8.93)[t]{$g$}
\FAVert(6.,10.){0}
\FAVert(14.,10.){1}
\end{feynartspicture}

\begin{feynartspicture}(300,300)(1,1)
\FADiagram{}
\FAProp(0.,15.)(6.,10.)(0.,){/Straight}{1}
\FAProp(0.,5.)(6.,10.)(0.,){/Straight}{-1}
\FALabel(2.,14.2213)[b]{$q$}
\FALabel(2.,5.77869)[t]{$q$}
\FAProp(20.,15.)(14.,10.)(0.,){/ScalarDash}{-1}
\FALabel(16.4877,13.2107)[br]{$\ba$}
\FAProp(20.,5.)(14.,10.)(0.,){/ScalarDash}{1}
\FALabel(17.5123,8.21074)[bl]{$\ba$}
\FAProp(6.,10.)(14.,10.)(0.,){/Cycles}{0}
\FALabel(10.,8.93)[t]{$g$}
\FAVert(14.,10.){0}
\FAVert(6.,10.){1}
\end{feynartspicture}

\begin{feynartspicture}(300,300)(1,1)
\FADiagram{}
\end{feynartspicture}

\caption{Virtual contributions to the process $q\bar{q}\to
  \ba\ba^\ast$. The diagrams result from EW insertions to tree-level
  QCD diagrams and from QCD insertions to tree-level EW
  diagrams. $V$, $S$, and $S^\pm$ are defined as in
  Figure~\ref{fig:feynman_ggvirt}. Crossed diagrams are not
  shown. The counterterms in the last two diagrams have to be
  evaluated at $\Order(\alpha)$.
 \label{fig:feynman_qqvirtEW}}
}

\FIGURE{
\footnotesize\unitlength=0.26bp%
\hspace*{20pt}
\begin{feynartspicture}(300,300)(1,1)
\FADiagram{}
\FAProp(0.,15.)(6.,13.5)(0.,){/Straight}{1}
\FALabel(3.37593,15.2737)[b]{$q$}
\FAProp(0.,5.)(6.,6.5)(0.,){/Straight}{-1}
\FALabel(3.37593,4.72628)[t]{$q$}
\FAProp(20.,15.)(12.,10.)(0.,){/ScalarDash}{-1}
\FALabel(15.6585,13.3344)[br]{$\ba$}
\FAProp(20.,5.)(12.,10.)(0.,){/ScalarDash}{1}
\FALabel(16.3415,8.3344)[bl]{$\ba$}
\FAProp(6.,13.5)(6.,6.5)(0.,){/Straight}{0}
\FALabel(5.18,10.)[r]{$\tilde g$}
\FAProp(6.,13.5)(12.,10.)(0.,){/ScalarDash}{1}
\FALabel(9.301,12.6089)[bl]{$\qa$}
\FAProp(6.,6.5)(12.,10.)(0.,){/ScalarDash}{-1}
\FALabel(9.301,7.39114)[tl]{$\qa$}
\FAVert(6.,13.5){0}
\FAVert(6.,6.5){0}
\FAVert(12.,10.){0}
\end{feynartspicture}
\begin{feynartspicture}(300,300)(1,1)
\FADiagram{}
\FAProp(0.,15.)(6.5,13.5)(0.,){/Straight}{1}
\FALabel(3.59853,15.2803)[b]{$q$}
\FAProp(0.,5.)(6.5,6.5)(0.,){/Straight}{-1}
\FALabel(3.59853,4.71969)[t]{$q$}
\FAProp(20.,15.)(13.5,13.5)(0.,){/ScalarDash}{-1}
\FALabel(16.4015,15.2803)[b]{$\ba$}
\FAProp(20.,5.)(13.5,6.5)(0.,){/ScalarDash}{1}
\FALabel(16.4015,4.71969)[t]{$\ba$}
\FAProp(6.5,13.5)(6.5,6.5)(0.,){/ScalarDash}{1}
\FALabel(5.43,10.)[r]{$\qa$}
\FAProp(6.5,13.5)(13.5,13.5)(0.,){/Straight}{0}
\FALabel(10.,14.32)[b]{$\tilde g$}
\FAProp(6.5,6.5)(13.5,6.5)(0.,){/Straight}{0}
\FALabel(10.,5.68)[t]{$\tilde g$}
\FAProp(13.5,13.5)(13.5,6.5)(0.,){/Straight}{-1}
\FALabel(14.57,10.)[l]{$b$}
\FAVert(6.5,13.5){0}
\FAVert(6.5,6.5){0}
\FAVert(13.5,13.5){0}
\FAVert(13.5,6.5){0}
\end{feynartspicture}
\begin{feynartspicture}(300,300)(1,1)
\FADiagram{}
\FAProp(0.,15.)(6.5,13.5)(0.,){/Straight}{1}
\FALabel(3.59853,15.2803)[b]{$q$}
\FAProp(0.,5.)(6.5,6.5)(0.,){/Straight}{-1}
\FALabel(3.59853,4.71969)[t]{$q$}
\FAProp(20.,15.)(13.5,13.5)(0.,){/ScalarDash}{-1}
\FALabel(16.4015,15.2803)[b]{$\ba$}
\FAProp(20.,5.)(13.5,6.5)(0.,){/ScalarDash}{1}
\FALabel(16.4015,4.71969)[t]{$\ba$}
\FAProp(6.5,13.5)(6.5,6.5)(0.,){/Straight}{1}
\FALabel(5.43,10.)[r]{$q$}
\FAProp(6.5,13.5)(13.5,13.5)(0.,){/Cycles}{0}
\FALabel(10.,15.27)[b]{$g$}
\FAProp(6.5,6.5)(13.5,6.5)(0.,){/Cycles}{0}
\FALabel(10.,5.43)[t]{$g$}
\FAProp(13.5,13.5)(13.5,6.5)(0.,){/ScalarDash}{-1}
\FALabel(14.57,10.)[l]{$\ba$}
\FAVert(6.5,13.5){0}
\FAVert(6.5,6.5){0}
\FAVert(13.5,13.5){0}
\FAVert(13.5,6.5){0}
\end{feynartspicture}
\hspace*{20pt}
\caption{Virtual QCD box contributions to the process $q\bar{q}\to
  \ba\ba^\ast$. Crossed diagrams are not shown.
  \label{fig:feynman_qqvirtQCD}}
}

\FIGURE{
\footnotesize
\unitlength=0.26bp%
\begin{feynartspicture}(300,300)(1,1)
\FADiagram{}
\FAProp(0.,15.)(5.5,10.)(0.,){/Cycles}{0}
\FALabel(2.18736,11.8331)[tr]{$g$}
\FAProp(0.,5.)(5.5,10.)(0.,){/Cycles}{0}
\FALabel(3.31264,6.83309)[tl]{$g$}
\FAProp(20.,17.)(12.5,10.)(0.,){/ScalarDash}{-1}
\FALabel(15.6724,14.1531)[br]{$\ba$}
\FAProp(20.,10.)(12.5,10.)(0.,){/ScalarDash}{1}
\FALabel(17.6,11.07)[b]{$\ba$}
\FAProp(20.,3.)(12.5,10.)(0.,){/Sine}{0}
\FALabel(15.6724,5.84686)[tr]{$\gamma$}
\FAProp(5.5,10.)(12.5,10.)(0.,){/Cycles}{0}
\FALabel(9.,8.93)[t]{$g$}
\FAVert(5.5,10.){0}
\FAVert(12.5,10.){0}
\end{feynartspicture}
\begin{feynartspicture}(300,300)(1,1)
\FADiagram{}
\FAProp(0.,15.)(10.,14.5)(0.,){/Cycles}{0}
\FALabel(5.11236,16.5172)[b]{$g$}
\FAProp(0.,5.)(10.,6.)(0.,){/Cycles}{0}
\FALabel(5.15423,4.43769)[t]{$g$}
\FAProp(20.,17.)(10.,14.5)(0.,){/ScalarDash}{-1}
\FALabel(14.6241,16.7737)[b]{$\ba$}
\FAProp(20.,10.)(10.,6.)(0.,){/ScalarDash}{1}
\FALabel(14.4243,8.95914)[b]{$\ba$}
\FAProp(20.,3.)(10.,6.)(0.,){/Sine}{0}
\FALabel(14.5546,3.49537)[t]{$\gamma$}
\FAProp(10.,14.5)(10.,6.)(0.,){/ScalarDash}{-1}
\FALabel(8.93,10.25)[r]{$\ba$}
\FAVert(10.,14.5){0}
\FAVert(10.,6.){0}
\end{feynartspicture}
\begin{feynartspicture}(300,300)(1,1)
\FADiagram{}
\FAProp(0.,15.)(10.,10.)(0.,){/Cycles}{0}
\FALabel(5.52623,14.0325)[bl]{$g$}
\FAProp(0.,5.)(10.,10.)(0.,){/Cycles}{0}
\FALabel(4.47377,9.03246)[br]{$g$}
\FAProp(20.,17.)(10.,10.)(0.,){/ScalarDash}{-1}
\FALabel(14.5911,14.2898)[br]{$\ba$}
\FAProp(20.,10.)(15.,6.5)(0.,){/ScalarDash}{1}
\FALabel(17.0911,9.03981)[br]{$\ba$}
\FAProp(20.,3.)(15.,6.5)(0.,){/Sine}{0}
\FALabel(17.9089,5.53981)[bl]{$\gamma$}
\FAProp(15.,6.5)(10.,10.)(0.,){/ScalarDash}{1}
\FALabel(12.0911,7.46019)[tr]{$\ba$}
\FAVert(15.,6.5){0}
\FAVert(10.,10.){0}
\end{feynartspicture}
\begin{feynartspicture}(300,300)(1,1)
\FADiagram{}
\FAProp(0.,15.)(5.5,10.)(0.,){/Cycles}{0}
\FALabel(2.18736,11.8331)[tr]{$g$}
\FAProp(0.,5.)(5.5,10.)(0.,){/Cycles}{0}
\FALabel(3.31264,6.83309)[tl]{$g$}
\FAProp(20.,17.)(11.5,10.)(0.,){/ScalarDash}{-1}
\FALabel(15.2447,14.2165)[br]{$\ba$}
\FAProp(20.,10.)(15.5,6.5)(0.,){/ScalarDash}{1}
\FALabel(17.2784,8.9935)[br]{$\ba$}
\FAProp(20.,3.)(15.5,6.5)(0.,){/Sine}{0}
\FALabel(18.2216,5.4935)[bl]{$\gamma$}
\FAProp(5.5,10.)(11.5,10.)(0.,){/Cycles}{0}
\FALabel(8.5,11.77)[b]{$g$}
\FAProp(11.5,10.)(15.5,6.5)(0.,){/ScalarDash}{-1}
\FALabel(12.9593,7.56351)[tr]{$\ba$}
\FAVert(5.5,10.){0}
\FAVert(11.5,10.){0}
\FAVert(15.5,6.5){0}
\end{feynartspicture}
\begin{feynartspicture}(300,300)(1,1)
\FADiagram{}
\FAProp(0.,15.)(10.,14.5)(0.,){/Cycles}{0}
\FALabel(5.11236,16.5172)[b]{$g$}
\FAProp(0.,5.)(10.,5.5)(0.,){/Cycles}{0}
\FALabel(5.0774,4.18193)[t]{$g$}
\FAProp(20.,17.)(10.,14.5)(0.,){/ScalarDash}{-1}
\FALabel(14.6241,16.7737)[b]{$\ba$}
\FAProp(20.,10.)(10.,5.5)(0.,){/ScalarDash}{1}
\FALabel(17.2909,9.58086)[br]{$\ba$}
\FAProp(20.,3.)(10.,10.)(0.,){/Sine}{0}
\FALabel(17.3366,3.77519)[tr]{$\gamma$}
\FAProp(10.,14.5)(10.,10.)(0.,){/ScalarDash}{-1}
\FALabel(8.93,12.25)[r]{$\ba$}
\FAProp(10.,5.5)(10.,10.)(0.,){/ScalarDash}{1}
\FALabel(8.93,7.75)[r]{$\ba$}
\FAVert(10.,14.5){0}
\FAVert(10.,5.5){0}
\FAVert(10.,10.){0}
\end{feynartspicture}
\begin{feynartspicture}(300,300)(1,1)
\FADiagram{}
\FAProp(0.,15.)(10.,14.5)(0.,){/Cycles}{0}
\FALabel(5.11236,16.5172)[b]{$g$}
\FAProp(0.,5.)(10.,7.)(0.,){/Cycles}{0}
\FALabel(5.30398,4.9601)[t]{$g$}
\FAProp(20.,17.)(10.,14.5)(0.,){/ScalarDash}{-1}
\FALabel(14.6241,16.7737)[b]{$\ba$}
\FAProp(20.,10.)(15.5,6.5)(0.,){/ScalarDash}{1}
\FALabel(17.2784,8.9935)[br]{$\ba$}
\FAProp(20.,3.)(15.5,6.5)(0.,){/Sine}{0}
\FALabel(17.2784,4.0065)[tr]{$\gamma$}
\FAProp(10.,14.5)(10.,7.)(0.,){/ScalarDash}{-1}
\FALabel(8.93,10.75)[r]{$\ba$}
\FAProp(10.,7.)(15.5,6.5)(0.,){/ScalarDash}{-1}
\FALabel(12.6097,5.68637)[t]{$\ba$}
\FAVert(10.,14.5){0}
\FAVert(10.,7.){0}
\FAVert(15.5,6.5){0}
\end{feynartspicture}


\begin{feynartspicture}(300,300)(1,1)
\FADiagram{}
\FAProp(0.,15.)(5.5,10.)(0.,){/Straight}{1}
\FALabel(2.18736,11.8331)[tr]{$q$}
\FAProp(0.,5.)(5.5,10.)(0.,){/Straight}{-1}
\FALabel(3.31264,6.83309)[tl]{$q$}
\FAProp(20.,17.)(12.5,10.)(0.,){/ScalarDash}{-1}
\FALabel(15.6724,14.1531)[br]{$\ba$}
\FAProp(20.,10.)(12.5,10.)(0.,){/ScalarDash}{1}
\FALabel(17.6,11.07)[b]{$\ba$}
\FAProp(20.,3.)(12.5,10.)(0.,){/Sine}{0}
\FALabel(15.6724,5.84686)[tr]{$\gamma$}
\FAProp(5.5,10.)(12.5,10.)(0.,){/Cycles}{0}
\FALabel(9.,8.93)[t]{$g$}
\FAVert(5.5,10.){0}
\FAVert(12.5,10.){0}
\end{feynartspicture}
\begin{feynartspicture}(300,300)(1,1)
\FADiagram{}
\FAProp(0.,15.)(5.5,10.)(0.,){/Straight}{1}
\FALabel(2.18736,11.8331)[tr]{$q$}
\FAProp(0.,5.)(5.5,10.)(0.,){/Straight}{-1}
\FALabel(3.31264,6.83309)[tl]{$q$}
\FAProp(20.,17.)(11.5,10.)(0.,){/ScalarDash}{-1}
\FALabel(15.2447,14.2165)[br]{$\ba$}
\FAProp(20.,10.)(15.5,6.5)(0.,){/ScalarDash}{1}
\FALabel(17.2784,8.9935)[br]{$\ba$}
\FAProp(20.,3.)(15.5,6.5)(0.,){/Sine}{0}
\FALabel(18.2216,5.4935)[bl]{$\gamma$}
\FAProp(5.5,10.)(11.5,10.)(0.,){/Cycles}{0}
\FALabel(8.5,11.77)[b]{$g$}
\FAProp(11.5,10.)(15.5,6.5)(0.,){/ScalarDash}{-1}
\FALabel(12.9593,7.56351)[tr]{$\ba$}
\FAVert(5.5,10.){0}
\FAVert(11.5,10.){0}
\FAVert(15.5,6.5){0}
\end{feynartspicture}
\begin{feynartspicture}(300,300)(1,1)
\FADiagram{}
\FAProp(0.,15.)(10.,13.)(0.,){/Straight}{1}
\FALabel(5.30398,15.0399)[b]{$q$}
\FAProp(0.,5.)(10.,5.5)(0.,){/Straight}{-1}
\FALabel(5.0774,4.18193)[t]{$q$}
\FAProp(20.,17.)(15.5,13.5)(0.,){/ScalarDash}{-1}
\FALabel(17.2784,15.9935)[br]{$\ba$}
\FAProp(20.,10.)(15.5,13.5)(0.,){/ScalarDash}{1}
\FALabel(18.2216,12.4935)[bl]{$\ba$}
\FAProp(20.,3.)(10.,5.5)(0.,){/Sine}{0}
\FALabel(15.3759,5.27372)[b]{$\gamma$}
\FAProp(10.,13.)(10.,5.5)(0.,){/Straight}{1}
\FALabel(8.93,9.25)[r]{$q$}
\FAProp(10.,13.)(15.5,13.5)(0.,){/Cycles}{0}
\FALabel(12.8903,12.1864)[t]{$g$}
\FAVert(10.,13.){0}
\FAVert(10.,5.5){0}
\FAVert(15.5,13.5){0}
\end{feynartspicture}

\begin{feynartspicture}(300,300)(1,1)
\FADiagram{}
\end{feynartspicture}

\caption{Real photon emission. The first six diagrams correspond to
  the $gg$ channel while the last three correspond to the $q\bar{q}$
  channel.
  \label{fig:feynman_ggRE}} }

\FIGURE{
\footnotesize\unitlength=0.26bp%
\begin{feynartspicture}(300,300)(1,1)
\FADiagram{}
\FAProp(0.,15.)(5.5,10.)(0.,){/Straight}{1}
\FALabel(2.18736,11.8331)[tr]{$q$}
\FAProp(0.,5.)(5.5,10.)(0.,){/Straight}{-1}
\FALabel(3.31264,6.83309)[tl]{$q$}
\FAProp(20.,17.)(12.5,10.)(0.,){/ScalarDash}{-1}
\FALabel(15.6724,14.1531)[br]{$\ba$}
\FAProp(20.,10.)(12.5,10.)(0.,){/ScalarDash}{1}
\FALabel(17.6,11.07)[b]{$\ba$}
\FAProp(20.,3.)(12.5,10.)(0.,){/Cycles}{0}
\FALabel(15.1948,5.33513)[tr]{$g$}
\FAProp(5.5,10.)(12.5,10.)(0.,){/Cycles}{0}
\FALabel(9.,8.93)[t]{$g$}
\FAVert(5.5,10.){0}
\FAVert(12.5,10.){0}
\end{feynartspicture}

\begin{feynartspicture}(300,300)(1,1)
\FADiagram{}
\FAProp(0.,15.)(5.5,10.)(0.,){/Straight}{1}
\FALabel(2.18736,11.8331)[tr]{$q$}
\FAProp(0.,5.)(5.5,10.)(0.,){/Straight}{-1}
\FALabel(3.31264,6.83309)[tl]{$q$}
\FAProp(20.,17.)(15.5,13.5)(0.,){/ScalarDash}{-1}
\FALabel(17.2784,15.9935)[br]{$\ba$}
\FAProp(20.,10.)(15.5,13.5)(0.,){/ScalarDash}{1}
\FALabel(18.2216,12.4935)[bl]{$\ba$}
\FAProp(20.,3.)(12.,10.)(0.,){/Cycles}{0}
\FALabel(14.9984,5.2867)[tr]{$g$}
\FAProp(5.5,10.)(12.,10.)(0.,){/Cycles}{0}
\FALabel(8.75,8.93)[t]{$g$}
\FAProp(15.5,13.5)(12.,10.)(0.,){/Cycles}{0}
\FALabel(13.134,12.366)[br]{$g$}
\FAVert(5.5,10.){0}
\FAVert(15.5,13.5){0}
\FAVert(12.,10.){0}
\end{feynartspicture}

\begin{feynartspicture}(300,300)(1,1)
\FADiagram{}
\FAProp(0.,15.)(5.5,10.)(0.,){/Straight}{1}
\FALabel(2.18736,11.8331)[tr]{$q$}
\FAProp(0.,5.)(5.5,10.)(0.,){/Straight}{-1}
\FALabel(3.31264,6.83309)[tl]{$q$}
\FAProp(20.,17.)(11.5,10.)(0.,){/ScalarDash}{-1}
\FALabel(15.2447,14.2165)[br]{$\ba$}
\FAProp(20.,10.)(15.5,6.5)(0.,){/ScalarDash}{1}
\FALabel(17.2784,8.9935)[br]{$\ba$}
\FAProp(20.,3.)(15.5,6.5)(0.,){/Cycles}{0}
\FALabel(18.2216,5.4935)[bl]{$g$}
\FAProp(5.5,10.)(11.5,10.)(0.,){/Cycles}{0}
\FALabel(8.5,11.77)[b]{$g$}
\FAProp(11.5,10.)(15.5,6.5)(0.,){/ScalarDash}{-1}
\FALabel(12.9593,7.56351)[tr]{$\ba$}
\FAVert(5.5,10.){0}
\FAVert(11.5,10.){0}
\FAVert(15.5,6.5){0}
\end{feynartspicture}

\begin{feynartspicture}(300,300)(1,1)
\FADiagram{}
\FAProp(0.,15.)(10.,13.)(0.,){/Straight}{1}
\FALabel(5.30398,15.0399)[b]{$q$}
\FAProp(0.,5.)(10.,5.5)(0.,){/Straight}{-1}
\FALabel(5.0774,4.18193)[t]{$q$}
\FAProp(20.,17.)(15.5,13.5)(0.,){/ScalarDash}{-1}
\FALabel(17.2784,15.9935)[br]{$\ba$}
\FAProp(20.,10.)(15.5,13.5)(0.,){/ScalarDash}{1}
\FALabel(18.2216,12.4935)[bl]{$\ba$}
\FAProp(20.,3.)(10.,5.5)(0.,){/Cycles}{0}
\FALabel(15.3759,5.27372)[b]{$g$}
\FAProp(10.,13.)(10.,5.5)(0.,){/Straight}{1}
\FALabel(8.93,9.25)[r]{$q$}
\FAProp(10.,13.)(15.5,13.5)(0.,){/Cycles}{0}
\FALabel(12.8903,12.1864)[t]{$g$}
\FAVert(10.,13.){0}
\FAVert(10.,5.5){0}
\FAVert(15.5,13.5){0}
\end{feynartspicture}

\begin{feynartspicture}(300,300)(1,1)
\FADiagram{}
\end{feynartspicture}
\\
\hspace*{95pt} (a)
\\
\begin{feynartspicture}(300,300)(1,1)
\FADiagram{}
\end{feynartspicture}


\begin{feynartspicture}(300,300)(1,1)
\FADiagram{}
\FAProp(0.,15.)(5.5,10.)(0.,){/Straight}{1}
\FALabel(2.18736,11.8331)[tr]{$q$}
\FAProp(0.,5.)(5.5,10.)(0.,){/Straight}{-1}
\FALabel(3.31264,6.83309)[tl]{$q$}
\FAProp(20.,17.)(12.5,10.)(0.,){/ScalarDash}{-1}
\FALabel(15.6724,14.1531)[br]{$\ba$}
\FAProp(20.,10.)(12.5,10.)(0.,){/ScalarDash}{1}
\FALabel(17.6,11.07)[b]{$\ba$}
\FAProp(20.,3.)(12.5,10.)(0.,){/Cycles}{0}
\FALabel(15.1948,5.33513)[tr]{$g$}
\FAProp(5.5,10.)(12.5,10.)(0.,){/Sine}{0}
\FALabel(9.,8.93)[t]{$V$}
\FAVert(5.5,10.){0}
\FAVert(12.5,10.){0}
\end{feynartspicture}

\begin{feynartspicture}(300,300)(1,1)
\FADiagram{}
\FAProp(0.,15.)(5.5,10.)(0.,){/Straight}{1}
\FALabel(2.18736,11.8331)[tr]{$q$}
\FAProp(0.,5.)(5.5,10.)(0.,){/Straight}{-1}
\FALabel(3.31264,6.83309)[tl]{$q$}
\FAProp(20.,17.)(11.5,10.)(0.,){/ScalarDash}{-1}
\FALabel(15.2447,14.2165)[br]{$\ba$}
\FAProp(20.,10.)(15.5,6.5)(0.,){/ScalarDash}{1}
\FALabel(17.2784,8.9935)[br]{$\ba$}
\FAProp(20.,3.)(15.5,6.5)(0.,){/Cycles}{0}
\FALabel(18.2216,5.4935)[bl]{$g$}
\FAProp(5.5,10.)(11.5,10.)(0.,){/Sine}{0}
\FALabel(8.5,11.07)[b]{$V$}
\FAProp(11.5,10.)(15.5,6.5)(0.,){/ScalarDash}{-1}
\FALabel(12.9593,7.56351)[tr]{$\ba$}
\FAVert(5.5,10.){0}
\FAVert(11.5,10.){0}
\FAVert(15.5,6.5){0}
\end{feynartspicture}

\begin{feynartspicture}(300,300)(1,1)
\FADiagram{}
\FAProp(0.,15.)(10.,13.)(0.,){/Straight}{1}
\FALabel(5.30398,15.0399)[b]{$q$}
\FAProp(0.,5.)(10.,5.5)(0.,){/Straight}{-1}
\FALabel(5.0774,4.18193)[t]{$q$}
\FAProp(20.,17.)(15.5,13.5)(0.,){/ScalarDash}{-1}
\FALabel(17.2784,15.9935)[br]{$\ba$}
\FAProp(20.,10.)(15.5,13.5)(0.,){/ScalarDash}{1}
\FALabel(18.2216,12.4935)[bl]{$\ba$}
\FAProp(20.,3.)(10.,5.5)(0.,){/Cycles}{0}
\FALabel(15.3759,5.27372)[b]{$g$}
\FAProp(10.,13.)(10.,5.5)(0.,){/Straight}{1}
\FALabel(8.93,9.25)[r]{$q$}
\FAProp(10.,13.)(15.5,13.5)(0.,){/Sine}{0}
\FALabel(12.8903,12.1864)[t]{$V$}
\FAVert(10.,13.){0}
\FAVert(10.,5.5){0}
\FAVert(15.5,13.5){0}
\end{feynartspicture}

\begin{feynartspicture}(300,300)(1,1)
\FADiagram{}
\end{feynartspicture}
\\(b)
\caption{Real gluon emission for the $q\bar{q}$ channel. (a) QCD
  based diagrams. (b) EW based diagrams.
  \label{fig:feynman_gluonRE}} }

\FIGURE{
\footnotesize\unitlength=0.26bp%
\begin{feynartspicture}(300,300)(1,1)
\FADiagram{}
\FAProp(0.,15.)(10.,5.5)(0.,){/Straight}{1}
\FALabel(2.89219,13.6512)[bl]{$q$}
\FAProp(0.,5.)(10.,14.)(0.,){/Cycles}{0}
\FALabel(2.98724,6.34418)[tl]{$g$}
\FAProp(20.,17.)(10.,14.)(0.,){/ScalarDash}{-1}
\FALabel(14.5546,16.5046)[b]{$\ba$}
\FAProp(20.,10.)(10.,14.)(0.,){/ScalarDash}{1}
\FALabel(15.5757,12.9591)[b]{$\ba$}
\FAProp(20.,3.)(10.,5.5)(0.,){/Straight}{-1}
\FALabel(15.3759,5.27372)[b]{$q$}
\FAProp(10.,5.5)(10.,14.)(0.,){/Cycles}{0}
\FALabel(11.07,9.75)[l]{$g$}
\FAVert(10.,5.5){0}
\FAVert(10.,14.){0}
\end{feynartspicture}

\begin{feynartspicture}(300,300)(1,1)
\FADiagram{}
\FAProp(0.,15.)(5.5,10.)(0.,){/Straight}{1}
\FALabel(2.18736,11.8331)[tr]{$q$}
\FAProp(0.,5.)(5.5,10.)(0.,){/Cycles}{0}
\FALabel(3.31264,6.83309)[tl]{$g$}
\FAProp(20.,17.)(15.5,13.5)(0.,){/ScalarDash}{-1}
\FALabel(17.2784,15.9935)[br]{$\ba$}
\FAProp(20.,10.)(15.5,13.5)(0.,){/ScalarDash}{1}
\FALabel(18.2216,12.4935)[bl]{$\ba$}
\FAProp(20.,3.)(12.,10.)(0.,){/Straight}{-1}
\FALabel(15.4593,5.81351)[tr]{$q$}
\FAProp(5.5,10.)(12.,10.)(0.,){/Straight}{1}
\FALabel(8.75,8.93)[t]{$q$}
\FAProp(15.5,13.5)(12.,10.)(0.,){/Cycles}{0}
\FALabel(13.134,12.366)[br]{$g$}
\FAVert(5.5,10.){0}
\FAVert(15.5,13.5){0}
\FAVert(12.,10.){0}
\end{feynartspicture}

\begin{feynartspicture}(300,300)(1,1)
\FADiagram{}
\FAProp(0.,15.)(10.,14.5)(0.,){/Straight}{1}
\FALabel(5.0774,15.8181)[b]{$q$}
\FAProp(0.,5.)(10.,5.5)(0.,){/Cycles}{0}
\FALabel(5.0774,4.18193)[t]{$g$}
\FAProp(20.,17.)(10.,10.)(0.,){/ScalarDash}{-1}
\FALabel(15.8366,15.2248)[br]{$\ba$}
\FAProp(20.,10.)(10.,5.5)(0.,){/ScalarDash}{1}
\FALabel(17.7693,10.1016)[br]{$\ba$}
\FAProp(20.,3.)(10.,14.5)(0.,){/Straight}{-1}
\FALabel(16.7913,4.81596)[tr]{$q$}
\FAProp(10.,14.5)(10.,10.)(0.,){/Cycles}{0}
\FALabel(8.93,12.25)[r]{$g$}
\FAProp(10.,5.5)(10.,10.)(0.,){/ScalarDash}{1}
\FALabel(8.93,7.75)[r]{$\ba$}
\FAVert(10.,14.5){0}
\FAVert(10.,5.5){0}
\FAVert(10.,10.){0}
\end{feynartspicture}

\begin{feynartspicture}(300,300)(1,1)
\FADiagram{}
\FAProp(0.,15.)(10.,5.5)(0.,){/Straight}{1}
\FALabel(3.19219,13.2012)[bl]{$q$}
\FAProp(0.,5.)(10.,13.)(0.,){/Cycles}{0}
\FALabel(3.17617,6.28478)[tl]{$g$}
\FAProp(20.,17.)(16.,13.5)(0.,){/ScalarDash}{-1}
\FALabel(17.4593,15.9365)[br]{$\ba$}
\FAProp(20.,10.)(16.,13.5)(0.,){/ScalarDash}{1}
\FALabel(17.4593,11.0635)[tr]{$\ba$}
\FAProp(20.,3.)(10.,5.5)(0.,){/Straight}{-1}
\FALabel(14.6241,3.22628)[t]{$q$}
\FAProp(10.,5.5)(10.,13.)(0.,){/Cycles}{0}
\FALabel(11.07,9.25)[l]{$g$}
\FAProp(10.,13.)(16.,13.5)(0.,){/Cycles}{0}
\FALabel(12.8131,15.0122)[b]{$g$}
\FAVert(10.,5.5){0}
\FAVert(10.,13.){0}
\FAVert(16.,13.5){0}
\end{feynartspicture}

\begin{feynartspicture}(300,300)(1,1)
\FADiagram{}
\FAProp(0.,15.)(10.,13.)(0.,){/Straight}{1}
\FALabel(5.30398,15.0399)[b]{$q$}
\FAProp(0.,5.)(10.,5.5)(0.,){/Cycles}{0}
\FALabel(5.0774,4.18193)[t]{$g$}
\FAProp(20.,17.)(15.5,13.5)(0.,){/ScalarDash}{-1}
\FALabel(17.2784,15.9935)[br]{$\ba$}
\FAProp(20.,10.)(15.5,13.5)(0.,){/ScalarDash}{1}
\FALabel(18.2216,12.4935)[bl]{$\ba$}
\FAProp(20.,3.)(10.,5.5)(0.,){/Straight}{-1}
\FALabel(15.3759,5.27372)[b]{$q$}
\FAProp(10.,13.)(10.,5.5)(0.,){/Straight}{1}
\FALabel(8.93,9.25)[r]{$q$}
\FAProp(10.,13.)(15.5,13.5)(0.,){/Cycles}{0}
\FALabel(12.8903,12.1864)[t]{$g$}
\FAVert(10.,13.){0}
\FAVert(10.,5.5){0}
\FAVert(15.5,13.5){0}
\end{feynartspicture}
\\
\hspace*{98pt}(a)
\\
\begin{feynartspicture}(300,300)(1,1)
\FADiagram{}
\FAProp(0.,15.)(10.,5.5)(0.,){/Straight}{1}
\FALabel(2.89219,13.6512)[bl]{$q$}
\FAProp(0.,5.)(10.,14.)(0.,){/Cycles}{0}
\FALabel(2.98724,6.34418)[tl]{$g$}
\FAProp(20.,17.)(10.,14.)(0.,){/ScalarDash}{-1}
\FALabel(14.5546,16.5046)[b]{$\ba$}
\FAProp(20.,10.)(10.,14.)(0.,){/ScalarDash}{1}
\FALabel(15.5757,12.9591)[b]{$\ba$}
\FAProp(20.,3.)(10.,5.5)(0.,){/Straight}{-1}
\FALabel(15.3759,5.27372)[b]{$q$}
\FAProp(10.,5.5)(10.,14.)(0.,){/Sine}{0}
\FALabel(11.07,9.75)[l]{$V$}
\FAVert(10.,5.5){0}
\FAVert(10.,14.){0}
\end{feynartspicture}

\begin{feynartspicture}(300,300)(1,1)
\FADiagram{}
\FAProp(0.,15.)(5.5,10.)(0.,){/Straight}{1}
\FALabel(2.18736,11.8331)[tr]{$q$}
\FAProp(0.,5.)(5.5,10.)(0.,){/Cycles}{0}
\FALabel(3.31264,6.83309)[tl]{$g$}
\FAProp(20.,17.)(15.5,13.5)(0.,){/ScalarDash}{-1}
\FALabel(17.2784,15.9935)[br]{$\ba$}
\FAProp(20.,10.)(15.5,13.5)(0.,){/ScalarDash}{1}
\FALabel(18.2216,12.4935)[bl]{$\ba$}
\FAProp(20.,3.)(12.,10.)(0.,){/Straight}{-1}
\FALabel(15.4593,5.81351)[tr]{$q$}
\FAProp(5.5,10.)(12.,10.)(0.,){/Straight}{1}
\FALabel(8.75,8.93)[t]{$q$}
\FAProp(15.5,13.5)(12.,10.)(0.,){/Sine}{0}
\FALabel(13.134,12.366)[br]{$V$}
\FAVert(5.5,10.){0}
\FAVert(15.5,13.5){0}
\FAVert(12.,10.){0}
\end{feynartspicture}

\begin{feynartspicture}(300,300)(1,1)
\FADiagram{}
\FAProp(0.,15.)(10.,14.5)(0.,){/Straight}{1}
\FALabel(5.0774,15.8181)[b]{$q$}
\FAProp(0.,5.)(10.,5.5)(0.,){/Cycles}{0}
\FALabel(5.0774,4.18193)[t]{$g$}
\FAProp(20.,17.)(10.,10.)(0.,){/ScalarDash}{-1}
\FALabel(15.8366,15.2248)[br]{$\ba$}
\FAProp(20.,10.)(10.,5.5)(0.,){/ScalarDash}{1}
\FALabel(17.7693,10.1016)[br]{$\ba$}
\FAProp(20.,3.)(10.,14.5)(0.,){/Straight}{-1}
\FALabel(16.7913,4.81596)[tr]{$q$}
\FAProp(10.,14.5)(10.,10.)(0.,){/Sine}{0}
\FALabel(8.93,12.25)[r]{$V$}
\FAProp(10.,5.5)(10.,10.)(0.,){/ScalarDash}{1}
\FALabel(8.93,7.75)[r]{$\ba$}
\FAVert(10.,14.5){0}
\FAVert(10.,5.5){0}
\FAVert(10.,10.){0}
\end{feynartspicture}

\begin{feynartspicture}(300,300)(1,1)
\FADiagram{}
\FAProp(0.,15.)(10.,13.)(0.,){/Straight}{1}
\FALabel(5.30398,15.0399)[b]{$q$}
\FAProp(0.,5.)(10.,5.5)(0.,){/Cycles}{0}
\FALabel(5.0774,4.18193)[t]{$g$}
\FAProp(20.,17.)(15.5,13.5)(0.,){/ScalarDash}{-1}
\FALabel(17.2784,15.9935)[br]{$\ba$}
\FAProp(20.,10.)(15.5,13.5)(0.,){/ScalarDash}{1}
\FALabel(18.2216,12.4935)[bl]{$\ba$}
\FAProp(20.,3.)(10.,5.5)(0.,){/Straight}{-1}
\FALabel(15.3759,5.27372)[b]{$q$}
\FAProp(10.,13.)(10.,5.5)(0.,){/Straight}{1}
\FALabel(8.93,9.25)[r]{$q$}
\FAProp(10.,13.)(15.5,13.5)(0.,){/Sine}{0}
\FALabel(12.8903,12.1864)[t]{$V$}
\FAVert(10.,13.){0}
\FAVert(10.,5.5){0}
\FAVert(15.5,13.5){0}
\end{feynartspicture}

\begin{feynartspicture}(300,300)(1,1)
\FADiagram{}
\end{feynartspicture}
\\(b)
\caption{Feynman diagrams contributing to real quark radiation. (a)
  QCD based diagrams. (b) EW based diagrams. 
  \label{fig:feynman_quarkRE}}
}

\clearpage

\bibliographystyle{JHEP}
\bibliography{references}

\providecommand{\href}[2]{#2}\begingroup\raggedright\begin{thebibliography}{10}

\bibitem{0954-3899-34-6-S01}
{\bf CMS} Collaboration, {\it Cms physics technical design report, volume ii:
  Physics performance},  {\em Journal of Physics G: Nuclear and Particle
  Physics} {\bf 34} (2007), no.~6 995--1579.

\bibitem{:2008zzm}
{\bf ATLAS} Collaboration, G.~Aad {\em et.~al.}, {\it {The ATLAS Experiment at
  the CERN Large Hadron Collider}},  {\em JINST} {\bf 3} (2008) S08003.

\bibitem{Atlas1}
{\bf ATLAS} Collaboration, {\it {Background studies to searches for long-lived
  stopped particles decaying out-of-time with LHC collisions}},  {\em
  ATLAS-CONF-2010-071}.

\bibitem{Atlas2}
{\bf ATLAS} Collaboration, {\it {Early supersymmetry searches with jets,
  missing transverse momentum and one or more leptons with the ATLAS
  Detector}},  {\em ATLAS-CONF-2010-066}.

\bibitem{Atlas3}
{\bf ATLAS} Collaboration, {\it {Early supersymmetry searches in channels with
  jets and missing transverse momentum with the ATLAS detector}},  {\em
  ATLAS-CONF-2010-065}.

\bibitem{Atlas4}
{\bf ATLAS} Collaboration, {\it {Early supersymmetry searches in events with
  missing transverse energy and b-jets with the ATLAS detector}},  {\em
  ATLAS-CONF-2010-079}.

\bibitem{CMS1}
{\bf CMS} Collaboration, {\it {Performance of Methods for Data-Driven
  Background Estimation in SUSY Searches}},  {\em CMS PAS SUS-10-001}.

\bibitem{Kane:1982hw}
G.~L. Kane and J.~P. Leveille, {\it {Experimental constraints on gluino masses
  and supersymmetric theories}},  {\em Phys. Lett.} {\bf B112} (1982) 227.

\bibitem{Harrison:1982yi}
P.~R. Harrison and C.~H. Llewellyn~Smith, {\it {Hadroproduction of
  supersymmetric particles}},  {\em Nucl. Phys.} {\bf B213} (1983) 223.

\bibitem{Reya:1984yz}
E.~Reya and D.~P. Roy, {\it {Supersymmetric particle production at $p~\bar{p}$
  collider energies}},  {\em Phys. Rev.} {\bf D32} (1985) 645.

\bibitem{Dawson:1983fw}
S.~Dawson, E.~Eichten, and C.~Quigg, {\it {Search for supersymmetric particles
  in hadron--hadron collisions}},  {\em Phys. Rev.} {\bf D31} (1985) 1581.

\bibitem{Baer:1985xz}
H.~Baer and X.~Tata, {\it {Component formulae for hadroproduction of
  left-handed and right-handed squarks}},  {\em Phys. Lett.} {\bf B160} (1985)
  159.

\bibitem{Beenakker:1996ch}
W.~Beenakker, R.~Hopker, M.~Spira, and P.~M. Zerwas, {\it {Squark and gluino
  production at hadron colliders}},  {\em Nucl. Phys.} {\bf B492} (1997)
  51--103, [\href{http://xxx.lanl.gov/abs/hep-ph/9610490}{{\tt
  hep-ph/9610490}}].

\bibitem{Beenakker:1997ut}
W.~Beenakker, M.~Kramer, T.~Plehn, M.~Spira, and P.~M. Zerwas, {\it {Stop
  production at hadron colliders}},  {\em Nucl. Phys.} {\bf B515} (1998) 3--14,
  [\href{http://xxx.lanl.gov/abs/hep-ph/9710451}{{\tt hep-ph/9710451}}].

\bibitem{Beenakker:1996ed}
W.~Beenakker, R.~Hopker, and M.~Spira, {\it {PROSPINO: A program for the
  PROduction of Supersymmetric Particles In Next-to-leading Order QCD}},
  \href{http://xxx.lanl.gov/abs/hep-ph/9611232}{{\tt hep-ph/9611232}}.

\bibitem{Langenfeld:2009eg}
U.~Langenfeld and S.-O. Moch, {\it {Higher-order soft corrections to squark
  hadro- production}},  {\em Phys. Lett.} {\bf B675} (2009) 210--221,
  [\href{http://xxx.lanl.gov/abs/0901.0802}{{\tt arXiv:0901.0802}}].

\bibitem{Kulesza:2008jb}
A.~Kulesza and L.~Motyka, {\it {Threshold resummation for squark-antisquark and
  gluino- pair production at the LHC}},  {\em Phys. Rev. Lett.} {\bf 102}
  (2009) 111802, [\href{http://xxx.lanl.gov/abs/0807.2405}{{\tt
  arXiv:0807.2405}}].

\bibitem{Kulesza:2009kq}
A.~Kulesza and L.~Motyka, {\it {Soft gluon resummation for the production of
  gluino-gluino and squark-antisquark pairs at the LHC}},  {\em Phys. Rev.}
  {\bf D80} (2009) 095004, [\href{http://xxx.lanl.gov/abs/0905.4749}{{\tt
  arXiv:0905.4749}}].

\bibitem{Beneke:2009nr}
M.~Beneke, P.~Falgari, and C.~Schwinn, {\it {Colour structure in threshold
  resummation and squark-antisquark production at NLL}},  {\em PoS} {\bf
  EPS-HEP2009} (2009) 319, [\href{http://xxx.lanl.gov/abs/0909.3488}{{\tt
  arXiv:0909.3488}}].

\bibitem{Beenakker:2009ha}
W.~Beenakker {\em et.~al.}, {\it {Soft-gluon resummation for squark and gluino
  hadroproduction}},  {\em JHEP} {\bf 12} (2009) 041,
  [\href{http://xxx.lanl.gov/abs/0909.4418}{{\tt arXiv:0909.4418}}].

\bibitem{Beenakker:2010nq}
W.~Beenakker {\em et.~al.}, {\it {Supersymmetric top and bottom squark
  production at hadron colliders}},  {\em JHEP} {\bf 08} (2010) 098,
  [\href{http://xxx.lanl.gov/abs/1006.4771}{{\tt arXiv:1006.4771}}].

\bibitem{Beneke:2010da}
M.~Beneke, P.~Falgari, and C.~Schwinn, {\it {Threshold resummation for pair
  production of coloured heavy (s)particles at hadron colliders}},  {\em Nucl.
  Phys.} {\bf B842} (2010) [\href{http://xxx.lanl.gov/abs/1007.5414}{{\tt
  arXiv:1007.5414}}].

\bibitem{Hollik:2007wf}
W.~Hollik, M.~Kollar, and M.~K. Trenkel, {\it {Hadronic production of
  top-squark pairs with electroweak NLO contributions}},  {\em JHEP} {\bf 02}
  (2008) 018, [\href{http://xxx.lanl.gov/abs/0712.0287}{{\tt
  arXiv:0712.0287}}].

\bibitem{Bornhauser:2007bf}
S.~Bornhauser, M.~Drees, H.~K. Dreiner, and J.~S. Kim, {\it {Electroweak
  contributions to squark pair production at the LHC}},  {\em Phys. Rev.} {\bf
  D76} (2007) 095020, [\href{http://xxx.lanl.gov/abs/0709.2544}{{\tt
  arXiv:0709.2544}}].

\bibitem{Hollik:2008vm}
W.~Hollik, E.~Mirabella, and M.~K. Trenkel, {\it {Electroweak contributions to
  squark--gluino production at the LHC}},  {\em JHEP} {\bf 02} (2009) 002,
  [\href{http://xxx.lanl.gov/abs/0810.1044}{{\tt arXiv:0810.1044}}].

\bibitem{Bornhauser:2009ru}
S.~Bornhauser, M.~Drees, H.~K. Dreiner, and J.~S. Kim, {\it {Rapidity gap
  events in squark pair production at the LHC}},  {\em Phys. Rev.} {\bf D80}
  (2009) 095007, [\href{http://xxx.lanl.gov/abs/0909.2595}{{\tt
  arXiv:0909.2595}}].

\bibitem{Arhrib:2009sb}
A.~Arhrib, R.~Benbrik, K.~Cheung, and T.-C. Yuan, {\it {Higgs boson enhancement
  effects on squark-pair production at the LHC}},  {\em JHEP} {\bf 02} (2010)
  048, [\href{http://xxx.lanl.gov/abs/0911.1820}{{\tt arXiv:0911.1820}}].

\bibitem{Germer:2010vn}
J.~Germer, W.~Hollik, E.~Mirabella, and M.~K. Trenkel, {\it {Hadronic
  production of squark-squark pairs: The electroweak contributions}},  {\em
  JHEP} {\bf 08} (2010) 023, [\href{http://xxx.lanl.gov/abs/1004.2621}{{\tt
  arXiv:1004.2621}}].

\bibitem{Bozzi:2005sy}
G.~Bozzi, B.~Fuks, and M.~Klasen, {\it {Non-diagonal and mixed squark
  production at hadron colliders}},  {\em Phys. Rev.} {\bf D72} (2005) 035016,
  [\href{http://xxx.lanl.gov/abs/hep-ph/0507073}{{\tt hep-ph/0507073}}].

\bibitem{Alan:2007rp}
A.~T. Alan, K.~Cankocak, and D.~A. Demir, {\it {Squark pair production in the
  MSSM with explicit CP violation}},  {\em Phys. Rev.} {\bf D75} (2007) 095002,
  [\href{http://xxx.lanl.gov/abs/hep-ph/0702289}{{\tt hep-ph/0702289}}].

\bibitem{Beccaria:2008mi}
M.~Beccaria, G.~Macorini, L.~Panizzi, F.~M. Renard, and C.~Verzegnassi, {\it
  {Stop-antistop and sbottom-antisbottom production at LHC: a one-loop search
  for model parameters dependence}},  {\em Int. J. Mod. Phys.} {\bf A23} (2008)
  4779--4810, [\href{http://xxx.lanl.gov/abs/0804.1252}{{\tt
  arXiv:0804.1252}}].

\bibitem{Hollik:2008yi}
W.~Hollik and E.~Mirabella, {\it {Squark anti-squark pair production at the
  LHC: the electroweak contribution}},  {\em JHEP} {\bf 12} (2008) 087,
  [\href{http://xxx.lanl.gov/abs/0806.1433}{{\tt arXiv:0806.1433}}].

\bibitem{Mirabella:2009ap}
E.~Mirabella, {\it {NLO electroweak contributions to gluino pair production at
  hadron colliders}},  {\em JHEP} {\bf 12} (2009) 012,
  [\href{http://xxx.lanl.gov/abs/0908.3318}{{\tt arXiv:0908.3318}}].

\bibitem{Paige:2003sh}
F.~E. Paige, {\it {SUSY signatures in ATLAS at LHC}},
  \href{http://xxx.lanl.gov/abs/hep-ph/0307342}{{\tt hep-ph/0307342}}.

\bibitem{Chiorboli:2004jk}
M.~Chiorboli and A.~Tricomi, {\it {Squark and gluino reconstruction in CMS}},
  {\em CMS-NOTE-2004-029}.

\bibitem{Kawagoe:2004rz}
K.~Kawagoe, M.~M. Nojiri, and G.~Polesello, {\it {A new SUSY mass
  reconstruction method at the CERN LHC}},  {\em Phys. Rev.} {\bf D71} (2005)
  035008, [\href{http://xxx.lanl.gov/abs/hep-ph/0410160}{{\tt
  hep-ph/0410160}}].

\bibitem{Heinemeyer:2004xw}
S.~Heinemeyer, W.~Hollik, H.~Rzehak, and G.~Weiglein, {\it {High-precision
  predictions for the MSSM Higgs sector at $\mathcal{O}(\alpha_b \alpha_s)$}},
  {\em Eur. Phys. J.} {\bf C39} (2005) 465--481,
  [\href{http://xxx.lanl.gov/abs/hep-ph/0411114}{{\tt hep-ph/0411114}}].

\bibitem{Heinemeyer:2010mm}
S.~Heinemeyer, H.~Rzehak, and C.~Schappacher, {\it {Proposals for bottom
  quark/squark renormalization in the complex MSSM}},
  \href{http://xxx.lanl.gov/abs/1007.0689}{{\tt arXiv:1007.0689}}.

\bibitem{Carena:1999py}
M.~S. Carena, D.~Garcia, U.~Nierste, and C.~E.~M. Wagner, {\it {Effective
  Lagrangian for the $\bar{t} b H^{+}$ interaction in the MSSM and charged
  Higgs phenomenology}},  {\em Nucl. Phys.} {\bf B577} (2000) 88--120,
  [\href{http://xxx.lanl.gov/abs/hep-ph/9912516}{{\tt hep-ph/9912516}}].

\bibitem{Hahn:2000kx}
T.~Hahn, {\it {Generating Feynman diagrams and amplitudes with FeynArts 3}},
  {\em Comput. Phys. Commun.} {\bf 140} (2001) 418--431,
  [\href{http://xxx.lanl.gov/abs/hep-ph/0012260}{{\tt hep-ph/0012260}}].

\bibitem{Hahn:2006qw}
T.~Hahn and M.~Rauch, {\it {News from FormCalc and LoopTools}},  {\em Nucl.
  Phys. Proc. Suppl.} {\bf 157} (2006) 236--240,
  [\href{http://xxx.lanl.gov/abs/hep-ph/0601248}{{\tt hep-ph/0601248}}].

\bibitem{Hahn:2001rv}
T.~Hahn and C.~Schappacher, {\it {The implementation of the minimal
  supersymmetric standard model in FeynArts and FormCalc}},  {\em Comput. Phys.
  Commun.} {\bf 143} (2002) 54--68,
  [\href{http://xxx.lanl.gov/abs/hep-ph/0105349}{{\tt hep-ph/0105349}}].

\bibitem{Baur:1998kt}
U.~Baur, S.~Keller, and D.~Wackeroth, {\it {Electroweak radiative corrections
  to $W$ boson production in hadronic collisions}},  {\em Phys. Rev.} {\bf D59}
  (1999) 013002, [\href{http://xxx.lanl.gov/abs/hep-ph/9807417}{{\tt
  hep-ph/9807417}}].

\bibitem{Hall:1993gn}
L.~J. Hall, R.~Rattazzi, and U.~Sarid, {\it {The top quark mass in
  supersymmetric SO(10) unification}},  {\em Phys. Rev.} {\bf D50} (1994)
  7048--7065, [\href{http://xxx.lanl.gov/abs/hep-ph/9306309}{{\tt
  hep-ph/9306309}}].

\bibitem{Hempfling:1993kv}
R.~Hempfling, {\it {Yukawa coupling unification with supersymmetric threshold
  corrections}},  {\em Phys. Rev.} {\bf D49} (1994) 6168--6172.

\bibitem{Carena:1994bv}
M.~S. Carena, M.~Olechowski, S.~Pokorski, and C.~E.~M. Wagner, {\it
  {Electroweak symmetry breaking and bottom - top Yukawa unification}},  {\em
  Nucl. Phys.} {\bf B426} (1994) 269--300,
  [\href{http://xxx.lanl.gov/abs/hep-ph/9402253}{{\tt hep-ph/9402253}}].

\bibitem{Pierce:1996zz}
D.~M. Pierce, J.~A. Bagger, K.~T. Matchev, and R.-j. Zhang, {\it {Precision
  corrections in the minimal supersymmetric standard model}},  {\em Nucl.
  Phys.} {\bf B491} (1997) 3--67,
  [\href{http://xxx.lanl.gov/abs/hep-ph/9606211}{{\tt hep-ph/9606211}}].

\bibitem{Nakamura:2010zzi}
{\bf Particle Data Group} Collaboration, K.~Nakamura {\em et.~al.}, {\it
  {Review of particle physics}},  {\em J. Phys.} {\bf G37} (2010) 075021.

\bibitem{CDFtopmass}
{\bf Tevatron Electroweak Working Group} Collaboration, {\it {Combination of
  CDF and D0 results on the mass of the top quark}},  2009.

\bibitem{AguilarSaavedra:2005pw}
J.~A. Aguilar-Saavedra {\em et.~al.}, {\it {Supersymmetry parameter analysis:
  SPA convention and project}},  {\em Eur. Phys. J.} {\bf C46} (2006) 43--60,
  [\href{http://xxx.lanl.gov/abs/hep-ph/0511344}{{\tt hep-ph/0511344}}].

\bibitem{Allanach:2001kg}
B.~C. Allanach, {\it {SOFTSUSY: A C++ program for calculating supersymmetric
  spectra}},  {\em Comput. Phys. Commun.} {\bf 143} (2002) 305--331,
  [\href{http://xxx.lanl.gov/abs/hep-ph/0104145}{{\tt hep-ph/0104145}}].

\bibitem{Martin:2004dh}
A.~D. Martin, R.~G. Roberts, W.~J. Stirling, and R.~S. Thorne, {\it {Parton
  distributions incorporating QED contributions}},  {\em Eur. Phys. J.} {\bf
  C39} (2005) 155--161, [\href{http://xxx.lanl.gov/abs/hep-ph/0411040}{{\tt
  hep-ph/0411040}}].

\bibitem{Stockinger:2006zn}
D.~Stockinger, {\it {The muon magnetic moment and supersymmetry}},  {\em J.
  Phys.} {\bf G34} (2007) R45--R92,
  [\href{http://xxx.lanl.gov/abs/hep-ph/0609168}{{\tt hep-ph/0609168}}].

\bibitem{Kniehl:2002wn}
B.~A. Kniehl, C.~P. Palisoc, and A.~Sirlin, {\it {Elimination of threshold
  singularities in the relation between on-shell and pole widths}},  {\em Phys.
  Rev.} {\bf D66} (2002) 057902,
  [\href{http://xxx.lanl.gov/abs/hep-ph/0205304}{{\tt hep-ph/0205304}}].

\end{thebibliography}\endgroup

\end{document}